\documentclass[twocolumn]{aastex631} 

\usepackage{amsmath}	
\usepackage{amssymb}	
\usepackage{longtable}
\usepackage{scalerel}
\usepackage{hyperref}

\mathchardef\mhyphen="2D

\newcommand{\oii}{O\,{\sc ii}}
\newcommand{\oiii}{O\,{\sc iii}}

\newcommand{\siiv}{Si\,{\sc iv}}

\newcommand{\cii}{[C\,{\sc ii}]}

\newcommand{\mgii}{Mg\,{\sc ii}}

\mathchardef\mhyphen="2D

\def\lya{Ly$\alpha$}
\def\ly{$\lambda$}
\def\ha{H$\alpha$}
\def\hb{H$\beta$}
\def\hg{H$\gamma$}

\def\hii{H\,{\sc ii}}

\def\cii{C\,{\sc ii}}

\def\nii{N\,{\sc ii}}

\def\oiii{O\,{\sc iii}}

\def\nai{Na\,{\sc i}}

\def\mgii{Mg\,{\sc ii}}

\def\siiv{Si\,{\sc iv}}
\def\Siii{Si\,{\sc ii}}

\def\siiii{Si\,{\sc iii}}

\def\sii{S\,{\sc ii}}

\def\Q0059{Q0059--2735}

\def\S2S3{S2S3}

\definecolor{blk}{rgb}{0.0,0.0,0.0}
\definecolor{red}{rgb}{0.75,0.0,0.0}
\definecolor{yel}{rgb}{0.65,0.65,0.0}
\definecolor{grn}{rgb}{0.0,0.75,0.0}
\definecolor{blu}{rgb}{0.0,0.0,0.75}
\definecolor{gry}{rgb}{0.75,0.75,0.75}

\def\nh{\ifmmode n_\mathrm{\scriptscriptstyle H} \else $n_\mathrm{\scriptscriptstyle H}$\fi}
\def\ne{\ifmmode n_\mathrm{\scriptstyle e} \else $n_\mathrm{\scriptstyle e}$\fi}
\def\Te{\ifmmode T_\mathrm{\scriptstyle e} \else $T_\mathrm{\scriptstyle e}$\fi}
\def\Qh{\ifmmode Q_\mathrm{\scriptstyle H} \else $Q_\mathrm{\scriptstyle H}$\fi}
\def\Uh{\ifmmode U_\mathrm{\scriptstyle H} \else $U_\mathrm{\scriptstyle H}$\fi}
\def\Nh{\ifmmode N_\mathrm{\scriptstyle H} \else $N_\mathrm{\scriptstyle H}$\fi}
\def\Nhi{\ifmmode N_\mathrm{\scriptstyle HI} \else $N_\mathrm{\scriptstyle HI}$\fi}
\def\Uhhp{\ifmmode U_\mathrm{\scriptstyle H,HP} \else $U_\mathrm{\scriptstyle H,HP}$\fi}
\def\Nhhp{\ifmmode N_\mathrm{\scriptstyle H,HP} \else $N_\mathrm{\scriptstyle H,HP}$\fi}
\def\Uhvhp{\ifmmode U_\mathrm{\scriptstyle H,VHP} \else $U_\mathrm{\scriptstyle H,VHP}$\fi}
\def\Nhvhp{\ifmmode N_\mathrm{\scriptstyle H,VHP} \else $N_\mathrm{\scriptstyle H,VHP}$\fi}
\def\Nion{\ifmmode N_\mathrm{\scriptstyle ion} \else $N_\mathrm{\scriptstyle ion}$\fi}

\def\Zsun{\ifmmode {\rm Z}_{\odot} \else $Z_{\odot}$\fi}
\def\Msun{\ifmmode {\rm M}_{\odot} \else M$_{\odot}$\fi}
\def\kms{\ifmmode {\rm km~s}^{-1} \else km~s$^{-1}$\fi}
\def\Lya{\ifmmode {\rm Ly}\alpha \else Ly$\alpha$\fi}
\def\Lyb{\ifmmode {\rm Ly}\beta \else Ly$\beta$\fi}
\def\Lyg{\ifmmode {\rm Ly}\gamma \else Ly$\gamma$\fi}
\def\Lyd{\ifmmode {\rm Ly}\delta \else Ly$\delta$\fi}
\def\neaod{\ifmmode n_\mathrm{\scriptscriptstyle AOD} \else $n_\mathrm{\scriptscriptstyle AOD}$\fi}
\def\necrit{\ifmmode n_\mathrm{\scriptstyle cr} \else $n_\mathrm{\scriptstyle cr}$\fi}
\def\ncr{\ifmmode n_\mathrm{\scriptstyle cr} \else $n_\mathrm{\scriptstyle cr}$\fi}
\def\nepi{\ifmmode n_\mathrm{\scriptscriptstyle PI} \else $n_\mathrm{\scriptscriptstyle PI}$\fi}
\def\gtorder{\mathrel{\raise.3ex\hbox{$>$}\mkern-14mu\lower0.6ex\hbox{$\sim$}}}
\def\ltorder{\mathrel{\raise.3ex\hbox{$<$}\mkern-14mu\lower0.6ex\hbox{$\sim$}}}

\def\vro{\ifmmode v_\mathrm{\scriptscriptstyle 1, \scriptstyle r} \else $v_\mathrm{\scriptscriptstyle 1, \scriptstyle r}$\fi}
\def\vrc{\ifmmode v_\mathrm{\scriptscriptstyle 2, \scriptstyle r} \else $v_\mathrm{\scriptscriptstyle 2, \scriptstyle r}$\fi}
\def\vzo{\ifmmode v_\mathrm{\scriptscriptstyle 1, \scriptstyle z} \else $v_\mathrm{\scriptscriptstyle 1, \scriptstyle z}$\fi}
\def\vzc{\ifmmode v_\mathrm{\scriptscriptstyle 2, \scriptstyle z} \else $v_\mathrm{\scriptscriptstyle 2, \scriptstyle z}$\fi}

\newcommand{\dotarcsec}{\rlap{.}\arcsec}
\newcommand{\FiCUS}{\textit{FICUS}}
\newcommand{\Prospector}{\textit{Prospector}}

\newcommand{\AB}{$A_{\text{b}}$}
\newcommand{\AN}{$A_{\text{n}}$}
\newcommand{\VC}{$v_{\text{c}}$}
\newcommand{\VCB}{$v_{\text{c,b}}$}
\newcommand{\VCN}{$v_{\text{c,n}}$}
\newcommand{\FWHM}{FWHM}
\newcommand{\HWHM}{HWHM}
\newcommand{\FWHMB}{FWHM$_{\text{b}}$}
\newcommand{\FWHMN}{FWHM$_{\text{n}}$}
\newcommand{\Vout}{$v^\text{out}$}
\newcommand{\Vaout}{$v_{a}^\text{out}$}

\newcommand{\Vnfout}{$v_{95}^\text{out}$}
\newcommand{\Vnf}{$v_{95}$}

\newcommand{\Asy}{$A_\text{90}$}

\newcommand{\Vnfoutha}{$v_{95, \text{H}\alpha}^\text{out}$}

\newcommand{\FWHMout}{FWHM$^\text{out}$}

\newcommand{\HWHMout}{HWHM$^\text{out}$}

\newcommand{\HWHMoutHa}{HWHM$_{\text{H}\alpha}^\text{out}$}

\newcommand{\Mstar}{M$_{\star}$}

\newcommand{\Mdot}{$\dot{M}^\text{out}$}
\newcommand{\MdotUV}{$\dot{M}^\text{out}_\text{\tiny UV}$}
\newcommand{\MdotOPT}{$\dot{M}^\text{out}_\text{\tiny OPT}$}
\newcommand{\MdotOut}{$\dot{M}^\text{out}$}

\newcommand{\rburst}{$r_{\star}$}
\newcommand{\rhalf}{r$_{50}$}

\newcommand{\Rout}{$R^\text{out}$}
\newcommand{\RoutUV}{$R^\text{out}_\text{\tiny UV}$}
\newcommand{\RoutOPT}{$R^\text{out}_\text{\tiny OPT}$}
\newcommand{\EBVint}{$E(B-V)_\text{int.}$}

\newcommand{\Newman}{N12}
\newcommand{\Freeman}{F19}
\newcommand{\SNR}{\textit{SNR}}

\def\ZStar{\ifmmode {\rm Z}_\text{stars} \else $Z_\text{stars}$\fi}
\def\ZGas{\ifmmode {\rm Z}_\text{gas} \else $Z_\text{gas}$\fi}

\revised{\today}
\shorttitle{Connections of Outflows seen in Absorption and Emission}
\shortauthors{Xu et al.}
\graphicspath{{./}{figures/}}

\begin{document}

\title{Shining a Light on the Connections between Galactic Outflows Seen in Absorption and Emission Lines}

\author[0000-0002-9217-7051]{Xinfeng Xu}
\affiliation{Department of Physics and Astronomy, Northwestern University,
2145 Sheridan Road, Evanston, IL, 60208, USA.}
\affiliation{Center for Interdisciplinary Exploration and Research in
Astrophysics (CIERA), 1800 Sherman Avenue,
Evanston, IL, 60201, USA.}
\correspondingauthor{Xinfeng Xu} 
\email{xinfeng.xu@northwestern.edu}

\author[0000-0002-6586-4446]{Alaina Henry}
\affiliation{Center for Astrophysical Sciences, Department of Physics \& Astronomy, Johns Hopkins University, Baltimore, MD 21218, USA}
\affiliation{Space Telescope Science Institute, 3700 San Martin Drive, Baltimore, MD 21218, USA}

\author[0000-0001-6670-6370]{Timothy Heckman}
\affiliation{Center for Astrophysical Sciences, Department of Physics \& Astronomy, Johns Hopkins University, Baltimore, MD 21218, USA}
\affiliation{School of Earth and Space Exploration, Arizona State University, Tempe, AZ 85287, USA}

\author[0000-0003-4166-2855]{Cody Carr}
\affiliation{Center for Cosmology and Computational Astrophysics, Institute for Advanced Study in Physics, Zhejiang University, Hangzhou 310058, China}
\affiliation{Institute of Astronomy, School of Physics, Zhejiang University, Hangzhou 310058, China}

\author[0000-0001-6369-1636]{Allison L. Strom}
\affiliation{Department of Physics and Astronomy, Northwestern University,
2145 Sheridan Road, Evanston, IL, 60208, USA.}
\affiliation{Center for Interdisciplinary Exploration and Research in
Astrophysics (CIERA), 1800 Sherman Avenue,
Evanston, IL, 60201, USA.}

\author[0000-0001-5860-3419]{Tucker Jones}
\affiliation{Department of Physics and Astronomy, University of California Davis, 1 Shields Avenue, Davis, CA 95616, USA}

\author[0000-0002-4153-053X]{Danielle A. Berg}
\affiliation{Department of Astronomy, The University of Texas at Austin, 2515 Speedway, Stop C1400, Austin, TX 78712, USA}

\author[0000-0002-0302-2577]{John Chisholm}
\affiliation{Department of Astronomy, The University of Texas at Austin, 2515 Speedway, Stop C1400, Austin, TX 78712, USA}

\author[0000-0001-9714-2758]{Dawn Erb}
\affiliation{The Leonard E. Parker Center for Gravitation, Cosmology and Astrophysics, Department of Physics, University of Wisconsin-Milwaukee, 3135 N Maryland Avenue, Milwaukee, WI 53211, USA}

\author[0000-0003-4372-2006]{Bethan L. James}
\affiliation{AURA for ESA, Space Telescope Science Institute, 3700 San Martin Drive, Baltimore, MD 21218, USA}

\author[0000-0002-6790-5125]{Anne Jaskot}
\affiliation{Department of Astronomy, Williams College, Williamstown, MA 01267, United States}

\author[0000-0001-9189-7818]{Crystal L. Martin}
\affiliation{Department of Physics, University of California, Santa Barbara, Santa Barbara, CA 93106, USA}

\author[0000-0003-2589-762X]{Matilde Mingozzi}
\affiliation{Space Telescope Science Institute, 3700 San Martin Drive, Baltimore, MD 21218, USA}

\author[0000-0002-9132-6561]{Peter Senchyna}
\affiliation{The Observatories of the Carnegie Institution for Science, 813 Santa Barbara Street, Pasadena, CA 91101, USA}

\author[0000-0002-4430-8846]{Namrata Roy}
\affiliation{Center for Astrophysical Sciences, Department of Physics \& Astronomy, Johns Hopkins University, Baltimore, MD 21218, USA}

\author[0000-0002-9136-8876]{Claudia Scarlata}
\affiliation{Minnesota Institute for Astrophysics, University of Minnesota, 116 Church Street SE, Minneapolis, MN 55455, USA}

\author[0000-0001-6106-5172]{Daniel P. Stark}
\affiliation{Steward Observatory, The University of Arizona, 933 N Cherry Ave, Tucson, AZ, 85721, USA}

\begin{abstract}

Galactic outflows provide important feedback effects to regulate the evolution of host galaxies. Two primary diagnostics of outflows are broad and/or blueshifted emission and absorption lines. Even though well-established methods exist to analyze these outflow signatures, connections between them are rarely studied and largely unknown. In this paper, we conduct such a study in a sample of 33 low-redshift starburst galaxies. Their UV absorption lines are detected by Hubble Space Telescope, and optical emission lines are observed by Keck or Very Large Telescope. We find outflow properties derived from emission and absorption lines are tightly correlated. These include outflow maximum velocity, line width, and radial extent. On average, in the same galaxy, the maximum velocity and line width of outflows measured from emission lines reach only 60 -- 70\% of those from the absorption lines. We also find outflow rates derived from emission lines are consistently lower than those from absorption lines by 0.2 -- 0.5 dex. These findings can be explained by a radial decline in density and a corresponding increase in outflow velocity, combined with the fact that emission line luminosity scales with the square of the density while absorption line depth scales linearly. We test both spherical and bi-conical outflow models, and find the same radial outflow velocity and density distributions can explain the observed correlations. These results provide novel calibration between galactic outflow properties measured from the two diagnostics and underscore the need for high‑fidelity UV and optical spectra to accurately assess galactic feedback effects in high-z galaxies.




\end{abstract}



\section{Introduction} \label{sec:intro}
Galactic-scale winds, driven by supernovae ejecta, hot stellar winds, and radiation pressure, have been proposed to explain various feedback effects. These include regulating the star formation (SF) inside the galaxy, enriching the intergalactic and circumgalactic medium with heavy metals, and explaining the ``overcooling problem" in cosmological simulations by reducing the baryon fractions in galactic discs \citep[see reviews in, e.g.,][]{Naab17, Heckman23}.

Galactic winds have been extensively studied in the low-redshift universe 
\citep[z $\lesssim$ 1, e.g.,][]{Heckman00, Martin05, Rupke13, Rubin14, Heckman15, Chisholm16a, Chisholm16b, Chisholm17, Sugahara17, Davis23, Guo23, Xu22a, Xu23c,Xu23a, Amorin24, Fisher24, Reichardt24} 
and are found to be more ubiquitous in galaxies at higher redshift 
\citep[e.g.,][]{Steidel10, Newman12a, Newman12b, Davies19, Freeman19, Avery22, Marasco23, Perrotta23, Zhang24, Kehoe24}.
Two major diagnostics of galactic winds are the blue-shifted absorption lines seen mostly in ultraviolet (UV) spectra and broad emission lines found in the optical and infrared spectra. For both types of spectroscopic features, well-established methods exist for deriving important galactic wind properties. These include the velocity gradient and velocity dispersion of the outflows, how dense the outflows are, and how much mass/energy/momentum is carried out by the outflows \citep[see reviews in, e.g.,][]{Veilleux20, Thompson24}. Hereafter, we focus on the warm ionized outflows (T $\sim$ 10,000 K). For this phase of the gas, the main outflow diagnostics are absorption lines studied in rest-UV, including \cii\ \ly 1334, the \Siii\ multiplet (\ly1190, 1193, 1260, 1304, and 1526), \siiii\ \ly 1206, and \siiv\ \ly\ly 1393, 1402, and emission lines studied in rest-optical, including [\nii] \ly\ly 6548, 6583, [\oii] \ly\ly 3727, 3729, [\oiii] \ly\ly 4959, 5007, [\sii] \ly\ly 6717, 6731, and the Balmer lines.

Even with the wide number of existing studies, it remains to be seen how the outflow measurements from emission and absorption features are related. Since emission line luminosity are weighted linearly by gas density squared ($n^{2}$) and absorption line column density are linearly weighted by $n$, these two types of outflow measurements should be sensitive to different parts of the outflow. Furthermore, given the fact that these emission and absorption lines likely trace the same warm-ionized gases in a given galaxy, a unified model (e.g., outflow velocity and density distributions as a function of radius, outflow geometry, etc) may explain both observations from emission and absorption lines.


However, to date, outflows have only been simultaneously studied in absorption and emission in rare, non-representative cases: ULIRGs \citep{Martin06, Martin15} and BALQSOs \citep{Xu20c}, or in single galaxies  \citep[e.g.,][]{Wood15, Martin-Fernandez16}. This problem is mainly due to the difficulty of retrieving a sample of galaxies with both high-quality rest-UV and rest-optical observations. On the one hand, UV bright galaxies cannot have very high gas densities, otherwise the dust attenuation would be too significant for us to observe the gas. On the other hand, galaxies have to be at specific redshifts so that their rest-UV bands are bright enough and covered by large ground-based optical telescopes (1 $\lesssim$ z $\lesssim$ 2) or UV space telescopes (z $\lesssim$ 0.4). This problem persists after the launch of the James Webb Space Telescope (JWST), for which most of the galactic outflow studies are based on emission lines due to the difficulty to detect faint UV continuum in individual galaxies at high redshifts \citep[3 $< z <$ 9, e.g.,][]{Carniani24, XuYi23, Zhang24}. Hence, we have yet to leverage the full suite of observational constraints to yield a consistent picture of outflows and their feedback effects to the hosts.


In this paper, we present this much-needed study, where we assembled a sample of 33 local starburst galaxies with high-SNR and moderate-resolution UV+optical spectra. For the first time, we present a systematic joint analysis of the two types of outflow diagnostics. Our galaxies are mainly low-mass starburst ($<$ 10$^{10}$ \Msun), where the majority show clear outflow features and may provide strong feedback to the host galaxies \citep[e.g.,][]{Heckman15}. Similar low-mass galaxies have also been found to host strong outflows at earlier universe \citep[3 $< z <$ 9, e.g.,][]{Carniani24}.




The structure of this paper is as follows. In Section \ref{sec:obs}, we introduce the observations and data reductions. In Section \ref{sec:analysis}, we describe how to extract outflow signatures from rest-UV absorption and rest-optical emission lines, respectively. In Section \ref{sec:compare}, we compare the outflow properties from absorption and emission lines. Finally, we discuss our findings in Section \ref{sec:discuss}, where we also construct models to explain the discovered correlations between two outflow diagnostics. We conclude the paper in Section \ref{sec:conclusion}.

\section{Observations and Data Reductions}
\label{sec:obs}

\subsection{Sample Selection}

We construct the sample in two steps. We first identify galaxies that have archival rest-UV spectra from at least one medium resolution grating of the Cosmic Origins Spectrograph (COS) onboard HST, including G130M and G160M. We require the spectra to cover at least the \Siii\ or \siiv\ absorption lines for outflow diagnostics. Then we cross-match the galaxies with the Keck and Very Large Telescope (VLT) archives to search for high-quality optical observations with spectral resolution ($Rs$) $\gtrsim$ 4000 and \ha\ peak \SNR\ $>$ 100. This choices of $Rs$ leads to a velocity bin $<$ 75 km s$^{-1}$, allowing sufficient resolution to extract outflow signatures from emission lines accurately (Section \ref{sec:GuassianFit}). Additionally, this resolution is comparable to that of our COS data ($\sim$ 9000), minimizing inconsistencies between datasets. A high \SNR\ is also necessary to achieve robust extractions of the broad outflow components (see discussion in Section \ref{sec:BFR}).

Overall, we gather a joint sample of 33 low-redshift starburst galaxies with redshifts between 0.002 -- 0.27. Our galaxies cover a wide range of stellar mass\footnote{It is possible that the lowest mass objects ($\sim$ 10$^{5}$ \Msun) in our sample are not the entire galaxy but the star-forming region(s). Since none of them show outflow signatures, they do not affect our main analyses in Section \ref{sec:compare}.} [log (\Mstar/\Msun) $\sim$ 5 -- 10, Section \ref{sec:SED}], star formation rate [log SFR /(\Msun yr$^{-1}$) $\sim$ --3 to +2 , Section \ref{sec:ancillary}], and oxygen abundances (12 + log(O/H) $\sim$ 7.4 -- 8.5). Thus, our galaxies are representative of a large population of local galaxies that have strong star-formation (i.e., starburst galaxies). As shown below, these galaxies are mainly composed of objects such as Green Peas and extreme emission line galaxies that are analogs to high-redshift SF galaxies instead of low redshift ones.

\begin{figure}
\center

	\includegraphics[page = 1,angle=0,trim={3.35cm 0.2cm 4.7cm 1.8cm},clip=true,width=1.0\linewidth,keepaspectratio]{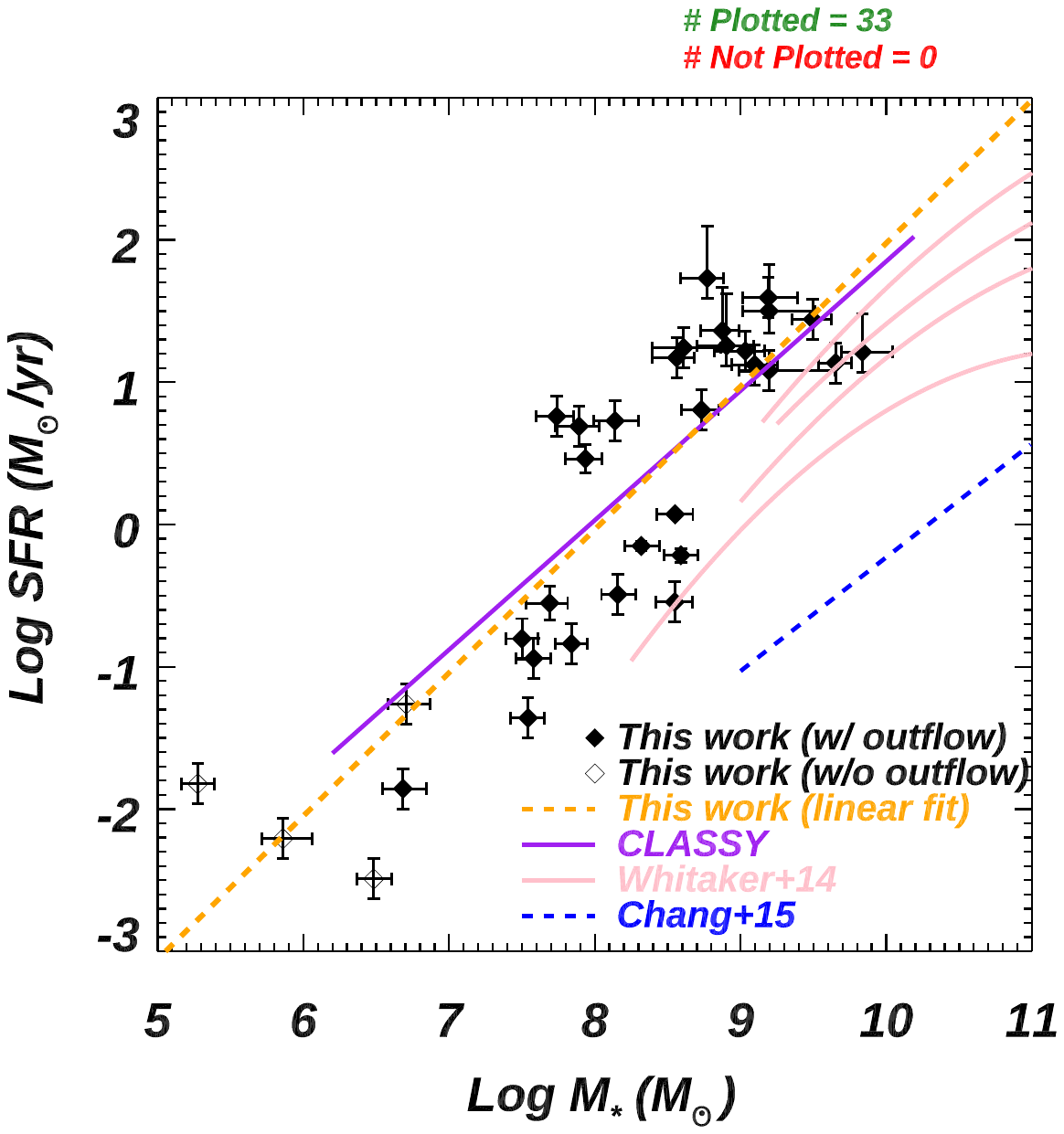}
\caption{\normalfont{Star formation rate vs stellar mass relationship (SFMR) for galaxies in our sample. Our galaxies with and without detected outflows are drawn as filled and hollow diamonds, respectively. The 1$\sigma$ errorbars are drawn as crosses around the symbols. The best linear fits to all of our galaxies are shown as the orange dashed line. For comparisons, we present the SFMR trends from z $\sim$ 0 SDSS+WISE sample from \cite{Chang15} as the blue dashed line, z $\sim$ 1 -- 2.5 SF galaxy sample from \cite{Whitaker14} as the pink lines, and the UV catalog of local SF galaxies from CLASSY survey in purple \citep{Berg22, James22}. Galaxies in our sample mostly represent low-mass starburst galaxies and their SFRs are offset to match better with galaxies at cosmic noon instead of with more typical galaxy populations at z $\sim$ 0.}  }
\label{fig:Mstar_SFR}
\end{figure}

In Figure \ref{fig:Mstar_SFR}, we show the parameter space (SFR vs \Mstar\ relationship, i.e., SFMR) spanned by galaxies in our sample. These values are measured within the aperture and the slit loss is discussed in Section \ref{sec:analysis}. The galaxies with (without) detected outflow features are drawn as  filled (hollow) symbols. The linear fit to all galaxies is shown as the orange dashed line. For a comparison, we also present SFMR from other surveys, including 1) the z $\sim$ 0 sample from 10$^6$ SDSS+WISE galaxies published in \cite{Chang15} as the blue dashed line; 2) the 0.5 $<$ z $<$ 2.5 samples from \cite{Whitaker14} that contain $\sim$ 4 $\times$ 10$^4$ SF galaxies in the CANDELS fields as the pink lines; and 3) a comprehensive, high-quality UV catalog built from the COS Legacy Archive Spectroscopic Survey \citep[CLASSY,][]{Berg22, James22} in the purple solid line. The scatter in all the correlations is $\sim$ 0.3 -- 0.4 dex.

In general, our galaxies follow the CLASSY SFMR trend. This is as expected since both our study and CLASSY select galaxies observed with the COS instrument, which favors UV-bright galaxies with high SFR surface densities. We note 7 out of 33 of our galaxies are also in CLASSY. Besides that, our galaxies have higher SFR than the extrapolated SFMR trend from SDSS at z $\sim$ 0, but they are comparable to the z $\sim$ 1 -- 2.5 galaxies in pink from \cite{Whitaker14}. This indicate that our galaxies have elevated SFR that are more commonly seen in galaxies around cosmic noon. Given these context, we refer to the galaxies in our sample as local starburst galaxies throughout the remainder of this paper.


\subsection{UV Observations}
\label{sec:UVData}
The UV spectra of galaxies in our sample were gathered from various HST programs. The original programs have diverse goals but most of the galaxies are selected to be SF or starburst galaxies and are analogs to high-redshift ones. These programs include studies of outflows and starbursts in Green Pea galaxies (PI-Henry: 12928; PI-Jaskot: 13293, 14080), studies of local Lyman Break Galaxies (PI-Heckman: 11727, 13017); analyses of extreme emission line galaxies (PI-Stark: 14168, 14679, 15185, 15646; PI-Senchyna: 17526), studies of galactic outflows in SF galaxies (PI-Chisholm: 15099), studies of Lyman continuum and Ly$\alpha$ in local SF galaxies (PI-Jaskot: 15626; PI-Gazagnes: 16643); and Far-UV catalog of nearby SF galaxies (PI-Berg: 15840). We download their reduced 1D spectra and COS acquisition images from the HST Mikulski Archive for Space Telescopes (MAST) for later analyses. The COS spectra for each galaxy are coadded and binned to be R $\sim$ 9000 (FWHM $\sim$ 33 km s$^{-1}$). The typical SNR of the UV continuum is from 5 to 10. The information for these observations and their corresponding references are summarized in Table \ref{tab:obs}.

\subsection{Optical Observations and Reductions}
\label{sec:OPTData}
Most of our galaxies (29/33) were observed by the Echellette Spectrograph and Imager (ESI) at the Cassegrain focus of the Keck II telescope \citep{Sheinis02}. ESI provides a broad wavelength coverage (3900 -- 11,000\AA) in a single exposure. The observations were conducted between 2016 and 2017 as part of several programs (PIs: Henry and Jones), using slit widths ranging from 0$\dotarcsec$75 to 1\arcsec. Given HST/COS has a 2.5\arcsec\ diameter aperture but only the central 0.8 -- 1\arcsec\ is unvignetted, the ESI slit widths align well with the COS unvignetted region. Given the slit sizes, the data have a spectral resolution of $\sim$ 56 -- 75 km s$^{-1}$ (FWHM). Typical seeing during the observations was $\sim$ 0$\dotarcsec$7. Detailed exposure times are listed in Table \ref{tab:obs}. 

We reduce the data following the \textit{ESIRedux} pipeline\footnote{\url{https://www2.keck.hawaii.edu/inst/esi/ESIRedux/index.html}} with several custom steps. First, we turn off the default removal of cosmic rays in the pipeline and do manual checks instead. This is because the default pipelines incorrectly identify and mask out the strong emission lines (e.g., [\oiii] 5007 and \ha) from our galaxies. We also find the sky was commonly over-subtracted around strong emission lines, especially when the lines are spatially extended. Thus, we add another input parameter to manually define the sky regions, which are measured from the 2D spectra around the strong lines for each galaxy. Finally, we note that in the co-added spectra, the adjacent diffraction orders in Echelle spectra can exhibit semi-sinusoidal features where the orders are stitched together. Thus, for each galaxy, we apply another flux calibration for these features. We conduct spline fits to the galaxy's continuum spectra from ESI and Sloan Digital Sky Survey (SDSS), separately. Then we scale the ESI spectrum by matching its spline fit to the SDSS one of the same galaxy. We have checked that this only has minor effects on our measurements of emission lines since they are scaled together with the local continuum levels. 





We also include four galaxies observed by the X-Shooter spectrograph \citep{Vernet11} mounted on the VLT as part of the ESO program ID 085.B-0784(A) and 096.B-0192(A) \citep[PI: Overzier,][]{Loaiza-Agudelo20}. These observations were done in slit mode (11\arcsec\ long slit) to obtain simultaneous spectra from the three arms, i.e., UVB (1\arcsec\ wide slit, R $\sim$ 5100), VIS (0$\dotarcsec$9 wide slit, R $\sim$ 8800), and NIR (0$\dotarcsec$9, R $\sim$ 5100). Our main emission line (\ha) fall into the VIS bands, which yield a resolution with FWHM $\sim$ 34 km s$^{-1}$, which is similar to our ESI data. The X-Shooter exposure times are also listed in Table \ref{tab:obs}. The detailed data reduction is done with ESO/X-Shooter pipeline \citep{Modigliani10} and the EsoRex command-line recipes and is discussed in  \cite{Loaiza-Agudelo20}. 



\begin{table*}
	\centering
	\caption{UV and Optical Observations for Galaxies in Our Sample}
	\label{tab:obs}
	\begin{tabular}{lllclllll} 
		\hline
		\hline
		ID & RA & Dec 	    & $z^{1}$   & T(G130M)$^{2}$ &	T(G160M)$^{2}$ & Ref.$^{3}$ & T(OPT)$^{2}$        &	OPT Instrument   \\
		\hline
		   &     &         	&           & (s)        & (s)     &     & (s) & \\
		\hline

            J0055--0021&00:55:27&	--00:21:48&	0.1672 &5040&2970     & (a)        &2560&VLT/X-Shooter \\
            J0150+1308&	01:50:28&	+13:08:58&	0.1464 &1327&1623     & (a)        &2560&VLT/X-Shooter \\
            J0808+1728&	08:08:41&	+17:28:56&	0.0442 &4451&4451     & (b)        &3600&Keck/ESI \\
            J0815+2156&	08:15:52&	+21:56:24&	0.1409 &0&6809        & (c)        &3600&Keck/ESI \\
            J0851+5840&	08:51:16&	+58:40:55&	0.0919 &12318&0       & (b)        &3300&Keck/ESI \\
            J0911+1831&	09:11:13&	+18:31:08&	0.2622 &2073&6530     & (d, e)     &1200&Keck/ESI \\
            J0926+4427&	09:26:00&	+44:27:37&	0.1807 &5640&6180     & (a, f)     &1500&Keck/ESI \\
            J0942+0928&	09:42:56&	+09:28:16&	0.0109 &0&2608        & (g)        &4800&Keck/ESI \\
            J0942+3547&	09:42:52&	+35:47:26&	0.0149 &10754&2664    & (g, h)     &8100&Keck/ESI \\
            J0944--0038&09:44:01&	--00:38:31&	0.0049 &6086&34582    & (f, g)     &7200&Keck/ESI \\
            J1024+0524&	10:24:29&	+05:24:50&	0.0332 &4296&2608     & (f, g)     &9000&Keck/ESI \\
            J1044+0353&	10:44:57&	+03:53:13&	0.0126 &6853&34871    & (f, i)     &2560&VLT/X-Shooter \\
            J1053+5237&	10:53:31&	+52:37:53&	0.2526 &824&2736      & (d, e)     &1200&Keck/ESI \\
            J1129+2034&	11:29:14&	+20:34:52&	0.0047 &6885&2588     & (f, g)     &5400&Keck/ESI \\
            J1133+6513&	11:33:04&	+65:13:41&	0.2414 &1232&4588     & (d, e)     &3900&Keck/ESI \\
            J1137+3524&	11:37:22&	+35:24:27&	0.1944 &1264&2339     & (e)        &1200&Keck/ESI \\
            J1148+2546&	11:48:27&	+25:46:10&	0.0451 &4552&2624     & (f, g)     &9000&Keck/ESI \\
            J1155+5739&	11:55:28&	+57:39:52&	0.0172 &0&2852        & (g)        &9000&Keck/ESI \\
            J1200+2719&	12:00:16&	+27:19:59&	0.0819 &4610&0        & (b)        &3300&Keck/ESI \\
            J1215+2038&	12:15:18&	+20:38:25&	0.0027 &0&2616        & (g)        &5400&Keck/ESI \\
            J1219+1526&	12:19:04&	+15:26:09&	0.1956 &716&2303      & (e)        &2100&Keck/ESI \\
            J1222+0434&	12:22:25&	+04:34:04&	0.0043 &0&2604        & (g)        &4200&Keck/ESI \\
            J1226+0415&	12:26:12&	+04:15:36&	0.0942 &11568&0       & (b)        &2700&Keck/ESI \\
            J1230+1202&	12:30:48&	+12:02:43&	0.0042 &0&2612        & (g)        &7200&Keck/ESI \\
            J1244+0215&	12:44:23&	+02:15:40&	0.2394 &2042&6507     & (e)        &2100&Keck/ESI \\
            J1248+1234&	12:48:35&	+12:34:03&	0.2634 &1644&6372     & (e)        &4200&Keck/ESI \\
            J1311--0038&13:11:31&	--00:38:44&	0.0811 &8426&0        & (b)        &3600&Keck/ESI \\
            J1416+1223&	14:16:12&	+12:23:40&	0.1228 &2379&4636     & (j)        &2560&VLT/X-Shooter \\
            J1424+4216&	14:24:06&	+42:16:46&	0.1848 &1208&0        & (e)        &1500&Keck/ESI \\
            J1448--0110&14:48:05&	--01:10:58&	0.0274 &9610&5192     & (b, f, g)  &1200&Keck/ESI \\
            J1457+2232&	14:57:35&	+22:32:02&	0.1486 &0&7029        & (c)        &3300&Keck/ESI \\
            J1509+3731&	15:09:34&	+37:31:46&	0.0326 &8095&7550     & (b)        &3600&Keck/ESI \\
            J1735+5703&	17:35:01&	+57:03:09&	0.0472 &5519&0        & (b)        &900&Keck/ESI \\
            \hline
            \hline
            
	\multicolumn{9}{l}{%
  	\begin{minipage}{17cm}%
	Note. --\\
    	(1)\ \ Redshift of the objects matched to the peak of the fitted narrow component of Balmer emission lines.\\
    	(2)\ \ Exposure times in seconds analyzed for each galaxy with COS G130M, COS G160M, and the optical instrument, respectively. Zero means no observations exist with that instrument by the time we did the analyses. \\
            (3)\ \ HST programs and the main references for each object: (a): HST-GO-11727, PI: Heckman \citep{Heckman15}; (b): HST-GO-14080, PI: Jaskot \citep{Jaskot17}; (c): HST-GO-13293, PI: Jaskot \citep{Jaskot14}; (d): HST-GO-15626, PI: Jaskot \citep{Flury22a,Flury22b}; (e): HST-GO-12928, PI: Henry \citep{Henry15}; (f): HST-GO-15840, PI: Berg \citep{Berg22, James22}; (g): HST-GO-14168, PI: Stark \citep{Senchyna17}; (h): HST-GO-15099, PI: Chisholm \citep{Chisholm19}; (i): HST-GO-15646, PI: Stark \citep{Senchyna22}; (j): HST-GO-13017, PI: Heckman \citep{Heckman15}; (g): HST-GO-15185, PI: Stark \citep{Senchyna21}.\\

  	\end{minipage}%
	}\\
	\end{tabular}
	\\ [0mm]
	
\end{table*}


\section{UV and Optical Measurements}
\label{sec:analysis}


\subsection{Spectral Energy Distribution Fitting}
\label{sec:SED}

To constrain the continuum level and remove the starlight contribution, we conduct SED fitting of our UV and optical spectra, separately. For the UV, we adopt the \FiCUS\ code\footnote{\url{https://github.com/asalda/FiCUS}} to fit the coadded HST/COS data for each galaxy. Detailed descriptions of the methods can be found in \cite{Chisholm19, Saldana-Lopez22}. For a brief summary, we fit each observed spectrum with a linear combination of
multiple bursts of single-metallicity and single-age stellar population models from Starburst99 \citep{Leitherer99, Leitherer10}. A nebular continuum was also included by self-consistently processing the stellar population synthesis models through the
cloudy v17.0 code \citep{Ferland17}. We assume attenuation law from \cite{Reddy16a}. In addition to the continuum level, we also get various important parameters from the fitting, including the galaxy's light-weighted metallicity, stellar wind absorption, and stellar ages and dust extinction. 

For the optical spectra, we conduct SED fitting using the \Prospector\ code \citep{Johnson21}. We follow the approaches in \cite{Xu22b} and assume a non-parametric star formation history (SFH) with a Kroupa IMF \citep{Kroupa01} and the attenuation law from \cite{Reddy15}. We adopt the isochrone library of MIST \citep{Choi16}, spectral library of C3K \citep{Conroy19}, and dust emission model from \cite{Draine07}. We also get the stellar mass of galaxies from the best SED fits (listed in Table \ref{tab:ancillary}).



Before the SED fitting, we masked out spectral regions $\pm$ 550 km~s$^{-1}$ around ISM absorption lines, nebular emission lines, Milky Way (MW) absorption lines, and Geocoronal emission lines. 
For especially broad lines, including \lya, [\ha], [\oiii] \ly 5007, we increase the mask region to $\pm$ 1500 km~s$^{-1}$.
Finally, for each galaxy, we normalize the spectra based on its best-fit SEDs. 



\begin{table*}
	\centering
	\caption{Measured Properties from the Absorption and Emission Lines$^{\textbf{(1)}}$}
	\label{tab:linemea}
         \begin{tabular}{llll|llr} 
		\hline
  		ID & \multicolumn{3}{c}{Absorption Lines} & \multicolumn{3}{c}{Emission Lines}   \\
		\hline
		 Param. & \Vnfout\ &	\FWHMout & log(\ne) & \Vnfoutha & \HWHMoutHa & log(\ne)   \\
		\hline
	Unit   &  (km s$^{-1}$)   &   (km s$^{-1}$)      & log(cm$^{-3}$)   & (km s$^{-1}$) & (km s$^{-1}$) & log(cm$^{-3}$) \\
		\hline

J0055--0021 & --412$\pm 48$ & 591$\pm 108$ & 1.0$\pm 0.4$ & --330$\pm 12$ & 212$\pm 1$ & 2.5$\pm 0.06$\\
J0150+1308 & --390$\pm 15$ & 345$\pm 63$ & 0.6$\pm 0.5$ & --204$\pm 12$ & 132$\pm 0.1$ & 2.6$\pm 0.03$\\
J0808+1728 & \multicolumn{1}{l}{No Outflow} & \dots & \dots & --124$\pm 5$ & 63$\pm 0.1$ & \dots\\
J0815+2156 & --401$\pm 164$ & 328$\pm 201$ & 1.5$\pm 0.3$ & --171$\pm 7$ & 93$\pm 3$ & \dots\\
J0851+5840 & --216$\pm 80$ & 280$\pm 93$ & 0.7$\pm 0.4$ & --142$\pm 8$ & 75$\pm 3$ & \dots\\
J0911+1831 & --440$\pm 53$ & 499$\pm 198$ & 1.1$\pm 0.4$ & --336$\pm 9$ & 215$\pm 8$ & \dots\\
J0926+4427 & --445$\pm 7$ & 402$\pm 130$ & 1.0$\pm 0.5$ & --327$\pm 7$ & 190$\pm 3$ & 2.0$\pm 0.05$\\
J0942+3547 & --280$\pm 70$ & 318$\pm 62$ & 0.4$\pm 0.2$ & --128$\pm 5$ & 75$\pm 0.1$ & \dots\\
J0942+0928 & --203$\pm 10$ & 207$\pm 10$ & $<$ 0.3 & --126$\pm 5$ & 60$\pm 0.1$ & 2.3$\pm 0.06$\\
J0944--0038 & --144$\pm 77$ & 163$\pm 54$ & $<$ 0.4 & --157$\pm 5$ & 85$\pm 2$ & \dots\\
J1024+0524 & --249$\pm 34$ & 286$\pm 54$ & 0.5$\pm 0.3$ & --157$\pm 5$ & 77$\pm 0.1$ & 1.6$\pm 0.05$\\
J1044+0353 & --143$\pm 18$ & 123$\pm 24$ & 0.3$\pm 0.2$ & --105$\pm 14$ & 50$\pm 0.1$ & \dots\\
J1053+5237 & --414$\pm 30$ & 534$\pm 155$ & 1.1$\pm 0.5$ & --331$\pm 7$ & 206$\pm 5$ & 2.0$\pm 0.05$\\
J1129+2034 & \multicolumn{1}{l}{No Outflow} & \dots & \dots & \multicolumn{1}{l}{No Outflow} & \dots & \dots\\
J1133+6513 & --453$\pm 23$ & 444$\pm 22$ & 0.7$\pm 0.5$ & --206$\pm 7$ & 117$\pm 3$ & \dots\\
J1137+3524 & --319$\pm 43$ & 345$\pm 79$ & 1.1$\pm 0.5$ & --330$\pm 7$ & 199$\pm 3$ & 1.9$\pm 0.05$\\
J1148+2546 & --202$\pm 30$ & 239$\pm 52$ & $<$ 0.5 & --168$\pm 5$ & 87$\pm 0.1$ & 2.6$\pm 0.006$\\
J1155+5739 & --154$\pm 13$ & 161$\pm 16$ & 0.7$\pm 0.2$ & --144$\pm 5$ & 73$\pm 0.1$ & 2.2$\pm 0.02$\\
J1200+2719 & --318$\pm 16$ & 315$\pm 16$ & 0.9$\pm 0.3$ & --143$\pm 5$ & 82$\pm 0.1$ & \dots\\
J1215+2038 & \multicolumn{1}{l}{No Outflow} & \dots & \dots & \multicolumn{1}{l}{No Outflow} & \dots & \dots\\
J1219+1526 & --459$\pm 27$ & 535$\pm 123$ & 1.8$\pm 0.3$ & --310$\pm 7$ & 182$\pm 2$ & 2.7$\pm 0.04$\\
J1222+0434 & \multicolumn{1}{l}{No Outflow} & \dots & \dots & \multicolumn{1}{l}{No Outflow} & \dots & \dots\\
J1226+0415 & --282$\pm 17$ & 368$\pm 68$ & 0.6$\pm 0.4$ & --125$\pm 5$ & 64$\pm 0.1$ & 1.8$\pm 0.05$\\
J1230+1202 & \multicolumn{1}{l}{No Outflow} & \dots & \dots & \multicolumn{1}{l}{No Outflow} & \dots & \dots\\
J1244+0215 & --322$\pm 74$ & 385$\pm 152$ & 1.1$\pm 0.6$ & --224$\pm 5$ & 140$\pm 1$ & 2.2$\pm 0.05$\\
J1248+1234 & --411$\pm 56$ & $<$ 395 & 1.2$\pm 0.5$ & --202$\pm 7$ & 110$\pm 2$ & 1.7$\pm 0.05$\\
J1311--0038 & --185$\pm 50$ & 147$\pm 56$ & 1.3$\pm 0.2$ & --148$\pm 5$ & 86$\pm 2$ & \dots\\
J1416+1223 & --441$\pm 10$ & 649$\pm 216$ & 1.5$\pm 0.3$ & --367$\pm 12$ & 212$\pm 1$ & 2.8$\pm 0.01$\\
J1424+4216 & --434$\pm 13$ & 449$\pm 106$ & 1.6$\pm 0.3$ & --340$\pm 7$ & 221$\pm 4$ & \dots\\
J1448--0110 & \multicolumn{1}{l}{No Outflow} & \dots & \dots & --233$\pm 5$ & 134$\pm 3$ & \dots\\
J1457+2232 & --160$\pm 114$ & 249$\pm 95$ & 1.2$\pm 0.4$ & --200$\pm 5$ & 114$\pm 2$ & 2.2$\pm 0.06$\\
J1509+3731 & --134$\pm 37$ & 126$\pm 50$ & 0.7$\pm 0.2$ & --112$\pm 5$ & 55$\pm 0.1$ & \dots\\
J1735+5703 & --247$\pm 56$ & 267$\pm 61$ & 0.3$\pm 0.3$ & --238$\pm 8$ & 140$\pm 4$ & \dots\\

            \hline
            \hline
            
	\multicolumn{7}{l}{%
  	\begin{minipage}{13cm}%
	Note. --\\
    	\textbf{(1)}\ \ For UV absorption lines, we show the maximum outflow velocity (\Vnfout), velocity widths (\FWHMout), and electron number density (\ne) by considering the median of all available UV lines that passed the F-test. All absorption lines include \Lyb, \cii\ \ly 1334, the \Siii\ multiplet (\ly1190, 1193, 1260, 1304, and 1526), \siiii\ \ly 1206, and \siiv\ \ly\ly 1393, 1402. For emission lines, we show similar properties, measured from \ha\ (for \Vnfout\ and \HWHMout) and [\sii] \ly\ly 6717, 6732 doublet (for \ne) when available. Galaxies with no detections of outflows are labelled as `No Outflow'. See details of the measurements in Sections  \ref{sec:linemea} and \ref{sec:density}.\\
    	\textbf{(2)}\ \ For J0851+5840 and J1735+5703, their \ha\ emission lines are in detector gaps, so we report their values from \hb.    	
  	\end{minipage}%
	}\\
	\end{tabular}
	\\ [0mm]
	
\end{table*}

\subsection{Spectral Line Fitting}
\label{sec:GuassianFit}
In the UV spectra of SF or starburst galaxies, outflow features appear as broad and blueshifted absorption lines \citep[e.g.,][]{Heckman15}. Similarly, outflows can show as broad emission lines in optical spectra\footnote{While Active Galactic Nucleus (AGN) activities can also generate the broad emission lines, we have checked that our galaxies do not fall into the AGN region in the BPT diagram \citep{Baldwin81}}.  \citep[e.g.,][]{Freeman19}. To compare the possible outflow signatures from UV and optical spectra, we need to first isolate them from the static interstellar medium (ISM) that is not moving out of the galaxy. To do so, we fit the diagnostic absorption and emission lines with multiple Gaussian profiles. In the following sections, we explain our line fitting techniques for UV and optical, separately. For each galaxy, we adopt the same redshift for its UV and optical lines, which is constrained by the peak of the narrow emission component.

\subsubsection{Fitting UV Absorption Lines}
\label{sec:GuassianFit:UV}

For diagnostic UV absorption lines, we follow the same methodology described in \cite{Xu22a} to fit them. These lines include \Lyb, \cii\ \ly 1334, the \Siii\ multiplet (\ly1190, 1193, 1260, 1304, and 1526), \siiii\ \ly 1206, and \siiv\ \ly\ly 1393, 1402, whenever they are cleanly detected. For a summary, we fit each UV absorption trough using a double-Gaussian profile, while we do not tie line velocities or widths for different troughs. The double-Gaussian profile includes a narrow Gaussian with its center fixed at zero velocity representing the static ISM and a broad, blueshifted Gaussian for the outflow component. We also attempted to fit the absorption trough using a single-Gaussian profile centered at zero velocity. We then compare the fitting results of single- and double-Gaussian profiles and apply an F-test to determine if the fitted absorption trough requires the additional broad component, i.e., if the trough contains outflow signatures \citep[see Equation (1) in][]{Xu22a}. We run F-test with $\alpha$ = 0.05, i.e., at 95\% confidence level. Examples of the fits are shown in Figure \ref{fig:oneobj}, while the fits for all galaxies are presented in Appendix \ref{app:sec:fit}. We find most of our galaxies (27/33) exhibit outflow signatures in the UV absorption lines.

\subsubsection{Fitting Optical Emission Lines}
\label{sec:GuassianFit:OPT}

For emission lines in the optical spectra, outflow features appear as broad wings and we fit them using the double-Gaussian fitting method described above. This approach has also been previously applied to analyses of broad emission lines in SF galaxies \citep{Newman12b, Freeman19} (hereafter referred to as \Newman\ and \Freeman). The major diagnostic emission lines include [\oii] \ly\ly 3726, 3729, [\oiii] \ly 4363, [\oiii] \ly\ly 4959, 5007, [\nii] \ly\ly 6549, 6585, [\sii] \ly\ly 6717, 6731 and the Balmer lines (\ha, \hb, and \hg). 

Given the high \SNR\ of our optical spectra, we do not enforce identical kinematics [i.e., velocity centers (\VC) and widths (\FWHM)] for all emission lines within a galaxy. Nonetheless, we have verified that the final fitted velocities and line widths for different strong lines (e.g., \ha\ and [\oiii] \ly5007) remain consistent within the error bars for a given galaxy, even though they are not fit simultaneously.

For emission lines linked by atomic physics, we constrain their parameters accordingly. This includes Balmer lines that share the same kinematics for the broad component (or the narrow component). Thus, we have two free parameters for each Balmer line, i.e., broad amplitude (\AB), and narrow amplitude (\AN), while there are four shared parameters over all Balmer lines, i.e., \VCB, \VCN, \FWHMB, and \FWHMN. Hereafter, b and n stand for broad and narrow components, respectively. A similar case applies to the [\sii] \ly\ly 6717, 6731 doublet. For [\oiii] \ly\ly 4959, 5007 or [\nii] \ly\ly 6549, 6585 doublets, they share the same \VCB, \VCN, \FWHMB, and \FWHMN, while their doublet line ratios are also constrained by the atomic physics \citep[2.98 and 2.93, respectively,][]{Osterbrock06}. For lines that are blended or weak, e.g., [\oii] \ly\ly 3726, 3729 and [\oiii] \ly 4363, the broad and narrow components cannot be reliably distinguished. Thus, their line kinematics are tied to the ones from [\oiii] \ly\ly 4959, 5007 doublet. Examples of the fits are shown in the bottom panels of Figure \ref{fig:oneobj}.

For approximately two-thirds of our galaxies, the double-Gaussian profile described above provides a sufficient fit to the optical emission lines. However, in the remaining galaxies, this method fails due to pronounced asymmetries in the emission lines. For this subset of galaxies, we add an additional narrow Gaussian component, which stands for the ISM gas in kinematically different \hii\ regions from the main ISM component. We have also adopted an F-test to determine if this additional Gaussian is necessary in such cases. 

We examine the properties of this subset of galaxies requiring this additional kinematic component and find no significant differences in outflow velocity, line width, broad to narrow flux ratio (BFR), SFR, or \Mstar\ compared to the rest of the sample. Interestingly, 4 out of 11 of them exhibit clear multiple cores—likely corresponding to distinct \hii\ regions—in their shallow HST/COS UV acquisition images. Future deeper imaging may clarify whether these features are common in all galaxies requiring additional kinematic components.

\subsubsection{Line Fitting Results}
\label{sec:GuassianFit:Results}

We present an example of the line fitting in Figure \ref{fig:oneobj} and the entire sample in Appendix \ref{app:sec:fit}. In each figure, the top two rows are for absorption lines in the UV, while the bottom two rows are for the optical emission lines of the same galaxy for a comparison. Data and uncertainties are shown as black and gray histograms. The best-fitted static ISM and outflow components are shown in green and blue lines. The additional narrow component is shown in orange lines when needed, and the sum of all models is shown in red lines. We also add the insets to show a zoom-in view of the extended emission-line wings.


At the first glance, by comparing the outflow components (in blue) of emission and absorption lines for each galaxy, the data strongly suggest that outflow signatures are consistently detected using both diagnostics, i.e., whenever an outflow component is required for the absorption line, the galaxy also needs the broad component to fit its emission line. This happens for 27 out of 33 galaxies. For a few cases when no clear outflow features are seen in the absorption, the optical lines are indeed narrower and show no broad wings (J1129+2034, J1215+2038, J1222+0434, J1230+1202). These four galaxies all have relatively low \Mstar\ ($<$ 10$^7$ \Msun) compared to other galaxies in our sample. 

There exist two exceptions: J0808+1728 and J1448–0110, whose \Siii\ and \cii\ absorption lines show no outflow signatures but whose \ha\ emission lines still require the broad component. For J1448–0110, however, we do detect blue-shifted absorption in its \siiv\ doublet lines. Thus, it is likely that galactic outflows in this galaxy are highly ionized, and undetectable in low-ionization lines \citep[see a few similar cases in][]{Grimes09, Martin15}. Unfortunately, no data exist for J0808+1728's \siiv\ region. It is also possible for geometric effects to explain the lack of absorption in these galaxies, in the case where the outflow emission is within the aperture, but not in front of the UV-bright sources \citep{Jaskot14}. Future larger samples with similar objects will provide more information.

Overall, there is clear evidence that the existence of broad components seen in emission and absorption lines are linked in the majority of our galaxies ($>$ 94\%) and they represent galactic outflows. We will discuss their correlations more quantitatively in Section \ref{sec:compare}.


\subsection{Line Measurements}
\label{sec:linemea}

After extracting the broad components from the absorption and emission lines, we conduct various measurements of the line properties. To make a fair comparison for outflows, we focus on the blue-shifted, broad component ($v$ $<$ 0 km s$^{-1}$) of the emission lines. This is because the absorption lines can only trace the gas along the line-of-sight (LOS), so we normally cannot detect the gas that is moving away from us (i.e., the redshifted side of the outflows). Thus, it is preferable to separate them when comparing with the blue-shifted, broad component in absorption lines. 


 
For each emission line, we first measure its best-fitted broad component flux and line width for the blue-shifted half. For the latter, we call it ``half-width-half-maximum (\HWHM)'' of the broad component. Then we measure the cumulative flux of this blue-shifted half as \citep[e.g.,][]{Whittle85}:

\begin{equation}
\label{Eq:cumFlux}
	\begin{aligned}
	\Phi(v)     &= \int_{v}^{0} F(v^{\prime}) dv^{\prime};\ (v < 0)
	\end{aligned}
\end{equation}
where $F(v^{\prime})$ is the line normalized flux at a velocity, and the integration is over blue-shifted velocities.
After that, we can define \Vaout\ representing the outflow velocity when $\Phi(v) = \frac{a}{100} \times \Phi(-\infty)$ \citep[][]{Liu13b}, i.e., the velocity at which the cumulative flux in the broad component reaches a certain percentage of the blue-shifted, broad component flux. Specifically, we calculate \Vaout\ for $a$ = 95, 90, 50, 10, and 05. We adopt \Vnfout\ to represent the maximum outflow velocity of the line. 


Additionally, we quantify the asymmetry of the broad component of emission lines with respect to v = 0. We first calculate another set of $v_{a}$ values as follows. For the best-fit full, broad component of each emission line, we integrate Equation (\ref{Eq:cumFlux}) again with the upper integration range to be $+\infty$. Then we calculate the \Asy\ parameter  \citep{Liu13b, Martin15}:

\begin{equation}
\label{Eq:SumFlux}
 	\begin{aligned}
	A_\text{90} &= \frac{(v_{90}^{f,b} - v_{50}^{f,b}) - (v_{50}^{f,b} - v_{10}^{f,b})}{v_{90}^{f,b} - v_{10}^{f,b}}
	\end{aligned}
\end{equation}
where a positive (negative) value of \Asy\ means the line contains more flux on the blue-shifted (red-shifted) side. The notation `f,b' means that we adopt the full, broad component of the emission line. We find all of our galaxies have quite small \Asy\ values ($\lesssim 0.01$), which suggests the broad component of emission lines representing the outflows are close to symmetric around $v$ = 0. As a comparison, for AGN outflows, strong blue-shifted emission lines are commonly observed due to the much dustier environment. Their broad components were found to have \Asy\ values ranging from 0.1 to 0.4 under our definition \citep{Liu13b}.


\begin{figure*}
\center

	\includegraphics[page = 1,angle=0,trim={0.1cm 0.1cm 0cm 8.5cm},clip=true,width=1.0\linewidth,keepaspectratio]{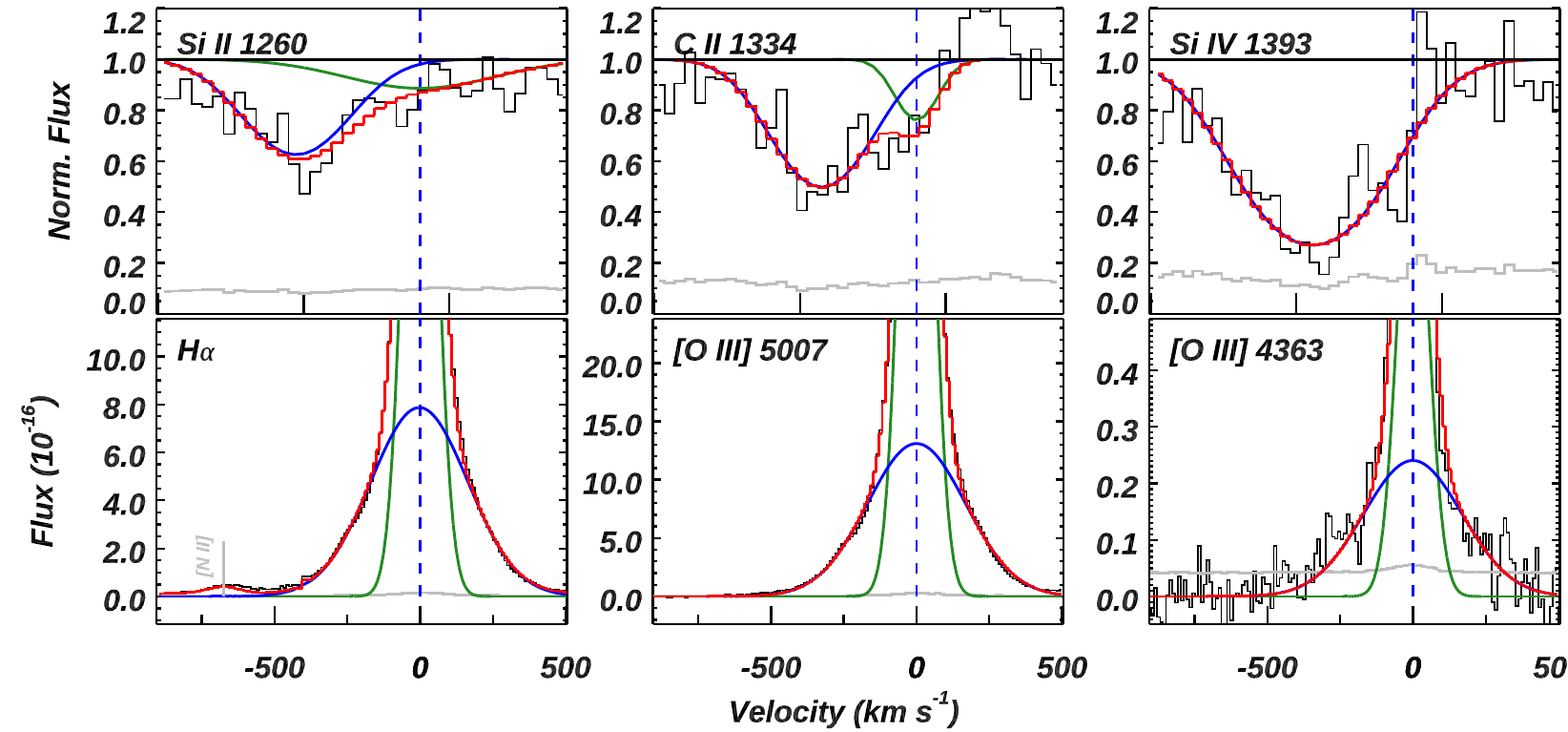}
\caption{\normalfont{Spectral line fitting for one galaxy (J0926+4427) in our sample. Other galaxies are shown in Appendix \ref{app:sec:fit}. The flux and uncertainties are shown in black and gray histograms, respectively. The flux unit for emission lines is ergs s$^{-1}$ cm$^{-2}$ \AA$^{-1}$, while absorption lines are shown in normalized flux. For each line, the best-fit narrow and broad components are shown in green and blue, respectively. The red line is the summation of all models. Line fitting details are discussed in Section \ref{sec:GuassianFit}.}}
\label{fig:oneobj}
\end{figure*}

For the above parameters, we propagate errors through a Monte Carlo (MC) simulation where we perturb the spectrum 10$^{3}$ times based on the observed 1$\sigma$ uncertainties (i.e., Poisson noise). These measurements and errors are listed in Table \ref{tab:linemea}. Given the high SNR of our optical observations, the errors for emission lines are not primarily dominated by Poisson noise but rather by the assumptions made in the multi-Gaussian fitting method. These systematic uncertainties are difficult to quantify and are therefore not included in the table. We also note the detection rates of broader but weaker outflow components in emission lines dependent on SNR (see discussion in Section \ref{sec:BFR}). Thus, in Table \ref{tab:linemea}, we report the measurements from one of our strongest lines, \ha. When \ha\ is contaminated by telluric features, we instead report measurements of \hb. For the remainder of this paper, we adopt these \ha\ measurements for comparison with those derived from UV absorption lines.


Similarly, for each diagnostic absorption line in the UV, we measure the \FWHMout\ and \Vaout\ values from the fitted broad component (Section \ref{sec:GuassianFit:UV}). For the latter, we integrate the optical depth instead of flux in Equation (\ref{Eq:cumFlux}) \citep{Xu20d}. For each galaxy, we calculate their median \FWHMout\ and \Vaout\, accounting for all absorption lines that passed the F-test. Since UV absorption lines typically have lower \SNR, and different lines can exhibit variations in line velocities and widths \citep{Xu22a}, median values provide a more robust representation of the overall UV line kinematics. We report these measurements in the 2nd and 3rd columns in Table \ref{tab:linemea}. We have already corrected the listed \FWHM\ and \HWHM\ values by subtracting the instrumental resolution in quadrature.





\begin{figure*}
\center

	\includegraphics[page = 2,angle=0,trim={2cm 0.1cm 4.7cm 1.8cm},clip=true,width=0.5\linewidth,keepaspectratio]{./V95_Comp.pdf}
	\includegraphics[page = 1,angle=0,trim={2cm 0.1cm 4.7cm 1.8cm},clip=true,width=0.5\linewidth,keepaspectratio]{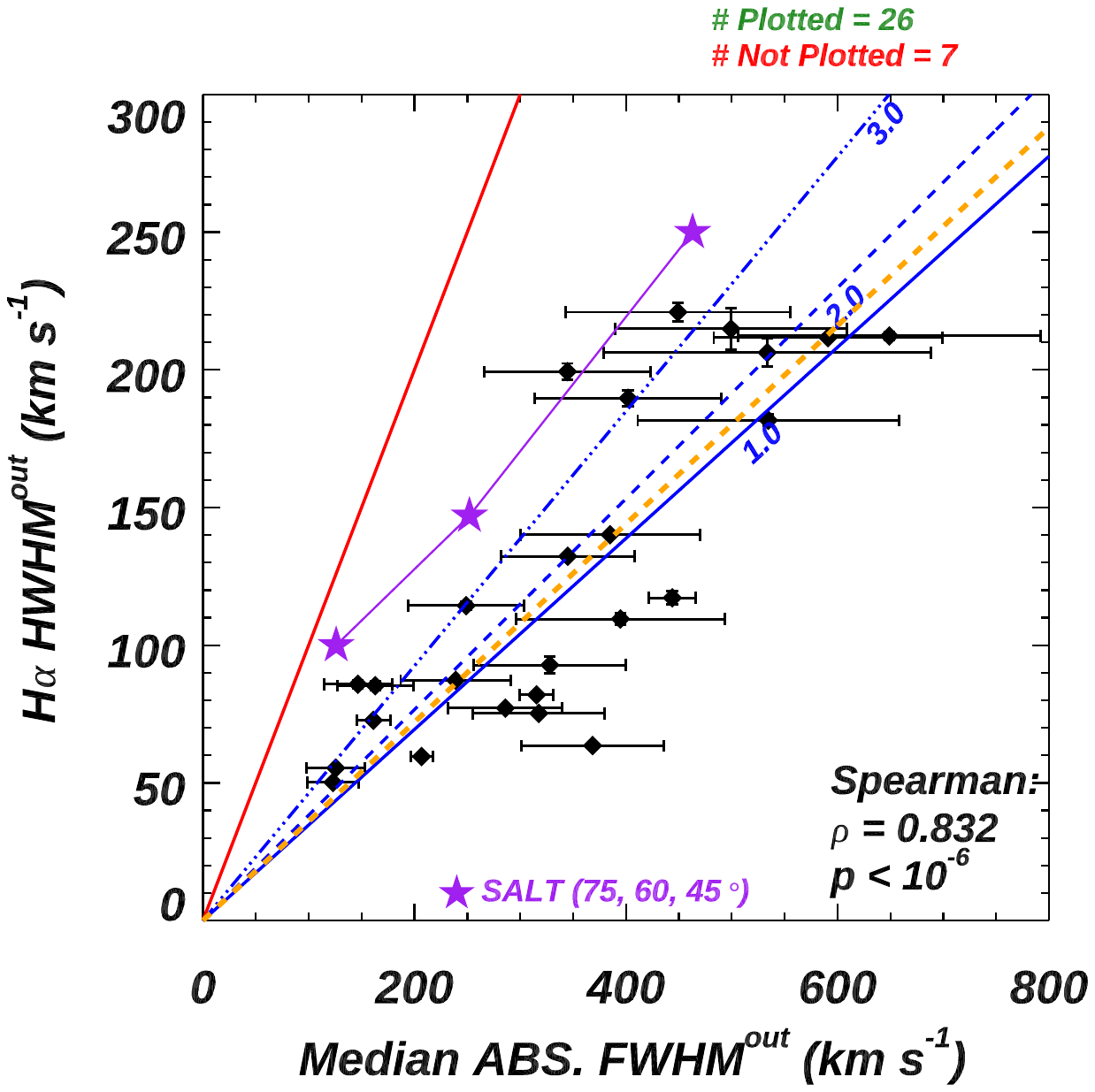}
\caption{\normalfont{Correlations of outflow kinematics derived from absorption and emission lines within individual galaxies across the sample. \textbf{Left:} For emission lines, we report \Vnfout\ from \ha\ (or \hb\ if \ha\ is in a detector gap). For absorption lines, we report the median \Vnfout\ measured from all absorption lines that contain outflowing features. Each symbol corresponds to one galaxy, and the bars indicate the error sizes. The red line denotes the 1:1 relationship. \textbf{Right:} Similarly as the left but for outflow velocity widths. For emission lines, we only measure the width of the blue-shifted portion of emission lines (i.e., \HWHMout) to make a fair comparison with that of blue-shifted absorption lines. In both panels, we show the best linear fit as the orange dashed lines. We overlap spherical outflow models with different outflow density distributions $n(r) \propto r^{-\gamma}$ in blue lines, while $\delta$ is fixed at 1.8 in Equation \ref{eq:n_r}.
We also overlay bi-conical outflow models from SALT \citep{Carr23} in purple stars given M 82 galactic outflow parameters, where the three stars represent different orientation angles listed in the parentheses from the left to right. We discuss the model details in Section \ref{sec:comp:model}.}  }
\label{fig:Kine}
\end{figure*}

\subsection{Ancillary Measurements}
\label{sec:ancillary}

To quantify the spatial extent of starburst regions, we measure galaxy starburst sizes (\rburst) from HST/COS data. We first calculate the UV half-light radius of each galaxy from its COS NUV acquisition images following \cite{Xu22a}. In short, we accumulate the net photon counts within a certain radius from the center of a galaxy until it reaches half of the total source counts. For galaxies that are more spatially extended than the unvignetted size of the COS aperture ($r \sim$ 0.4\arcsec), we treat their measured \rhalf\ as lower limits. We adopt these measured \rburst\ to derive UV outflow spatial extension in Section \ref{sec:size}.

For each galaxy, we find their Milky Way dust extinction using the Galactic Dust Reddening and Extinction Map \citep{Schlafly11} from the NASA/IPAC Infrared Science Archive. We then calculate the internal dust extinction (\EBVint) for each galaxy from a set of Balmer lines (i.e., \ha+\hb+\hg) following the methods in \cite{Xu22b}.

For each galaxy, we have done SED fitting to constrain its Balmer absorption and mass (see Section \ref{sec:SED}). Since SFR estimates from SED fitting covary with other fit parameters such as metallicity, dust, and age, we instead derive SFR based on the full \ha\ emission line, which stands for SFR within the slit. The slit loss of our galaxies are estimated using the 2D spectra and are about 10 -- 50\%. 

After correcting for Balmer absorption using the SED fitting results and the dust extinction, we adopt a metallicity-dependent calibration for inferring SFR from \ha\ luminosity \citep[][and Korhonen Cuestas et al. in prep.]{Strom17}. When \ha\ is not available, we adopt the dust-corrected \hb\ luminosity $\times$ 2.86 into the calculations. All ancillary parameters are reported in Table \ref{tab:ancillary} of Appendix \ref{app:sec:ancillary}.



\section{Connecting Outflow Properties from UV and Optical diagnostics}
\label{sec:compare}

\subsection{General Justification}


Absorption and emission lines provide us with distinct approaches to investigate the outflowing gas. From the last two decades,  well-established methods have been developed to measure galactic outflow properties from  absorption \citep[e.g.,][]{Rupke02, Martin05, Steidel10, Heckman15, Chisholm18, Prusinski21, Xu22a} and emission lines \citep[e.g.,][]{Genzel11, Newman12b, Soto12, Freeman19, Swinbank19, Perna20, Perrotta21, Amorin24, Carniani24, Weldon24}. However, it is surprising that we are still unclear if and how the outflow properties derived from the two diagnostics are connected. 

Theoretically, the ions studied here using absorption and emission lines (Section \ref{sec:analysis}) probe similar temperatures of warm-ionized gas (T $\sim$ 10$^{4}$ K). Since the absorption line's column density scales with gas density ($\propto n$) and the emission line's luminosity scales with density squared ($\propto n^2$), these two types of outflow measurements are thought to be weighted towards different regions of the outflow. If both outflows are dominated by similar driving mechanisms (in our case, by the energy and momentum from SF regions), we would hypothesize that the outflow properties from the two diagnostics should be connected.





Given the high-quality UV + optical data for the galaxies in our sample, we next investigate this hypothesis and the possible connections between outflow properties measured from the two different diagnostics. We start by comparing the outflow kinematics in Section \ref{sec:comp:kine}. Then we present the links between outflow sizes in Section \ref{sec:size} and outflow densities in Section \ref{sec:density}.




\subsection{Outflow Kinematics}
\label{sec:comp:kine}

For galaxies with detected outflows in both emission and absorption lines, we start by comparing the maximum outflow velocity (\Vnfout) derived from the two diagnostics. We choose to use \Vnfout\ instead of the central velocity (\VC) of a line. This is because \VC\ is normally quite close to zero for emission lines, given the fact that we can detect emission from both blue- and red-shifted sides of the outflows. On the contrary, \VC\ $\ll$ 0 for all absorption lines since we can only detect them in the foreground of the galaxies.



The comparison of \Vnfout\ is shown in the left panel of Figure \ref{fig:Kine}. The Spearman's correlation results in $\rho$ = 0.795 and p $<$ 10$^{-4}$, which suggest there is a strong positive correlation. The \Vnfout\ values measured from emission lines are similar to the ones measured from absorption lines for weak outflows (100 $<$ \Vnfout\ $<$ 200 km s$^{-1}$), but are smaller for stronger outflows. We show the best linear fit in the orange dashed line as:
\begin{equation}
    \label{eq:V95}
    v_{95}^\text{out} ( \text{H}\alpha ) = (0.68\pm0.09)\times v_{95}^\text{out} (\text{ABS.}) 
\end{equation}
which yields a shallower slope compared to the 1:1 correlation in the red solid line. Several factors may contribute to the relatively low outflow velocity observed in emission lines. First, emission lines are more sensitive to high-density regions near the starburst, where gas velocities are intrinsically lower. Second, the observed velocity of emission and absorption lines are affected differently by projection effects \citep[see, e.g.,][]{Scarlata15, Carr18}; variations in the inclination of the outflowing cone can alter the measured line-of-sight (LOS) velocity. We return to this point later when we present models of outflows in Section \ref{sec:comp:model}.

In the right panel of Figure \ref{fig:Kine}, we show the line widths measured from outflow emission and absorption lines in the same manner. As illustrated in Section \ref{sec:linemea}, we adopt the \HWHMout\ of emission lines in the comparison since this represents the blue-shifted portion of the emission that coincides with the blue-shifted absorption lines. In this case, we also find a tighter correlation with a p-value $<$ 10$^{-6}$. \HWHMout\ from emission lines are typically smaller than the \FWHMout\ measured from absorption lines, where the best linear fit yields:
\begin{equation}
    \label{eq:HWHM95}
    \begin{aligned}
    \text{HWHM}_{95}^\text{out} (\text{H}\alpha ) &= (0.36 \pm 0.07) \times \text{FWHM}_{95}^\text{out} (\text{ABS.}) 
    \end{aligned}
\end{equation}


In Figure \ref{fig:Kine}, we also overlay two different sets of model predictions. This includes the spherical models with different outflow density distributions (see Equations \ref{eq:n_r}) in blue lines and bi-conical outflow models given M 82 parameters in purple stars. In short, we find our models can well explain the observed correlations.  We discuss these models in more detail in Section \ref{sec:comp:model}.



\begin{figure*}
\center

	\includegraphics[page = 1,angle=0,trim={2.2cm 0.1cm 4.7cm 1.8cm},clip=true,width=0.5\linewidth,keepaspectratio]{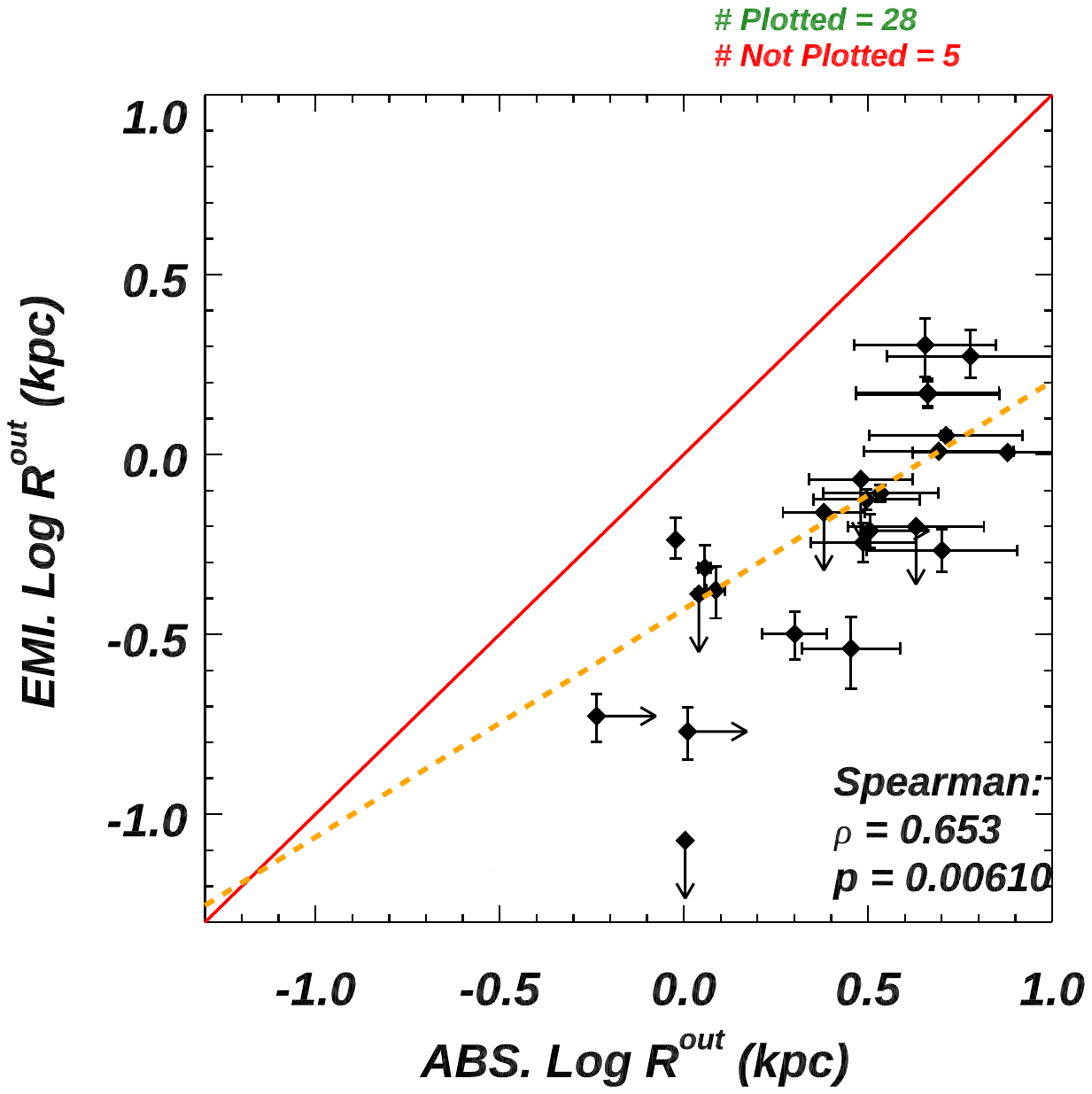}
 	\includegraphics[page = 1,angle=0,trim={2.2cm 0.1cm 4.7cm 1.8cm},clip=true,width=0.5\linewidth,keepaspectratio]{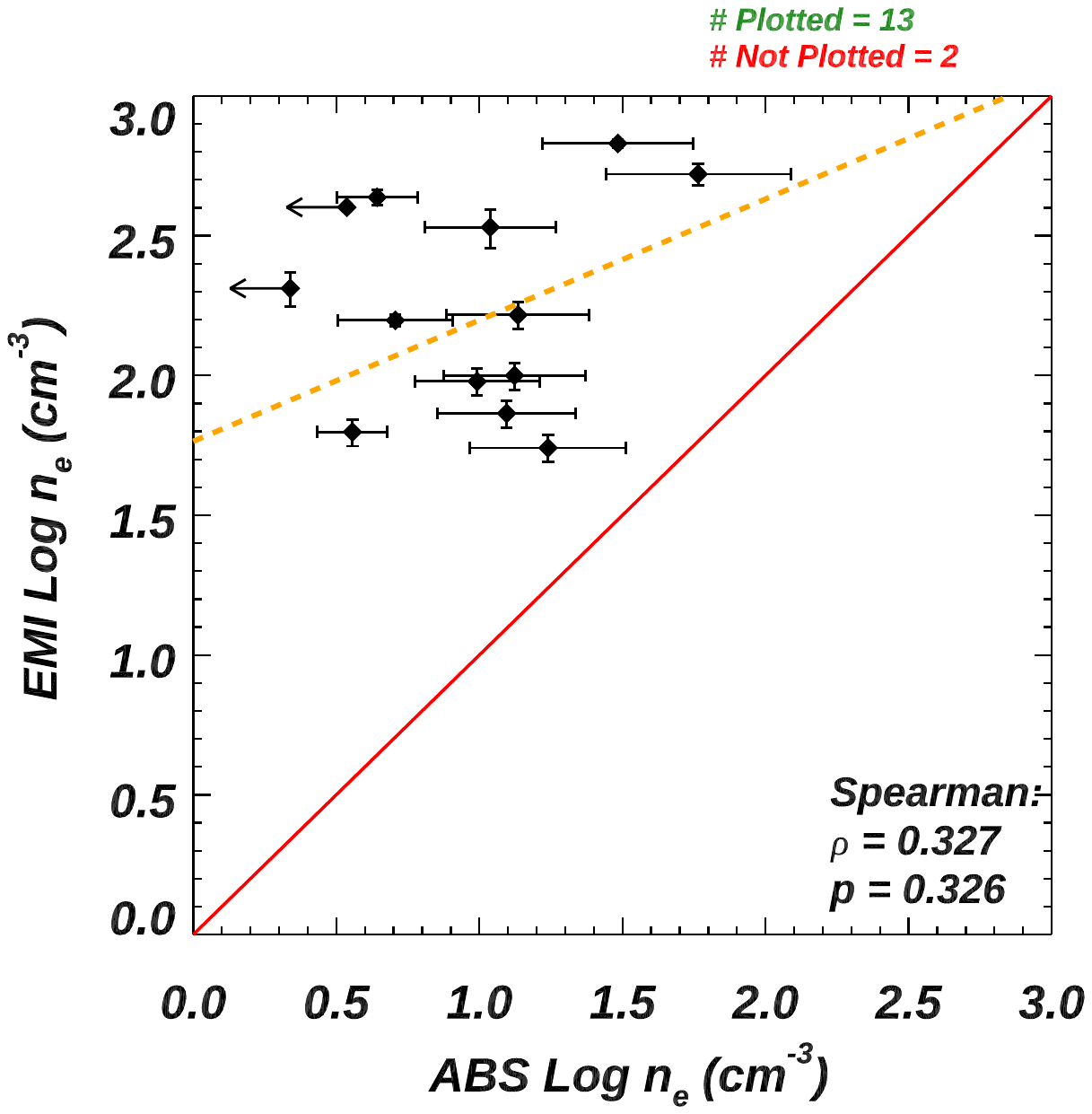}
  
\caption{\normalfont{Same as Figure \ref{fig:Kine} but for comparing the outflow spatial extensions (\textbf{Left}) and outflow density (\textbf{Right}). Values in both axes are in logarithm scales. The red line is the 1:1 relationship to guide the eyes. It is evident that the outflow components in emission lines are weighted towards smaller $R^{out}$ regions and higher \ne\ than that of absorption lines in the same galaxy.}  }
\label{fig:Size}
\end{figure*}

\subsection{Outflow Spatial Extensions}
\label{sec:size}

It is critical to understand whether outflows seen in absorption or emission lines originate from similar spatial scales. Thus, we estimate the outflow sizes (\Rout) from UV and optical spectra separately. For UV absorption lines, given the measured starburst size (\rburst) from the COS NUV acquisition image (Section \ref{sec:ancillary}) for each galaxy, we calculate \RoutUV\ based on the correlation given by \cite{Xu23a}:

\begin{equation}
    \label{eq:RoutP}
    \log(R^\text{out})  = 0.822 \times \log(r_{\star}) + 0.583
\end{equation}
where \Rout\ and \rburst\ are in units of kpc. \cite{Xu23a} established this correlation by showing that specific UV lines (such as \Siii\ and \Siii*) can be used to estimate the electron number density, which in turn is used to calculate \Rout\ and has been shown to scale with \rburst.


For galaxies with Keck/ESI long-slit spectra, we measure outflow sizes from the broad \ha\ emission lines as follows. We first extract the spectra in thin boxes at different locations along the cross-dispersion direction, starting from the center of the trace. Each thin box is chosen to cover the \ha\ + [\nii] wavelength regions with a width of 2 pixels in the cross-dispersion direction. Then we conduct the Gaussian spectral line fitting to the spectra from each box, adopting the same methodology in Section \ref{sec:GuassianFit:OPT}. This allows us to calculate the relative strengths of the broad and narrow components at different spatial locations in the cross-dispersion direction. In turn, we can construct the 2D modeled spectra of the broad and narrow components separately. After that, we convert the measured \ha\ flux in the broad component to surface brightness (SB) and calculate the radius enclosing half of the total integrated SB (i.e., $\int SB(r) \times 2 \pi r dr$). This assumes the outflow SB(r) measured from the broad components are spherically symmetric. Finally, we subtract half of the average seeing (0$\dotarcsec$7/2 = 0$\dotarcsec$35) in quadrature from the resulting radius. In cases where we find radius is less than 0$\dotarcsec$35, we consider the outflows unresolved and report the measured outflow radius (\Rout) as an upper limit.

In Figure \ref{fig:Size}, we compare the two sets of \Rout\ derived above, where we find a strong positive correlation with a p-value $\sim 6\times10^{-3}$ . The lower and upper limits are shown as arrows, while the 1:1 relationship is in red to guide the eyes. The best linear fit is shown as the orange dashed line, which has the form as:

\begin{equation}
    \label{eq:Rout}
    \begin{aligned}
    \log [R^\text{out}_\text{\tiny OPT}] &= (0.63 \pm 0.2) \times \log [R^\text{out}_\text{\tiny UV}] \\
    &\quad - (0.43 \pm 0.10) \ \text{kpc}
    \end{aligned}
\end{equation}

On average, we find \RoutOPT\ is $\sim$ 2 -- 5 times smaller than \RoutUV. This is consistent with the fact that emission line strengths are weighted towards higher density regions, i.e., $\propto$ $n^2$ versus $\propto$ $n$ for absorption lines. Thus, the observed emission line fluxes should be dominated by gas that is closer to the centers of the galaxies. We also find this matches our model predictions described in Section \ref{sec:comp:model}.\\


\subsection{Outflow Density}
\label{sec:density}
Outflow electron number density (\ne) is another critical parameter that determines the observed line profiles of the outflowing gas. For emission lines, we follow the commonly adopted method to constrain \ne\ of the low-ionization zone \citep[e.g.,][]{Mingozzi22}, based on the density-sensitive lines of [\sii] \ly\ly 6717, 6731. Here, we are able to isolate the outflow \ne\ by measuring only the blueshifted, broad component of emission lines. We adopt \textit{PyNeb} and calculate electron temperature (T$_\text{e}$) from [\oiii] 4363/([\oiii] 4959 + 5007) simultaneously \citep[e.g.,][]{Berg22}. The errors are calculated using MC simulations by perturbing the input line flux 10$^{3}$ times according to the observed 1$\sigma$ uncertainties. In total, we have 15 galaxies that have a clean, measurable [\sii] doublets.

For outflows detected in absorption lines, \cite{Xu23a} have shown that \Rout\ and \ne\ are correlated in local SF galaxies. Given earlier assumption that we take in Section \ref{sec:size}, i.e., detected outflows in UV are in pressure equilibrium with the hot wind fluid, we have:

\begin{equation}
\label{eq:neoutP}
    n_\text{e, UV} = \frac{\dot{p}_\text{SFR}} {8 \pi (R_\text{UV}^\text{out})^2  k T} 
\end{equation}
where $\dot{p}_\text{SFR}$ is the total momentum flux of the hot wind that can be estimated given SFR \citep{Heckman15}, $k$ is the Boltzmann constant, and $T$ = 10$^4$ K is the assumed temperature. The measured \ne\ values are listed in Table \ref{tab:linemea}. There exist five galaxies that already have direct \ne\ measurements in \cite{Xu23a} and they are consistent with what we measured here within errorbars.

In the right panel of Figure \ref{fig:Size}, we compare the two sets of outflow densities derived from absorption and emission lines. It is evident that \ne\ from the [\sii] emission lines is consistent with being larger than \ne\ from absorption lines in all cases. This is again a direct expectation from the $\propto n^{2}$ scaling for emission lines versus $\propto n$ scaling for absorption lines. However, we don't find a strong linear correlation between the two sets of \ne. This is likely because \ne\ is spatially dependent and the measured densities from unresolved spectra have more uncertainties due to, e.g., orientation of the outflows to the LOS and the slope of $n(r)$ distribution. Future spatially resolved observations and simulations of galactic outflows will provide more insights on this problem.

\begin{figure}
\center

	\includegraphics[page = 1,angle=0,trim={2.2cm 0.1cm 4.7cm 1.8cm},clip=true,width=1.0\linewidth,keepaspectratio]{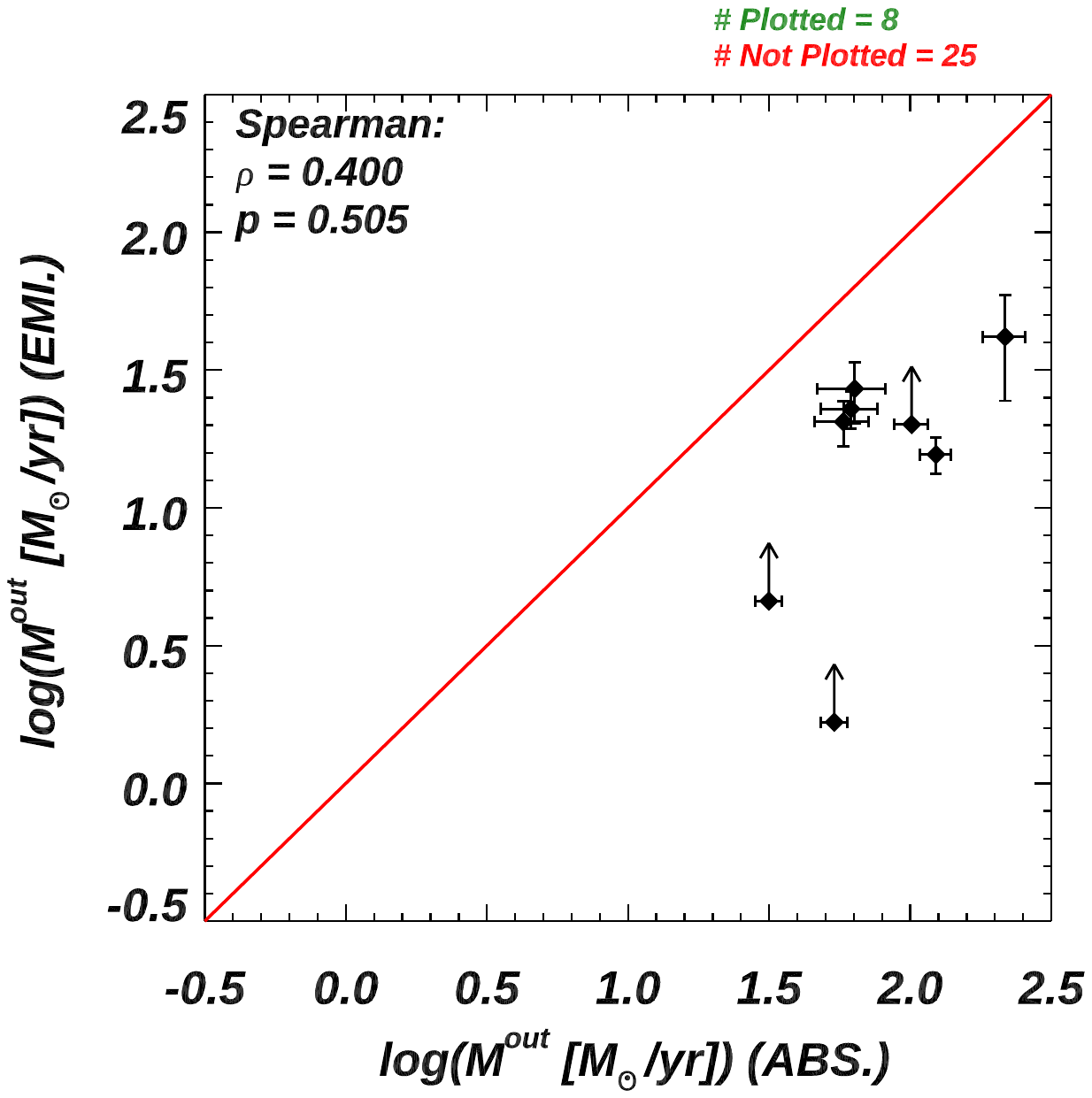}

\caption{\normalfont{Same as Figure \ref{fig:Kine} but for comparing the mass outflow rates derived from the two diagnostics. Since we need to estimate both the column density and number density for a galaxy, there are far fewer objects with such data in our sample. The outflow rates from the two diagnostics show positive correlations. Moreover, the one derived from emission lines is about 0.5 dex lower than the one from absorption lines for the same galaxy. See more discussion in Section \ref{sec:rate}.}}  
\label{fig:OutflowRate}
\end{figure}

\subsection{Outflow Rates}
\label{sec:rate}

To constrain the strength of feedback effects from galactic outflows, the key parameter to calculate is the how much mass they carry per unit time. To do so, we estimate mass outflow rates (\Mdot) from optical-emission and UV-absorption lines, separately, then compare them as follows.


For emission lines, we follow the outflow models described in \cite{Genzel11, Newman12b}, which yield:

\begin{equation}
\label{eq:Mdot_out_OPT}
    \dot{M}^\text{out}_\text{\tiny OPT} = \frac{1.36 m_p}{\gamma_{\text{H}\alpha} n_\text{e,\tiny{OPT}}} \left( L_{\text{H}\alpha, \text{Broad}} \right) \frac{v^\text{out}_\text{\tiny OPT}}{R^\text{out}_{\text{\tiny OPT}}} 
\end{equation}
where $m_p$ is the proton mass, $\gamma_{\text{H}\alpha} = 3.56 \times 10^{-25}$ erg cm$^{-3}$ s$^{-1}$ is the H$\alpha$ emissivity for photoionized gas at the electron temperature $T = 10^4$ K, $ n_\text{e,\tiny{OPT}}$ is the electron density of the outflowing gas, $L_{\text{H}\alpha,\text{Broad}}$ is the extinction-corrected, slit-loss corrected H$\alpha$ luminosity measured from the broad component, and $v^\text{out}_{\text{\tiny OPT}}$ and $R^\text{out}_{\text{\tiny OPT}}$ are the outflow velocity and size measured from optical-emission lines. 

To calculate \MdotUV, we follow the methodology in \cite{Xu22a}, which yield:

\begin{equation}\label{eq:Mdot_out_UV}
    \begin{aligned}
        \dot{M}^\text{out}_\text{\tiny UV} \simeq \Omega \mu m_p R^\text{out}_{\text{\tiny UV}} \int \frac{dN_\text{H}}{dv^\text{out}_\text{\tiny UV}}\times v^\text{out}_\text{\tiny UV}dv^\text{out}_\text{\tiny UV}
    \end{aligned}
\end{equation}
where $\Omega$ is the solid angle of outflow, which has been found to be $\sim$ 4$\pi$ for local starburst galaxies \citep{Xu22a}, $R^\text{out}_{\text{\tiny UV}}$ is the outflow sizes measured in UV (Section \ref{sec:size}), $v^\text{out}_\text{\tiny UV}$ is the outflow velocity measured in UV, and \Nh\ is the outflow's total hydrogen column density, which is dependent on $v$.

For these calculations, a galaxy must exhibit clear detections of \Siii\ and \siiv\ doublet absorption lines to measure \Nh, as well as \sii\ doublet emission lines to determine $n_\text{e,\tiny{OPT}}$. These criteria are met by only a small subset of our sample (8 out of 33 galaxies), and their results are presented in Figure \ref{fig:OutflowRate}. We find  \MdotOPT\ is consistently lower than \MdotUV\ by approximately 0.2 -- 0.5 dex. This may reflect a combination of uncertainty in the correction for aperture loss and the weighting of the mean density for the emission-line gas, which will preferentially sample dense gas with relatively small mass. In addition, for these galaxies, we have checked that the outflowing gas traced by absorption lines are close to be fully ionized (median \Nhi/\Nh $\sim$ 0.2\%). This is consistent with the assumptions made to calculate \MdotOPT\ in Equation \ref{eq:Mdot_out_OPT}. Thus, the differences of ionization corrections between UV-absorption and optical-emission lines are minor.

This discrepancy in \Mdot\ qualitatively agrees with previous studies of local SF galaxies \citep{Wood15}, which report that \MdotOPT\ is $\sim$ 10 times smaller than \MdotUV. Furthermore, this difference may explain the generally low \MdotOut\ derived from emission lines in low-mass SF galaxies at cosmic noon \citep[e.g.,][]{Davies19, Concas22}. We discuss more about this point in Section \ref{sec:scale}. Future larger samples will also be helpful to understand the discrepancy better.


\section{Discussion}
\label{sec:discuss}

\subsection{Outflows in Emission and Absorption Lines from Analytical Models}

In this subsection, we explore analytical models of galactic outflows to interpret the observed emission and absorption line profiles using unified density and velocity distributions within a single galaxy. We begin with a simple spherical outflow model (Section \ref{sec:comp:simplemod}) and then adopt a more realistic bi-conical outflow model (Section \ref{sec:comp:modelSALT}). While numerous analytical and numerical outflow models exist \citep[see reviews in][]{Somerville15,Naab17, Thompson24}, our analysis aims to provide an initial assessment of how the models can explain the emission and absorption outflow observations simultaneously.

\label{sec:comp:model}


\subsubsection{Simple Analytical Models of an M82-like \\Galactic Wind}
\label{sec:comp:simplemod}

In an idealized case, we assume the outflows consist of spherically expanding thin shells. Inspired by \cite{Scarlata15} and \cite{Carr18}, we derive emission and absorption line profiles as follows. We first assume velocity and density fields, i.e., $v(r)$ and $n(r)$, that match the ones measured in M 82 galactic outflows \citep{Xu23c}:
\begin{equation}\label{eq:v_r}
    \begin{aligned}
    v(r)    &= v_\infty \left (1 - R_\text{SF}/r  \right )^{\beta}
    \end{aligned}
\end{equation}
where v$_\infty$ is the terminal velocity of outflows, $\beta$ is the power index, and R$_\text{SF}$ is the launch radius of the outflow, which we assume it equals the size of the star-forming region. We start with v$_\infty$ = 800 km s$^{-1}$, R$_\text{SF}$ = 300 pc, and $\beta = 1$ to match the observed ones seen in M 82. Similarly for $n(r)$, we take:
\begin{equation}
    \begin{aligned} \label{eq:n_r}
    n(r)    &= n_{0} \left (R_\text{SF}/r  \right )^\gamma\\
    f\!f(r) &= f\!f_{0} \left (R_\text{SF}/r  \right )^\delta 
    \end{aligned}
\end{equation}
where $n_{0}$ = 372 cm$^{-3}$ is the observed outflow density at R$_\text{SF}$ and $f\!f(r)$ is the volume-filling factor, which represents how much volume filled by the \ha\ emitting gas. Spatially resolved studies of galactic outflows in M~82 yield $\gamma$ = 1.17, $f\!f_{0}$ = 0.378\% and $\delta$ = 1.8 \citep{Xu23c}. Then the predicted flux of \ha\ from each shell is:

\begin{equation}\label{eq:dF}
    \begin{aligned}
    dF_{H\alpha}(r)    &\propto n(r)^{2} \cdot f\!f(r) \cdot d\text{Vol}(r) \\
                    &= n(r)^{2} \cdot f\!f(r) \cdot 4\pi r^2 dr\\ 
    \end{aligned}
\end{equation}
where $\text{Vol}(r)$ is the volume of a shell at $r$. It can also be shown that, for a spherically symmetric outflow shell, this flux is evenly distributed between [--$v(r)$, +$v(r)$] due to velocity projection effects \citep{Peng24}. Thus, we get:

\begin{equation}\label{eq:dF_dv}
    \begin{aligned}
    \frac{dF_{H\alpha}(r)}{dv} = \frac{n(r)^{2} \cdot f\!f(r) \cdot 4\pi r^2 dr}{2v(r)}
    \end{aligned}
\end{equation}

\begin{figure*}
\center

	\includegraphics[page = 1,angle=0,trim={0.35cm 0.1cm 0.1cm 0.0cm},clip=true,width=0.5\linewidth,keepaspectratio]{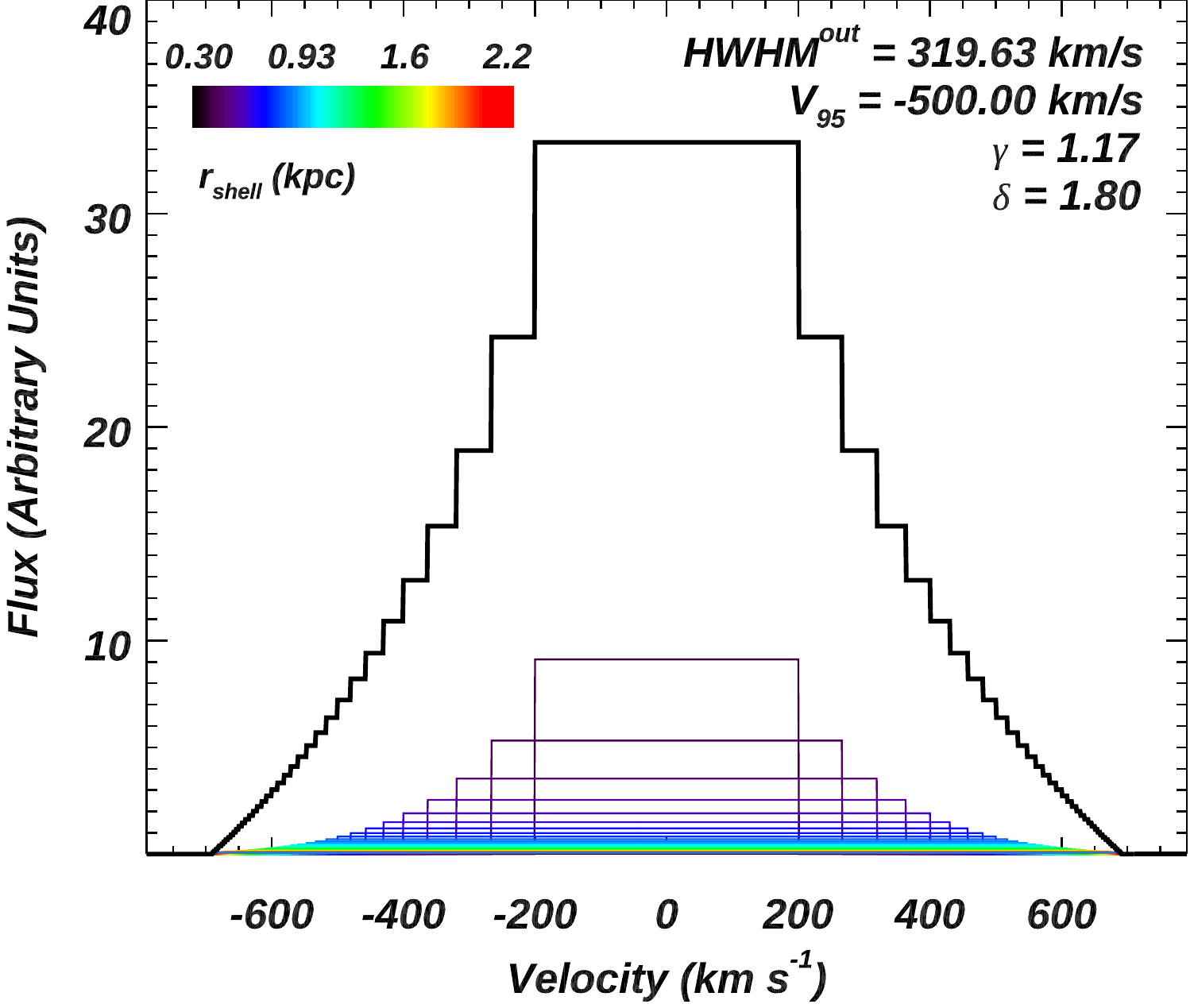}
    \includegraphics[page = 1,angle=0,trim={0.35cm 0.1cm 0.1cm 0.0cm},clip=true,width=0.5\linewidth,keepaspectratio]{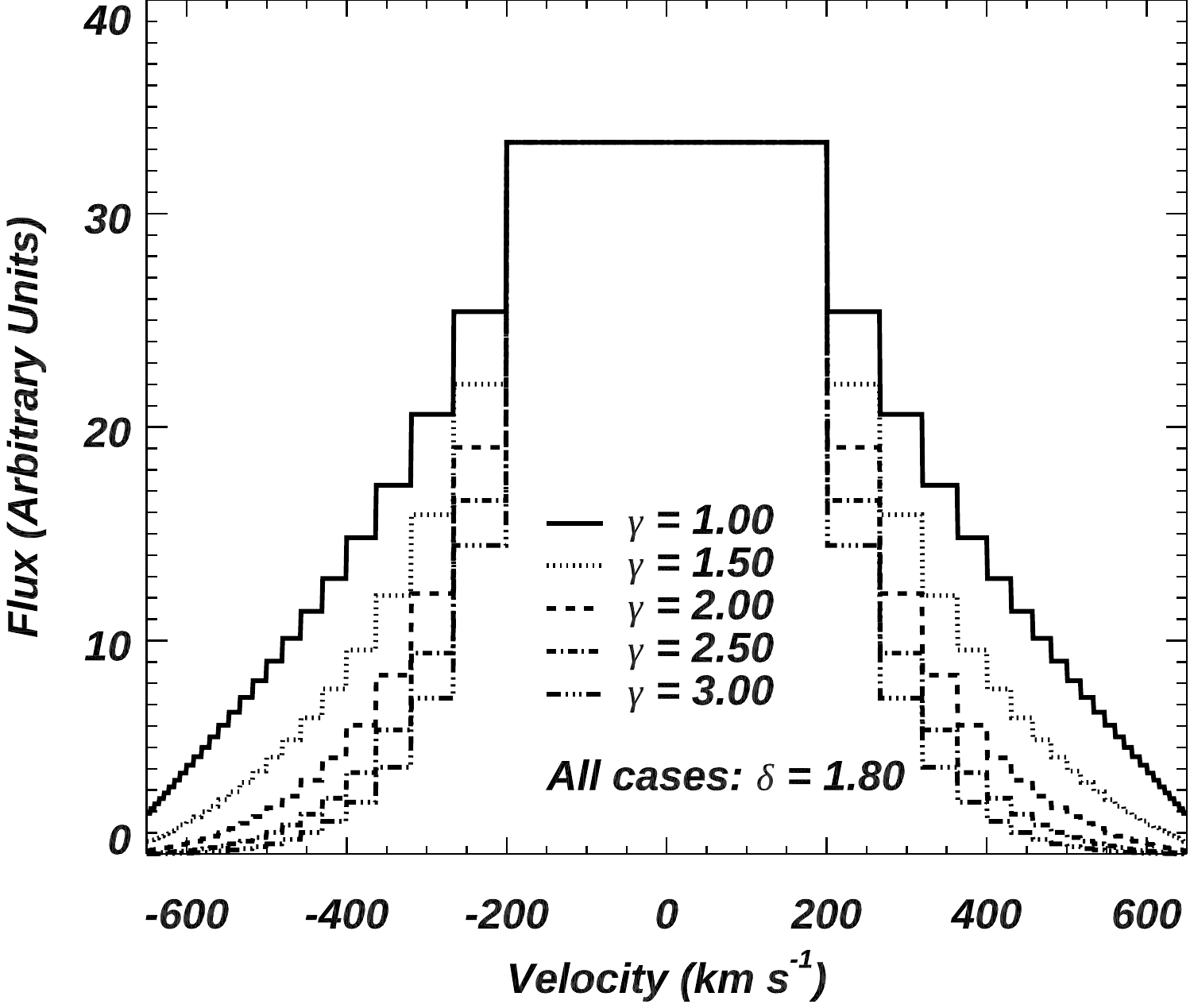}
	
\caption{\normalfont{Modelled \ha\ outflow emission line profiles assuming spherical expanding shells for galactic outflows. \textbf{Left:} We adopt the observed radial distributions of velocity, density, and volume filling factor that are measured from M 82 galactic outflows \citep{Xu23c}. The \ha\ emission from each shell (with thickness of 0.1 pc) is shown as a rectangle with different colors, while the sum of emission from all shells is shown in a thick black line. The measured half-width-half-maximum (\HWHMout, see Section \ref{sec:comp:kine}) of the line and V$_{95}$ of the blue-shifted emission are listed at the top-right corner. We do not include the static ISM component in these models. \textbf{Right:} Same as the left panel but we compare the results given different outflow density profiles $n(r) \propto r^{-\gamma}$. A steeper $n(r)$ distributions yield narrower emission line profiles. See more discussions in Section \ref{sec:comp:simplemod}.}  }
\label{fig:Model1-EMI}
\end{figure*}

The $dF/dv$ values for each shell are shown as the colored lines in the left panel of Figure \ref{fig:Model1-EMI} for M 82 outflows. The sum of the emission from all shells is shown in the thick black line.  The summation goes from $r$ = 0.3 -- 2.2 kpc to match the range adopted in \cite{Xu23c} for M 82. We present the measured \HWHMout\ and $v_{95}$ in the top-right corner. It can be seen that the \ha\ emission line flux is clearly dominated by the shells at small radius (i.e., in purple colors). This is as expected since these shells have the highest density and the \ha\ flux is proportional to $n^2\cdot f\!f(r)$.

As Equation \ref{eq:dF_dv} shows, the emission-line profiles will be determined by the radial drop in both density ($n(r)$) and volume filling factor ($f\!f(r)$). 
In the right panel of Figure \ref{fig:Model1-EMI}, we compare models with different choices of $\gamma$ from 1.0 to 3.0, standing for $n(r) \propto r^{-1}$ to $r^{-3}$. In all cases we fix $f\!f(r) \propto r^{-1.8}$. It is apparent that the modeled line profiles become more concentrated towards the lower velocities given higher $\gamma$ values. The measured \HWHMout\ also decreases from 180 km s$^{-1}$ to 100 km s$^{-1}$ and $v_{95}$ drops from --520 km s$^{-1}$ to --310 km s$^{-1}$ given the increasing $\gamma$ values.  Note that in all cases, we do not include the static ISM components, which can be comparable to, or stronger than, the outflow components.


\begin{figure*}
\center

	\includegraphics[angle=0,trim={0.35cm 0.07cm 0.1cm 0.1cm},clip=true,width=0.5\linewidth,keepaspectratio]{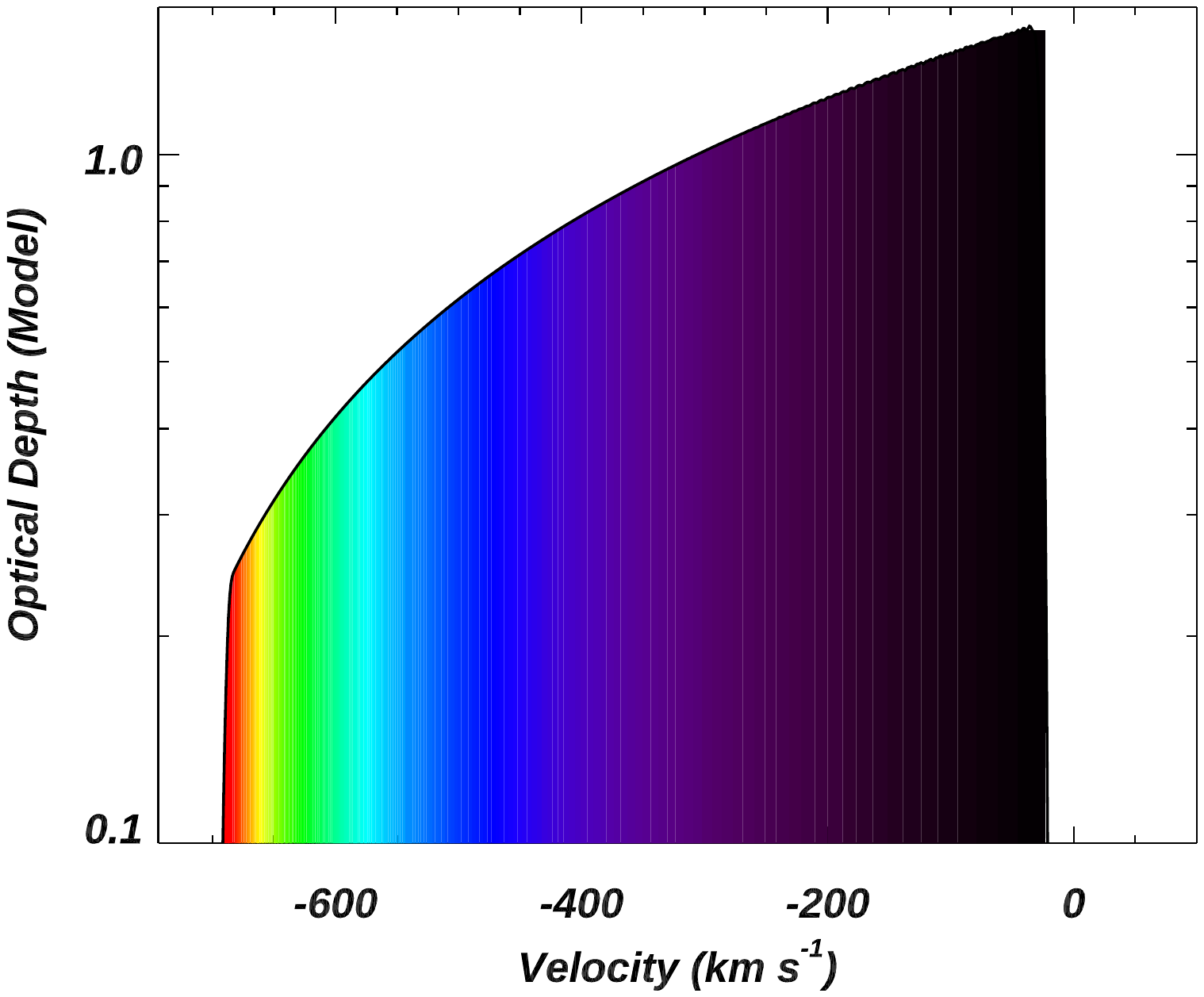}
	\includegraphics[page = 2,angle=0,trim={0.35cm 0.24cm 0.1cm 0.0cm},clip=true,width=0.5\linewidth,keepaspectratio]{./M82ABSMod_MoreRbins_MoreModelSiII_1260.42_ga1.17.pdf}

\caption{\normalfont{Absorption line profiles assuming the same spherical outflow settings as the left panel of Figure \ref{fig:Model1-EMI}. \textbf{Left:} Absorption line optical depths at each velocity. The color-codes are the same as Figure \ref{fig:Model1-EMI}, illustrating contributions from different shells at different radii.  \textbf{Right:} Normalized flux calculated given the optical depth in the left panel. It is clear that the gas produce outflow absorption lines with broader line width and larger maximum velocity than those in emission lines. These trends are consistent with our observed correlations in Section \ref{sec:compare}. See discussion in Section \ref{sec:comp:simplemod}.}  }
\label{fig:Model1-ABS}
\end{figure*}

Under the same assumptions, we can also estimate the outflow absorption lines profiles. In this sample model, we only consider the absorption but not the scattering. Then we can write down the cross section of an absorption line as:

\begin{equation}\label{eq:sigma_v}
    \begin{aligned}
    \sigma(v,r)    &= \sigma_{0}\phi(\nu,r)
                 &= \frac{\pi e^{2} }{m_e c^2} \cdot f \cdot\phi(\nu,r)
    \end{aligned}
\end{equation}
where $e$ is the electron charge, $m_e$ is the electron mass, $c$ is the speed of light, $f$ is the oscillator strength of the transition (which we assume \Siii\ \ly 1260 hereafter), and $\phi(\nu,r)$ is the line profile function, which is dependent on both the frequency $\nu$\ and shell radius $r$. Then we take the Doppler broadening in a Gaussian profile as:

\begin{equation}\label{eq:phi_v}
\phi(\nu,r) = \frac{1}{\sqrt{\pi} \Delta \nu_D} \exp \left( -\frac{(\nu - \nu_0)^2}{(\Delta \nu_D)^2} \right)
\end{equation}
where $\nu_0$ is the central frequency of the line\footnote{The frequency $\nu$ is mapped to the wavelength given $\lambda = c/\nu$. For a certain line, $\lambda$ is then mapped to the velocity ($v$) according to the relativistic Doppler effect.}, and $\Delta \nu_D$ = $\nu_0/c \sqrt{2 k_B T/m_\text{ion}}$ is the Doppler width assuming thermal broadening. For \Siii\ \ly 1260 with T = 10,000 K, this $\Delta \nu_D$ can be converted to $\sim$ 0.01\AA\ or 2.4 km s$^{-1}$. Then we can get the optical depth velocity distribution as:

\begin{equation}\label{eq:tau_v}
    \begin{aligned}
    \tau(v)    &= \int f\!f(r) \cdot n_\text{ion}(r) \cdot \sigma(v,r) dr
    \end{aligned}
\end{equation}
where we adopt the same integration range ( $r = 0.3 - 2.2$ kpc) as we used for calculating the emission lines profiles, and $n_\text{ion}(r)$ is the ion number density.  For \Siii, we adopt $n_\text{\Siii}/n_\text{H}$ $\sim$ 10$^{-6.5}$ in local SF galaxy outflows \citep[][]{Xu22a}. The integration only considers the part of the shells that are in front of the starburst \citep[e.g.,][]{Scarlata15}.
We include $f\!f(r)$ here since we assume the ion is not filling the whole volume as seen in M 82 (Equation \ref{eq:n_r}), but in the calculations, $n_\text{ion}(r)$ and $f\!f(r)$ are degenerate.

Finally, we can calculate the normalized flux for absorption lines assuming a full coverage model:

\begin{equation}\label{eq:I_v}
    \begin{aligned}
    I/I_{0} = e^{-\tau(v)}
    \end{aligned}
\end{equation}
The resulting optical depth and absorption line profiles are shown in Figure \ref{fig:Model1-ABS}. For $\tau(v)$, we also draw the contributions from each shell in the same color schemes as emission lines in the left panel of Figure \ref{fig:Model1-EMI}.  We find that the absorption lines are clearly broader and reach higher velocities than the emission lines. This is because we can still detect significant absorption at higher velocities that are dominated by outer shells. On the contrary, the emission from these outer shells are negligible as seen in Figure \ref{fig:Model1-EMI}. This difference is mainly because emission lines are naturally dominated by the higher density regions than absorption lines.

\begin{figure}
\center

	\includegraphics[page = 1,angle=0,trim={0.5cm 0.2cm 0.1cm -0.2cm},clip=true,width=1.0\linewidth,keepaspectratio]{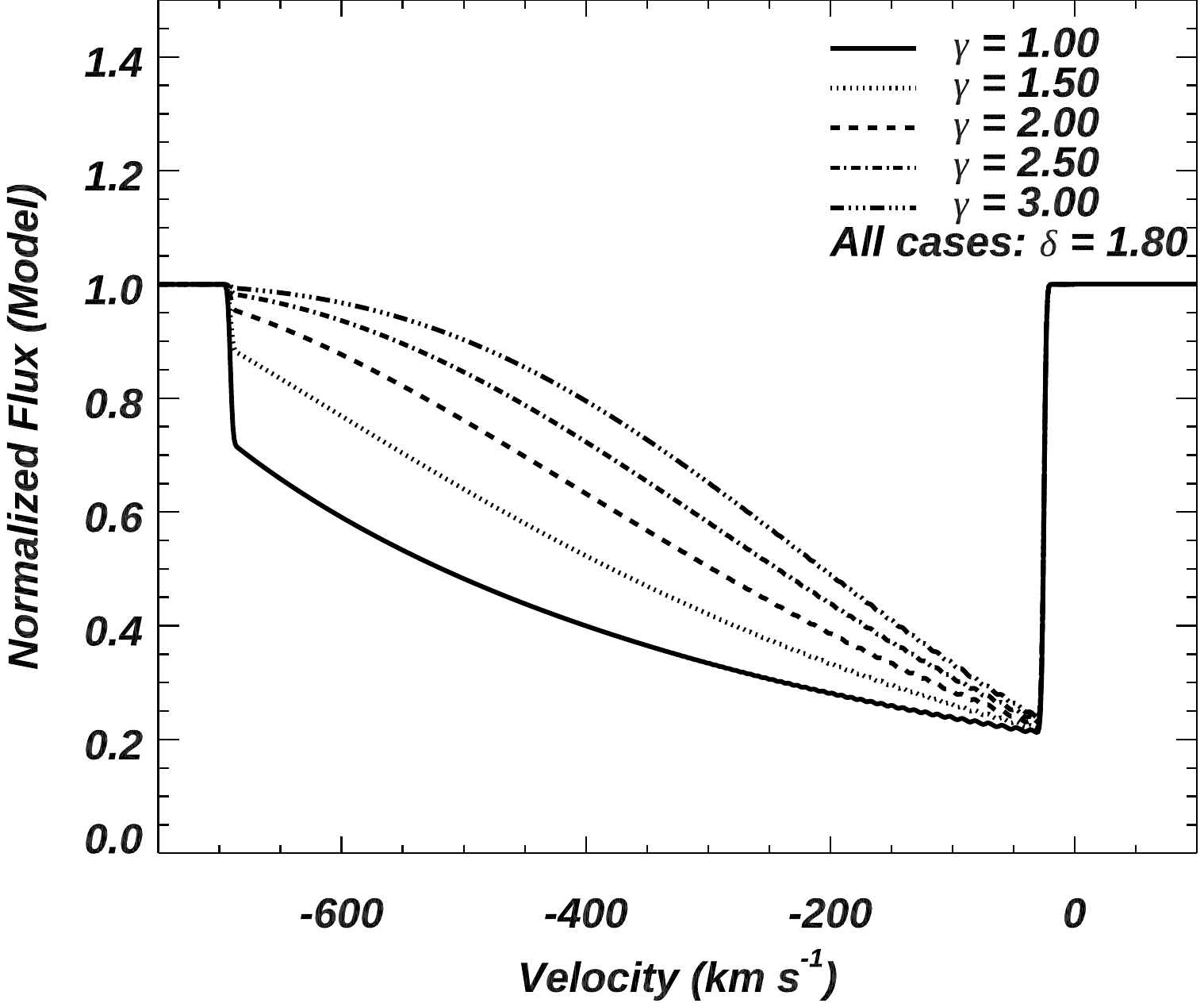}

\caption{\normalfont{Same as Figure \ref{fig:Model1-ABS} but we compare the outflow absorption line profiles given different outflow density profiles $n(r) \propto r^{-\gamma}$ and the same volume filling factor, $f\!f(r) \propto r^{-1.8}$ (see Equation \ref{eq:n_r}). A steeper $n(r)$ distributions yield more column densities piled up at lower velocity.  See discussion in Section \ref{sec:comp:simplemod}.}  }
\label{fig:Model1-ABS2}
\end{figure}

Additionally, we compare modeled outflow absorption lines profiles with different $\gamma$ values in Figure \ref{fig:Model1-ABS2}. The model with a larger $\gamma$ results in more column densities at lower velocities. To further quantify the similarities between observations (Section \ref{sec:compare}) and models of outflow absorption and emission lines, we overlay the modeled results into Figure \ref{fig:Kine} as the blue lines. Each line represents a set of models with the same $\gamma$ but different $v_\infty$ values. We find the model with $\gamma \sim $ 2.0 ($\delta$ fixed at 1.8) is on-average the best match to both the overall correlations in maximum outflow velocity and line widths. But there are also galaxies consistent with significantly smaller ($\sim$ 1) or larger ($\sim$ 3) $\gamma$ values. This is likely because the simplicity of our models here, which have ignored several factors, e.g., outflow opening angle and orientation. We consider them in a more sophisticated model in Section \ref{sec:comp:modelSALT}. 



%


Our simple spherical shell models qualitatively reproduce the observed kinematic correlations between outflows seen in emission and absorption lines. Specifically, we find that the same outflow radial and density distributions result in narrower emission-line widths and lower maximum velocities compared to absorption lines, a consequence of density weighting and the line-of-sight velocity integration. 
Additionally, our models indicate that emission lines originate preferentially from denser, inner shells, while absorption lines can probe materials at larger radii and lower densities (see Figure \ref{fig:Size}).

Furthermore, we confirm that these simple spherical models yield results consistent with more complex bi-conical outflow models (see Section \ref{sec:comp:modelSALT} below) when the cones’ half-opening angle is set to $90^{\circ}$. Finally, we highlight that the correlations we derive (e.g., Equations \ref{eq:V95}, \ref{eq:HWHM95}, \ref{eq:Rout}, and \ref{eq:FWHM95}) can be used to estimate the properties of outflow absorption lines from emission lines and vice versa, providing a valuable tool for interpreting observational data.

\subsubsection{SALT Models of Galactic Outflows}
\label{sec:comp:modelSALT}

In reality, we know the hot galactic wind and warm outflows in M 82 are not totally spherical symmetric but more close to bi-cones \citep[e.g.,][]{Shopbell98, Xu23c}. This is not uncommon among SF galaxies \citep{Burchett21,Carr21}. Furthermore, \cite{Xu22a} found the high detection rates of outflows in local starburst galaxies suggest they cover large fractions ($\sim$ 90\%) of the galaxy's solid angle. These do not conflict with the bi-cone geometry, but instead suggest the opening angle of the cone should be large for local starburst galaxies.

As such, wind geometry can also affect the appearance of the emission and absorption line components. For instance, a bi-cone oriented perpendicular or parallel to the line of sight can result in an emission or an absorption profile, respectively, and the observed velocities are projected \citep{Carr18}. Here we explore the implications of the bi-conical wind geometry observed in M 82 on the resulting line profiles using an adaptation of the Semi-analytical Line Transfer (SALT) model presented in \cite{Carr23}. We note SALT already considered resonantly scattered emission. In addition, we adapt SALT to allow for the velocity field described in Equation~\ref{eq:v_r} and develop a new feature in SALT to account for \emph{in situ} \ha\ emission emanating from the bi-cone. In all calculations in SALT, we also adopt the same density profile and initial parameters (e.g., v$_\infty$, R$_\text{SF}$, and n$_\text{0}$) as the model described in Section \ref{sec:comp:simplemod}. The detailed models are illustrated in Appendix \ref{app:sec:SALT}.

\begin{figure}
\center

	\includegraphics[page = 1,angle=0,trim={0.0cm 1.4cm 0.1cm 0.2cm},clip=true,width=1.0\linewidth,keepaspectratio]{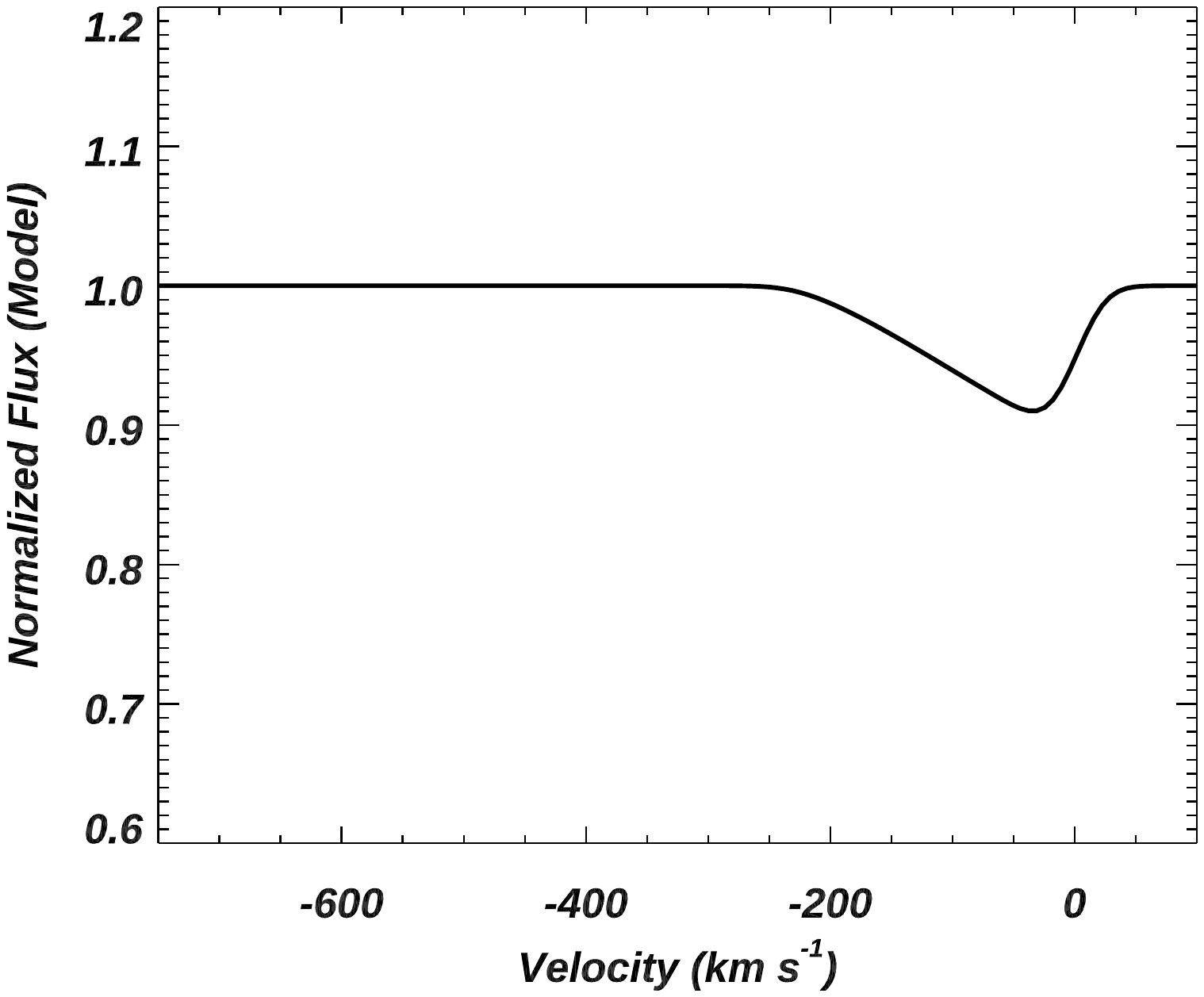}

	\includegraphics[page = 2,angle=0,trim={0.0cm 0.2cm 0.1cm 0.2cm},clip=true,width=1.0\linewidth,keepaspectratio]{SALT_model_deg75.pdf}
 
\caption{\normalfont{SALT model predictions of M 82-like galactic outflows given the same density profile and initial parameters (e.g., v$_\infty$, R$_\text{SF}$, and n$_\text{0}$) as the model described in Section \ref{sec:comp:simplemod}.  The top panel shows the absorption profile for Si II 1260\AA\ line.  Since M 82's bi-conical outflow is oriented away from the line of sight by $\sim$ $75^{\circ}$ \citep{Shopbell98}, the covering fraction is small resulting in a shallow absorption well.  The bottom panel shows the H$\alpha$ line emanating from the outflow.  Here, the orientation of the outflow results in a line profile with a single peak centered around zero velocity. See discussion in Section \ref{sec:comp:modelSALT}. }  }
\label{fig:SALT_model_profile}
\end{figure}

We show the resulting absorption and emission line profiles in Figure~\ref{fig:SALT_model_profile}, where we assume the outflow has a half-opening angle of $37^{\circ}$ and the bi-cone is oriented at $75^{\circ}$ away from the line of sight as observed in M 82 \citep{Shopbell98, Xu23c}. This orientation results in a low covering fraction and shallow absorption trough. On the other hand, the orientation favors a prominent $\rm H\alpha$ emission spike roughly centered on velocity = 0. We then measure these line profiles using the same method in Section \ref{sec:linemea} and overlay the results as the purple stars in Figure \ref{fig:Kine}. We similarly show results with two additional bi-cone orientations ($60^{\circ}$ and $45^{\circ}$) with other parameters the same as M 82. The order of them is listed in the parentheses.

We find SALT models show quite consistent results with our observed galaxies in the cases of $60^{\circ}$ and $45^{\circ}$. When the bi-cone is oriented $75^{\circ}$ away from the LOS (i.e., the true value for M 82), the modeled outflow velocity and widths are close to the margin of our observed galaxies. This is not surprising since none of our galaxies should have bi-cone outflow orientations close to that of M 82, which is too perpendicular to the LOS, so there is no UV background light to produce the absorption lines when observed from Earth. 

This test provides an example where the outflow geometry can alter the observed outflow line profiles (especially absorption) for a specific galaxy, but measured outflow properties from emission and absorption lines are still in agreement with the empirical correlations found in Section \ref{sec:compare}. More robust fits of the models to each of our galaxies are beyond the scope of the paper, but we plan to investigate this topic further in a future publication.
 

\begin{figure}
\center

	\includegraphics[page = 1,angle=0,trim={2cm 0.1cm 4.7cm 1.8cm},clip=true,width=1.0\linewidth,keepaspectratio]{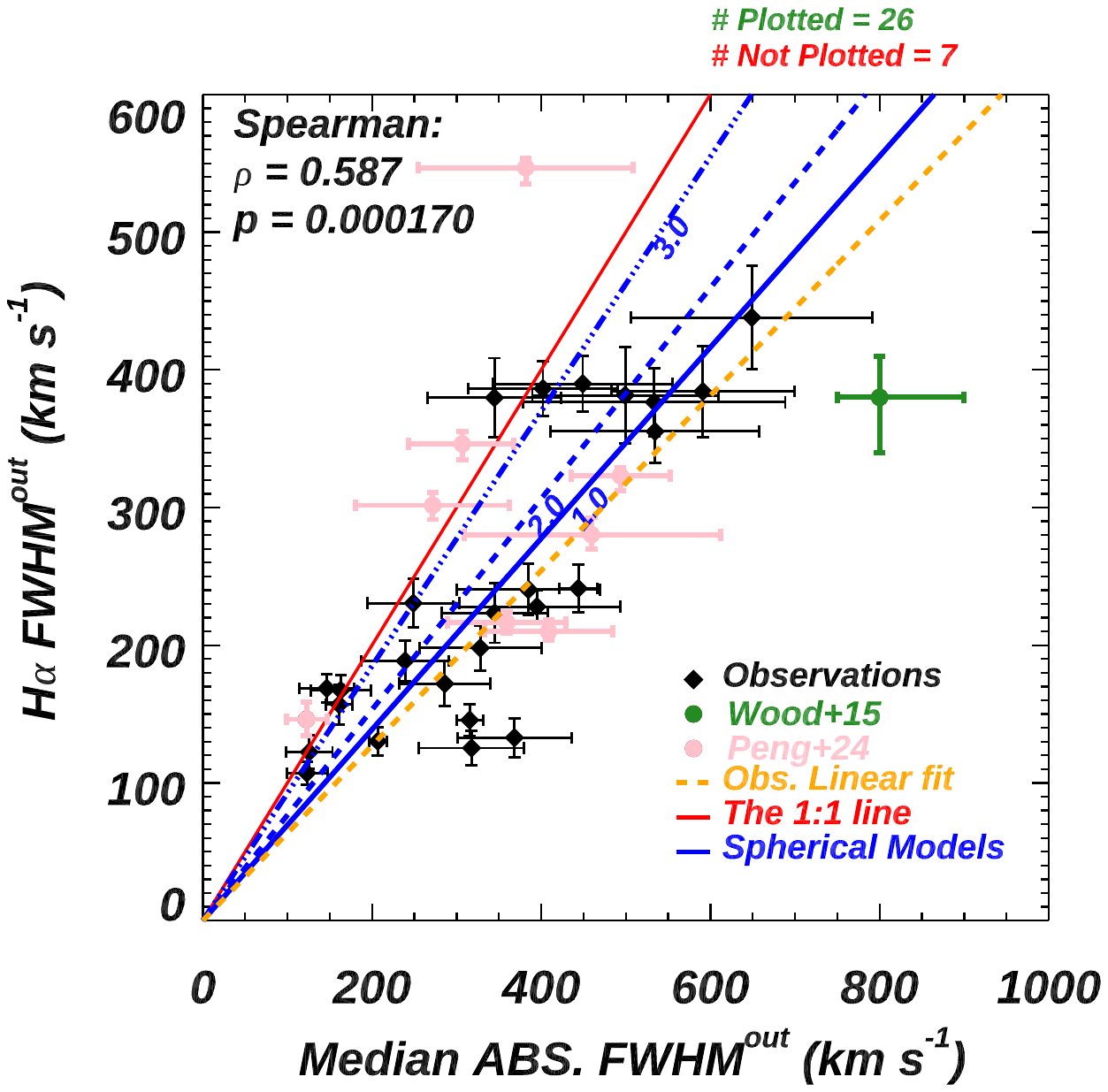}
\caption{\normalfont{Same as the right panel of Figure \ref{fig:Kine} but the y-axis is \FWHMout\ instead of \HWHMout. We also overlay multi-diagnostic warm-ionized outflow detections from the literature in different colors. See discussion in Section \ref{sec:comp_other}.}  }
\label{fig:Kine2}
\end{figure}

\subsection{Other Studies of Comparing Outflow Diagnostics}
\label{sec:comp_other}
In Figure \ref{fig:Kine2}, we compare our results with previous studies of warm-ionized galactic outflows given both the optical emission and UV absorption line diagnostics. In this case, we show \FWHMout\ (instead of \HWHMout\ in Figure \ref{fig:Kine}), which is to match the common line width measurements adopted in the literature. All symbols and lines are the same as Figure \ref{fig:Kine} except the green and pink ones from the literature as follows.

We first show NGC 7552 reported in \cite{Wood15} in the green cross. It is a barred spiral galaxy with log \Mstar = 10.52 \Msun\ and SFR = 13 -- 47 \Msun/yr \citep{Sheth10, Pan13}. \cite{Wood15} find the broad components in \ha\ (\FWHMout\ up to 380 km s$^{-1}$) is much smaller than that from UV absorption lines, such as \Siii, \siiii, and \siiv\ ($\sim$ 800 $\pm$ 100 km s$^{-1}$). This yields a bit lower \ha\ \FWHMout\ than our trend, which suggests that \ha\ \FWHMout\ would be $\sim$ 500 km s$^{-1}$ if UV \FWHMout\ = 800 km s$^{-1}$. One difference is that \cite{Wood15} compared only a small region ($<$ 500 pc) of a nearby face-on galaxy, so that the velocity projection effects may differ from those in our study.

We also overlay several galaxies from \cite{Peng24} in pink colors in Figure \ref{fig:Kine2}. They are low-mass starburst galaxies selected from the CLASSY sample with follow-up optical observations. In general, we find galaxies from the literature exhibiting a consistent tight correlation with ours.  Considering all low-z galaxies shown in Figure \ref{fig:Kine2}, we have for warm-ionized outflows:
\begin{equation}
    \label{eq:FWHM95}
    \begin{aligned}
    \text{FWHM}_{95}^\text{out} (\text{H}\alpha ) &= (0.64 \pm 0.11) \times \text{FWHM}_{95}^\text{out} (\text{ABS.}) 
    \end{aligned}
\end{equation}

Besides the warm-ionized outflows, there exist studies of less ionized and neutral outflowing gases. \cite{Soto12} studied a sample of 39 ultraluminous infrared galaxies (ULIRGs) at z = 0.043 -- 0.163 using Keck ESI observations. They find that broad optical emission lines tracing galactic outflows span a velocity range quite similar to that of the blueshifted \nai\ absorption troughs seen in the same galaxies \citep{Martin05, Martin06}. Unlike the ionized outflows studied in our work, \nai\ traces the neutral, dustier gas. Thus, their work suggests that the neutral outflowing gas detected from absorption lines may correlate with ionized gas seen in emission lines.

\cite{Perrotta21} also briefly compared galactic outflows detected from emission and absorption lines in a sample of 14 massive (\Mstar\ $\sim$ 10$^{11}$ \Msun), compact starburst galaxies ($\Sigma_\text{SFR}$ = 5 -- 1750 \Msun/yr/kpc$^{2}$) at z = 0.4 -- 0.7. They find the maximum outflow velocity (in the form of $V_{98}$) measured from the [\oii] emission line is always less than or equal to the ones measured from \mgii\ absorption lines in the same galaxy. This again is consistent with what we find in Figure \ref{sec:comp:kine}. They do not provide further discussions.


%

\begin{figure*}
\center

	\includegraphics[page = 1,angle=0,trim={0.35cm 0.1cm 4.7cm 1.8cm},clip=true,width=0.5\linewidth,keepaspectratio]{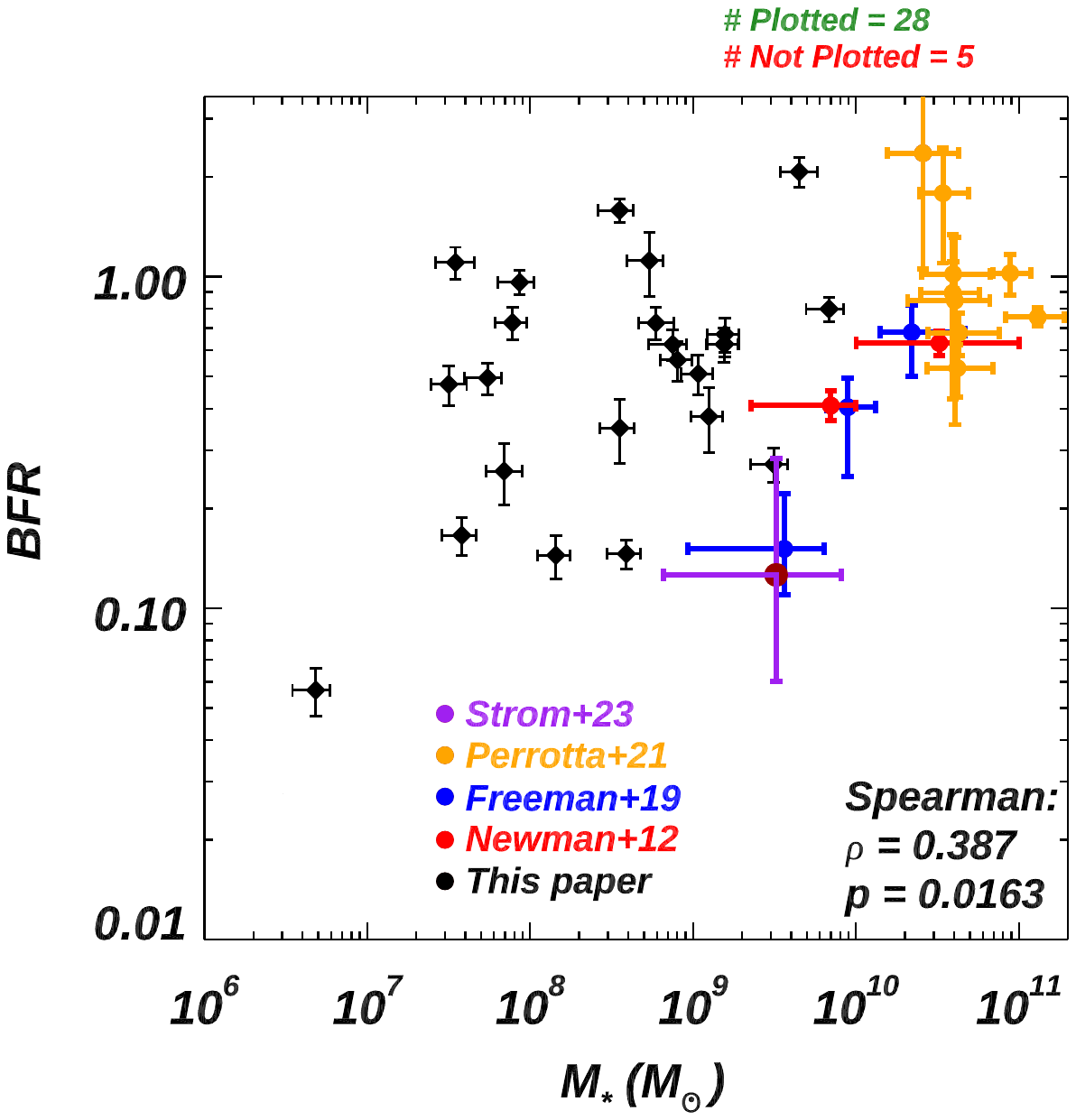}
	\includegraphics[page = 3,angle=0,trim={0.35cm 0.1cm 4.4cm 1.8cm},clip=true,width=0.5\linewidth,keepaspectratio]{./BFR_Figure.pdf}
\caption{\normalfont{Comparisons of the broad to narrow flux ratio (BFR) with galaxy stellar mass \textbf{(left)} and star-forming rate surface density \textbf{(right)}. In each panel, galaxies in our sample are shown as black diamonds. We also present results from the literature as colored circles with crosses as the errorbars. These include the composite of 23 star-forming galaxies at z = 2 -- 3 from JWST/CECILIA survey in purple \citep{Strom23}, stacks of 27 star-forming galaxies at z $\sim$ 2 from the SINS and zC-SINF surveys in red (\Newman), stacks of 113 galaxies at z $\sim$ 2 from the MOSDEF survey in blue (\Freeman), and 9 individual compact starburst galaxies at z = 0.4 -- 0.7 from \cite{Perrotta21} in orange. For the z $\sim$ 2 samples, the detections of of outflows in emission lines are mainly from massive galaxies due to the selection effects (i.e., hard to detect the broad components in lower SNRs spectra in dwarf galaxies). See discussion in Section \ref{sec:BFR}.   }  }
\label{fig:BFR}
\end{figure*}






Overall, we find that although connections between outflow signatures seen in optical emission and UV metal absorption lines have been noticed previously, no systematic studies were made until our current study. Our observations and models are consistent with all previous works.




\subsection{Broad to Narrow Flux Ratio}
\label{sec:BFR}

In this subsection, we discuss another important property of the emission line: the broad to narrow flux ratio (BFR), which is defined as the measured flux from the broad to narrow components. Note here we measure the flux of the whole broad or narrow components in contrast to measuring the flux of only the blueshifted component in Section \ref{sec:linemea}. These choices are made to be consistent with previous studies of BFRs in the literatures (e.g., \Newman\ and \Freeman). 

In the left panel of Figure \ref{fig:BFR}, we compare BFR with the stellar mass (\Mstar) of the galaxy. Objects in our sample are shown as black diamonds. We also present various emission-line-based outflow detections from the literature, including stacks of 27 star-forming galaxies at z $\sim$ 2 from the SINS and zC-SINF surveys in red (\Newman), stacks of 113 galaxies at z $\sim$ 2 from the MOSDEF survey in blue (\Freeman), 9 individual compact starburst galaxies at z = 0.4 -- 0.7 from \cite{Perrotta21} in orange, and the stack of 23 star-forming galaxies at z = 2 -- 3 from JWST/CECILIA survey in purple \citep{Strom23}. We note that like our analysis, all of these surveys adopt similar Gaussian fitting methods to extract the broad components from the emission lines. Most of the observations have similar spectral resolution (R $\sim$ 3000 -- 5000) for \ha\ regions except that \cite{Strom23} has R $\sim$ 1000, \cite{Perrotta21} has R $\sim$ 2000 and four of our galaxies have X-Shooter spectra with R $\sim$ 8800. Thus, the major differences between galaxies should be their intrinsic properties (e.g., different \Mstar, SFR, and age) and redshifts.

Overall, there is a possible positive correlation ($\sim$ 1.7$\sigma$) between BFR and \Mstar\ for all objects in the left panel of Figure \ref{fig:BFR}, albeit with significant scatter. This trend was previously suggested in \Newman\ and \Freeman, but rejected by \cite{Perrotta21}. While all of the previous measurements have \Mstar\ $>$ 10$^{9}$ \Msun, the inclusion of low-mass galaxies from our sample shows that the trend likely extends below  $10^8$ \Msun.  In addition, our galaxies show higher BFR at the same \Mstar\ compared with other samples. This is likely due to our selections of highly star-forming systems (Section \ref{sec:obs}).

In the right panel of Figure \ref{fig:BFR}, we compare the BFR with the surface brightness of SFR ($\Sigma_\text{SFR}$). \Newman\ find this is the galaxy property that is best correlated with BFR (see red crosses for stacks from them), while \cite{Perrotta21} suggest there is no clear correlation. When we consider all their measurements and our new data, we find all objects are consistent with a positive correlation for $\Sigma_\text{SFR} < $ 100 \Msun/yr/kpc$^{2}$, but reach a plateau beyond that. The inhomogeneity of these compiled samples makes it difficult to draw firm conclusions.





\begin{figure}
\center

	\includegraphics[page = 5,angle=0,trim={1.7cm 0.1cm 4.3cm 1.8cm},clip=true,width=1.0\linewidth,keepaspectratio]{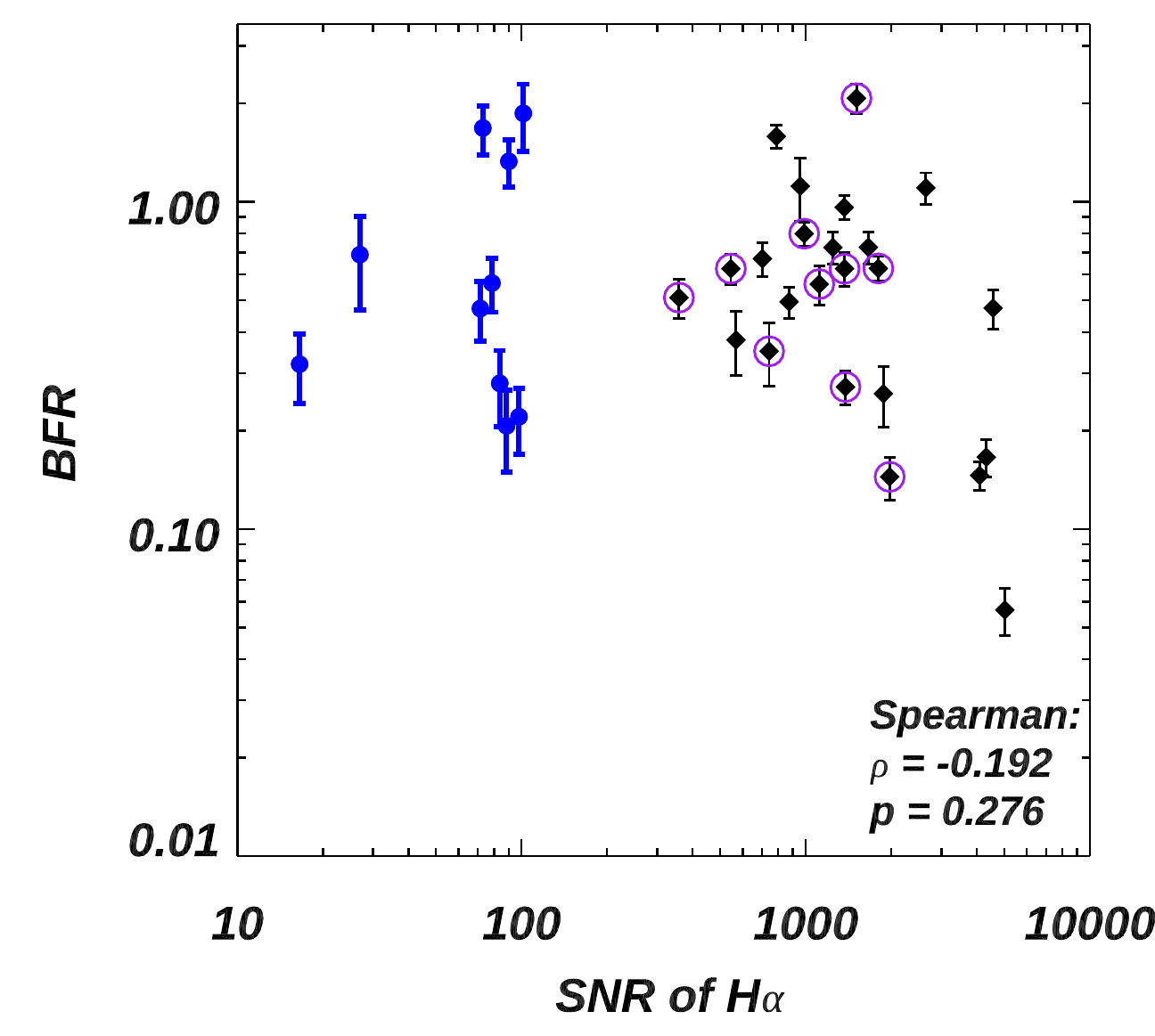}
\caption{\normalfont{Comparisons of BFR and SNR measured from \ha\ emission lines. Galaxies in our sample are shown in black diamonds, while individual z $\sim$ 2 galaxies in \cite{Freeman19} with $\sim$ 3$\sigma$ significance detections of the broad components are shown as blue dots with 1$\sigma$ errorbars. Our galaxies with \ha\ broad component FWHM $>$ 275 km s$^{-1}$ are marked in purple circles, which match the selection criteria chosen by \cite{Freeman19}.   }  }
\label{fig:BFR2}
\end{figure}

Another instructive test is the effects of signal-to-noise\footnote{Hereafter, the SNR of a spectral line is defined as its fitted flux divided by the error of the flux. This is to be consistent with the definition in \Freeman.}. \Freeman\ found that the measured BFR is dependent on SNR since it is hard to disentangle the broad and narrow components in low-SNR spectra. Thus, we show BFR versus SNR measured from the \ha\ emission line in Figure \ref{fig:BFR2}. We find there is a tentative negative correlation ($\sim 2 \sigma$), which has also been noticed in \Freeman. We overlay individual galaxies from their MOSDEF sample with a broad component detection of $\sim$ 3$\sigma$ signiﬁcance as the blue symbols. While their galaxies typically have SNRs $\lesssim$ 100, the measured BFRs are consistent with the low SNR end of our galaxies.


Furthermore, to test the impact of SNRs on BFR, we choose ten galaxies that match \Freeman's selection criteria of broad components (\FWHMout\ $>$ 275 km s$^{-1}$, marked in purple circles) and downgrade the observed spectra into SNR  = 100, 50, and 20. Then we refit our galaxies using the method described in Section \ref{sec:GuassianFit:OPT}. For each galaxy, we find their spectra with SNR $>$ 100 yield the same BFR within errorbars. When we reach SNR = 50, the extracted BFRs begin to have small to moderate fluctuations (0 -- 50\%). Finally, when we reach SNR = 20, we find most of the tested galaxies (8 out of 10) do not pass the F-test in spectral line fittings, i.e., one loses the information to isolate the broad components from emission lines (see Section \ref{sec:GuassianFit}). Thus, if the intrinsic emission lines in \Freeman\ are the same as ours, their detections of the broad, outflow components in the MOSDEF survey may be incomplete due to the limits of SNR in individual galaxy spectra \citep[average $\sim$ 25,][]{Weldon24}. Overall, this test suggests that the accurate extractions of broad components from emission lines require SNR to be at least 50 and ideally larger than 100 for moderate resolution spectra (R $\sim$ 3000 -- 5000).






\subsection{Comparisons of Scaling Relationships from Emission and Absorption Lines}
\label{sec:scale}

Outflow properties have been observed to scale with various galaxy characteristics, such as SFR and \Mstar. For low-redshift star-forming and starburst galaxies, these correlations have been extensively studied \citep[e.g.,][]{Martin05, Rupke05, Heckman15, Chisholm15, Chisholm16a, Sugahara17,Xu22a, Reichardt24}. However, a direct comparison of these scaling relationships derived from emission and absorption lines within the same sample remains unexplored. 

\begin{figure*}
\center

	\includegraphics[page = 5, angle=0,trim={2.2cm 3.5cm 4.7cm 1.8cm},clip=true,width=0.5\linewidth,keepaspectratio]{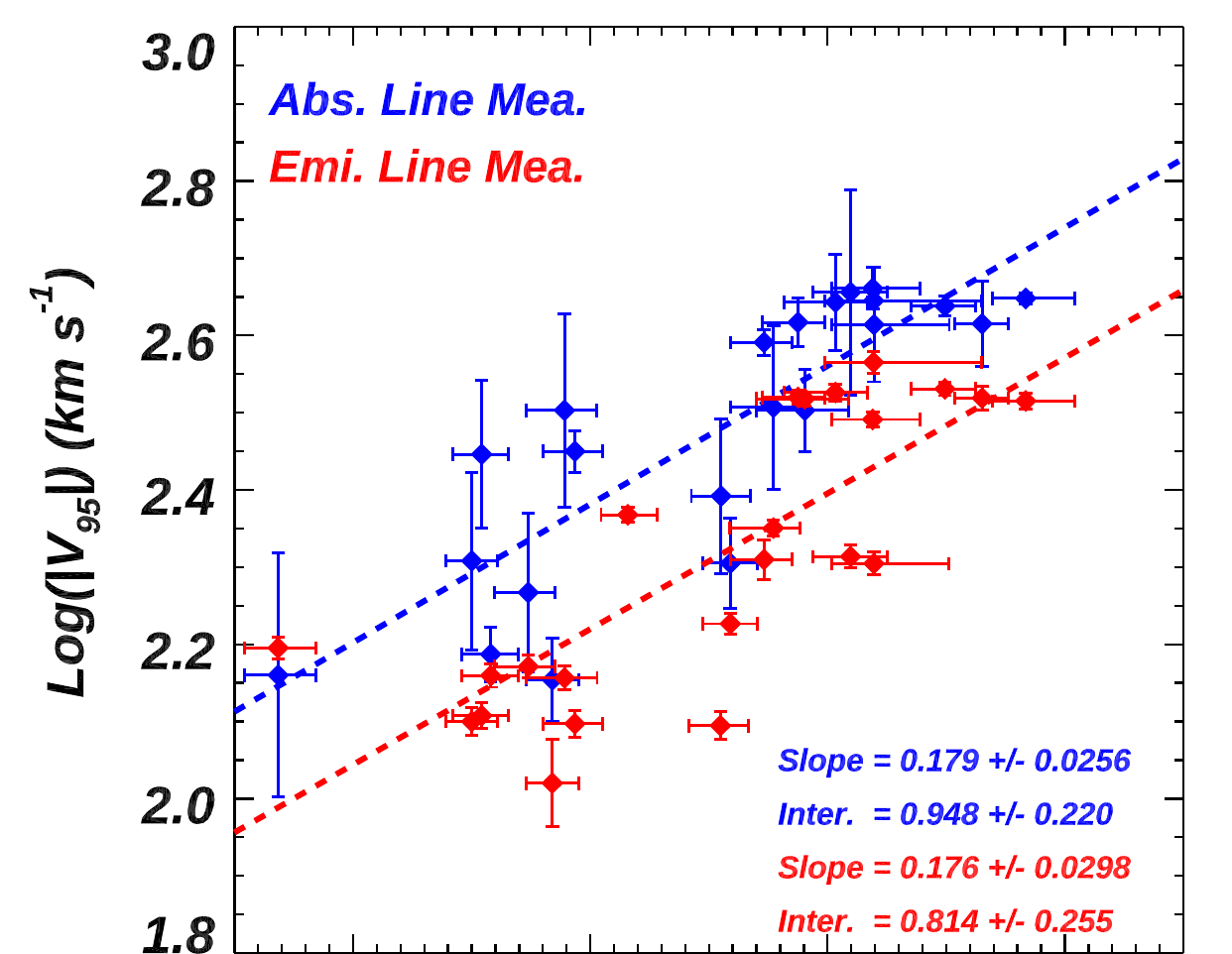}
	\includegraphics[page = 6, angle=0,trim={2.2cm 3.5cm 4.7cm 1.8cm},clip=true,width=0.5\linewidth,keepaspectratio]{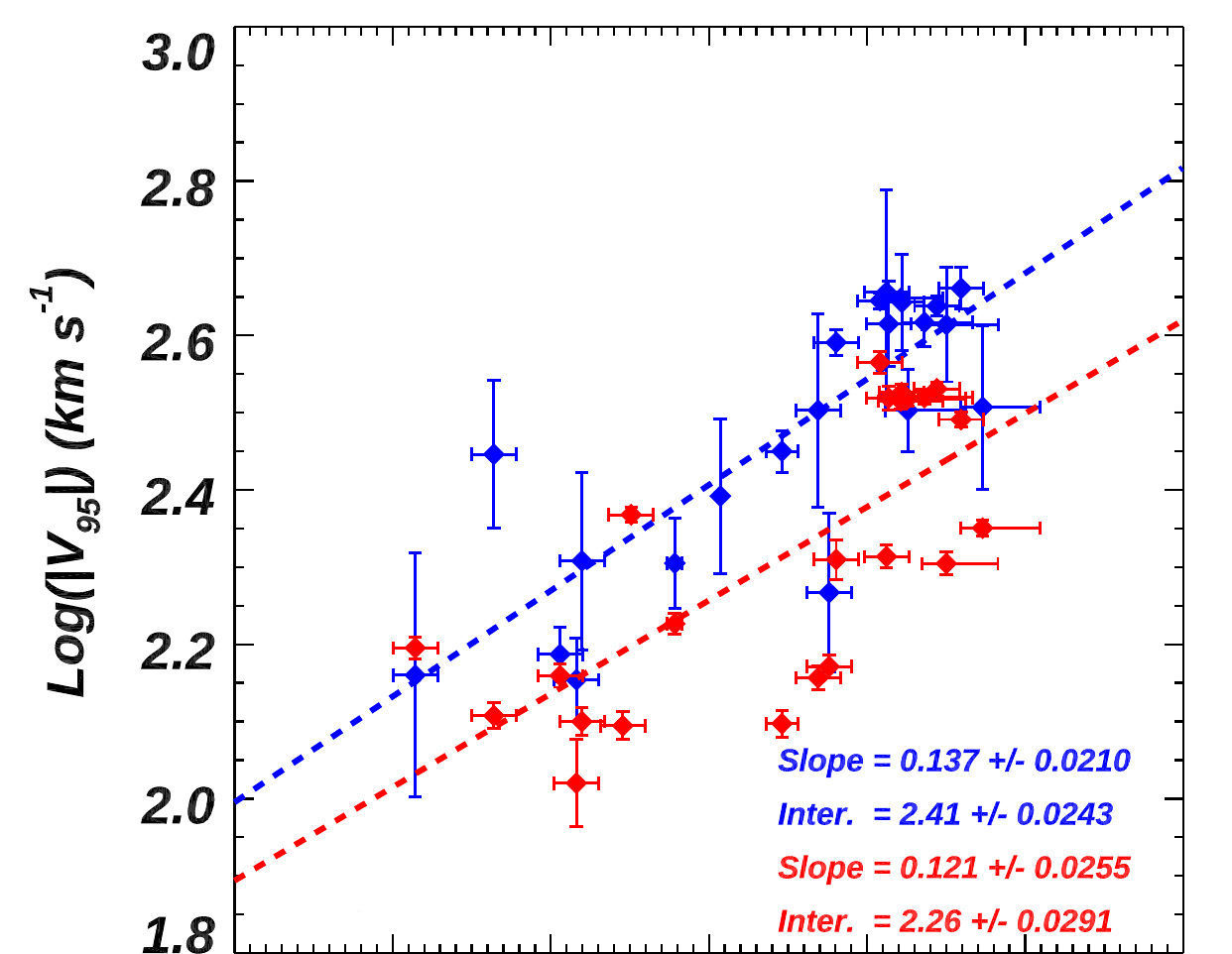}

	\includegraphics[page = 9, angle=0,trim={2.2cm 0.1cm 4.7cm 2.0cm},clip=true,width=0.5\linewidth,keepaspectratio]{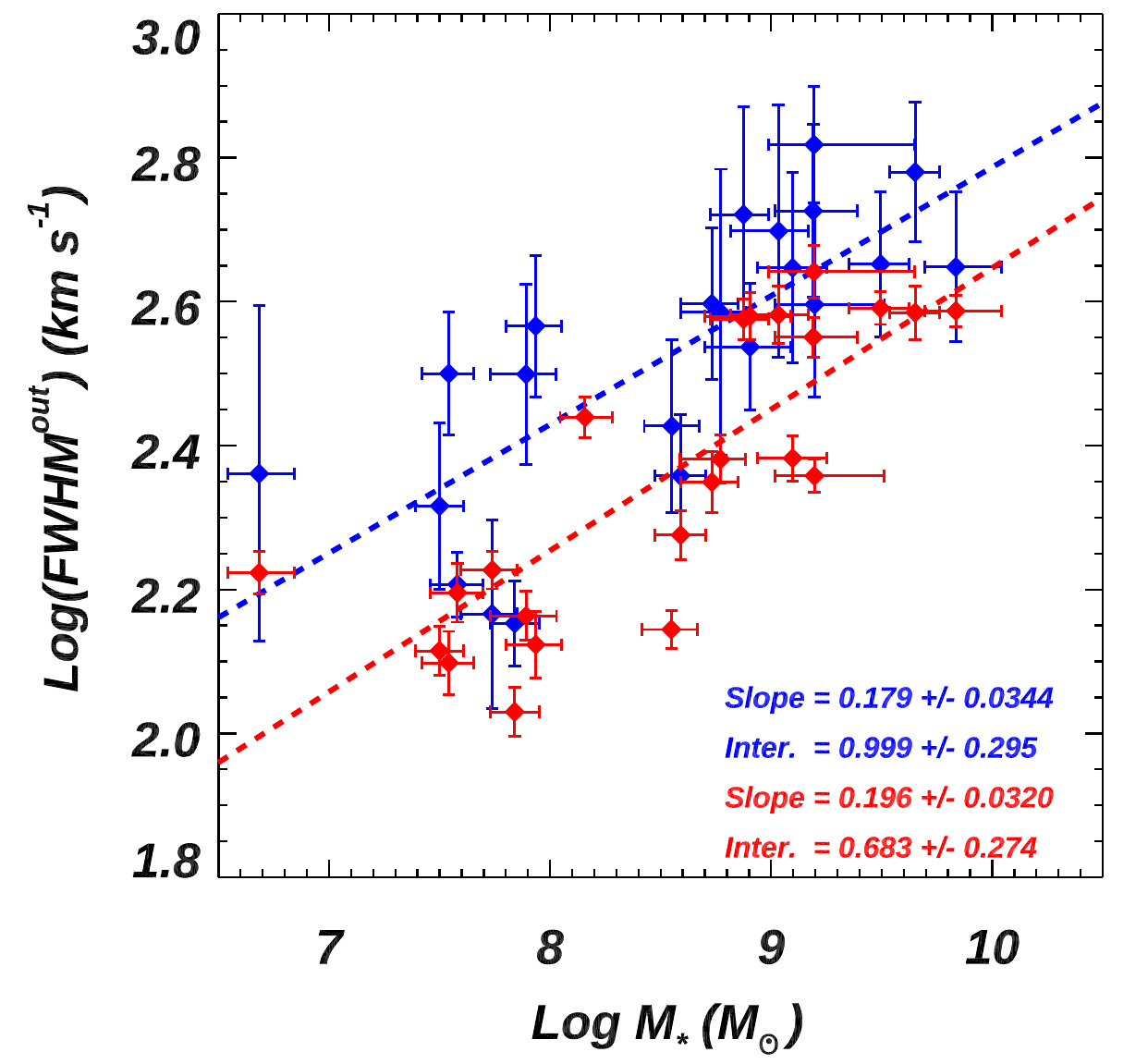}
	\includegraphics[page = 10, angle=0,trim={2.2cm 0.1cm 4.7cm 2.0cm},clip=true,width=0.5\linewidth,keepaspectratio]{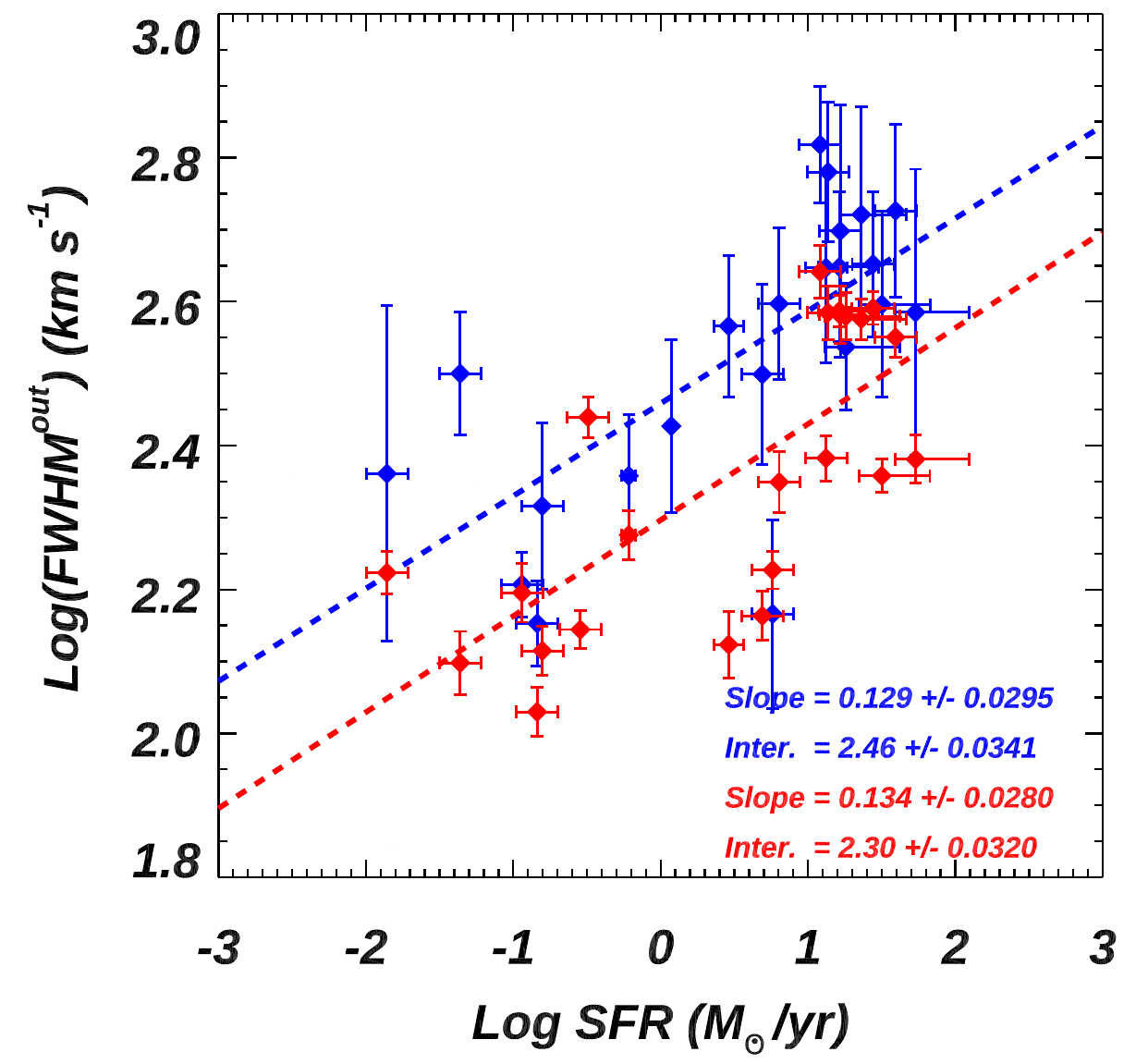}

\caption{\normalfont{Scaling relationships for outflow maximum velocities (\Vnf) and full-width-half-maximum (\FWHMout) measured independently from the two outflow diagnostics, i.e., UV absorption lines (blue) or optical emission lines (red). Each galaxy in our sample is represented as a point, with error bars indicated by crosses. The best-fit linear regressions for absorption and emission measurements are depicted by blue and red dashed lines, respectively. Notably, while both diagnostics exhibit similar slopes, the absorption line measurements systematically yield intercepts that are higher by 0.1 -- 0.2 dex compared to those from emission lines.}  }
\label{fig:scaling}
\end{figure*}

In Figure \ref{fig:scaling}, we provide such a comparison, where outflow measurements derived from absorption and emission lines are represented by blue and red symbols, respectively. Across all panels, the two diagnostics exhibit similar slopes. However, absorption line measurements systematically yield intercepts that are higher by 0.1–0.2 dex relative to those from emission lines. These results align with the expectations from Figures \ref{fig:Kine} and \ref{fig:Kine2}, where absorption lines exhibit $\sim$ 30–40\% higher \Vnf\ and \FWHMout\ than their emission-line counterparts. 

Our findings help explain the ongoing debate over the impact of galactic outflow feedback in high redshift galaxies, especially when there only exist emission-line diagnostics. At cosmic noon, \cite{Davies19} analyzed 28 SF galaxies from the SINFONI project and found strong correlations between outflow velocities and $\Sigma_\text{SFR}$. However, \cite{Weldon24} found no such correlations in MOSDEF survey and suggest that nebular emission-line-traced outflows are negligible. Similarly, \cite{XuYi23}, using JWST/NIRSpec observations of galaxies at z $\sim$ 3–9, found weak to no correlation between outflow velocities and either SFR or \Mstar. They concluded that the outflow velocities are typically too low to escape the galaxy, implying fountain-type outflows. In contrast, \cite{Carniani24} examined 52 galaxies from the JWST Advanced Deep Extragalactic Survey (JADES) and found that their median velocity and mass-loading factor ($\eta$) are 1.5–100 times higher than those of local dwarf galaxies, suggesting that outflows can significantly regulate star formation in low-mass galaxies within the first 2 Gyr of the universe. 

Given the fact these studies adopt purely emission line diagnostics, their discrepancies are not unexpected under what we have so far: (1) the robust decomposition of the broad components from emission lines require high SNR and spectral-resolution data. If not, the detection rates of broad, outflow components can be underestimated (see details in Section \ref{sec:BFR}), and (2) as shown in Figures \ref{fig:OutflowRate} and \ref{fig:scaling}, outflows in emission lines generally exhibit lower velocities, velocity dispersions, and outflow rates than in absorption lines, complicating their detections in distant galaxies. Overall, estimations of outflow properties through emission line diagnostics can be simply biased given insufficient SNR, spectral resolution, and intrinsically smaller measurable quantitatives. Our results underscore the need for high-SNR, high-resolution spectra, especially in UV, to accurately characterize galactic outflows and their feedback effects for high-z galaxies.


\subsection{Implications for Future Galactic Outflow Studies}


In the current work, we present how to jointly interpret the warm-ionized outflows seen from rest-UV absorption and optical emission lines for the same galaxy. These results can provide various implications for future studies of warm-ionized galactic outflows. 

First of all, for warm ionized outflows, one could reliably predict the line kinematics of outflow absorption lines (e.g., \Siii, \siiv) given the observed outflow emission lines (e.g., \ha, \hb) and vice versa. This will be especially helpful when one of the outflow diagnostics is not observed or difficult to access. For example, given the large samples of optically detected star-forming galaxies at low redshift, many show galactic outflow features in emission lines \citep[e.g., $\sim$ 160,000 from SDSS,][]{Cicone16}. However, only a very small portion have dedicated rest-UV coverage that allow us to study ionized outflow properties through absorption lines \citep[e.g., $\sim$ 100 from FUSE or HST/COS,][]{Heckman15,Xu22a}. On the other hand, for galaxies at z $\sim$ 2 -- 4, when the rest-UV spectra of galaxies can be directly accessed using large ground-based optical telescopes \citep[e.g., from Keck and VLT,][]{Steidel14,Urrutia19}, their rest-optical spectral ranges can be inaccessible from the ground\footnote{Due to telluric absorption of Earth's atomsphere or limited spectral coverage of ground-based spectrographs.} and JWST observations are expensive. Thus, our current work can provide guidelines for 1) designing future spectroscopic surveys by making realistic predictions of one type of outflow diagnostics using another, and 2) making fair comparisons of outflow properties across the cosmic times using different diagnostics.


Furthermore, our work indicates that models and simulations of galactic outflows can now have additional constraints by matching outflow features from both absorption and emission lines for the same galaxy. This can effectively double the number of constraints and could potentially break degeneracy of many models that have been adopted to explain one of the outflow diagnostics \citep[e.g.,][]{Cottle18, Cruz21, Li24, Huberty24}. We have plans to study more sophisticated models in future papers, e.g., using SALT \citep{Carr23}, to vary the density and velocity distributions to fit the observed absorption and emission lines for each galaxy.

Finally, our work has a few caveats or limitations: 1) The current sample is not large. While we find a tight correlation in kinematics of the two outflow diagnostics, we do not have enough objects to reveal a clear correlation in outflow density and spatial extend.3) Outflow properties are spatially dependent. Our LOS integrated observations missed the spatial variations, which can provide additional constraints on the interpretation and models of galactic outflows. Thus, future integral field unit (IFU) observations covering both rest-UV and rest-optical spectral regions of the same galaxy will provide useful insights toward this direction. In addition, IFU observations will also be useful for more accurately removing the kinematically complex narrow components.  3) Without sufficient spatial information, our estimates of outflow rates are crude since they rely on single values of \Rout, \ne, and \Vout.  4) Our current models assume a simple mapping between velocity, density, and radius. But high resolution simulations of cloud acceleration show that the reality can be more complex \citep[e.g.,][]{Nikolis24,Strawn24}.  5) Galaxies and their outflows evolve over cosmic times. Our work only focuses on starburst galaxies at low-redshift. Similar analyses should also be conducted for large samples of galaxies at the cosmic noon and in the early universe.


\section{Conclusion}
\label{sec:conclusion}

In this paper, we shine a light on the connection between galactic outflow properties measured from two different diagnostics, i.e., rest-UV absorption and rest-optical emission lines. We have collected a sample of 33 low-redshift starburst galaxies that cover a wide range of stellar mass, star-formation rate, and metallicity. For each galaxy, we combine their UV observations from HST/COS and optical observations from Keck/ESI or VLT/X-Shooter. We adopt multiple Gaussian profiles to fit the observed emission and absorption lines and isolate the outflow components. Then, we compare the properties of the two sets of outflow diagnostics in various ways. The principal results are as follows:

\begin{enumerate}
    \item For all galaxies with an outflow component detected in its absorption lines, we also detect it in its emission lines.

    \item The maximum outflow velocity (or line width) derived from the emission and absorption lines are well-correlated with moderate scatter. On average, outflows detected in emission lines reach $\sim$ 68\% of the maximum velocity and 64\% of the line widths than those of absorption lines for the same galaxy. 

    \item Outflow sizes derived from emission and absorption lines are also well-correlated. Outflows seen in emission lines are preferentially detected in the inner regions of the outflow and have higher gas densities than those of absorption lines detected in the same galaxy.

    \item We find the mass outflow rates estimated from emission lines are consistently lower than those from absorption lines by 0.2 -- 0.5 dex. This discrepancy is consistent with previous studies of SF galaxies, which commonly present low outflow rates measured from emission lines.
    
    \item We constructed analytical galactic wind models giving M 82 radial velocity, density, and volume filling factor profiles. We find the spherical outflow models with a density distribution $n (r)f\!f(r) \propto r^{-3.8}$ on-average matches well with the observed correlations of outflow velocity (or widths) between the two diagnostics. Bi-conical outflow models from SALT shows that the outflow geometry can affect the observed line profiles but predictions are still consistent with the observed correlations.

    

    \item We study the broad to narrow flux ratios (BFRs) of our galaxies and compare them with various other outflow studies. We find there are possible correlations between BFR and galaxy stellar mass and with SFR surface density. We also find that signal-to-noise larger than 50 is required to accurately extract BFR from emission lines with spectral resolution $\sim$ 3000.

    \item By comparing the scaling relationships derived from emission and absorption lines independently, we find they yield similar slopes but emission-lines scaling relationships have 0.1--0.2 dex lower intercepts. Based on these, we discuss the challenges faced by high-z galactic outflow studies using purely emission line spectra and underscore the needs for high-fidelity UV and optical spectra to accurately characterize the feedback effects of high-z galaxies.
    
    We also highlight a few implications for future studies. Given the correlations found between the two diagnostics, one can now reliably predict the line kinematics of outflow absorption lines given the observed outflow emission lines, and vice versa. This should help design future spectroscopic surveys and make comparisons between heterogeneous outflow samples. Furthermore, our work indicates that future models and simulations can now have additional constraints by considering both outflow absorption and emission lines for the same galaxy.
\end{enumerate}

\begin{acknowledgements}
X.X. and A.H. acknowledge supports from NASA STScI grant HST-AR-17042. X.X. thank T. Thompson for helpful discussions. X.X., A.H., and T.J. thank M. Rafelski and R. Sanders for Keck/ESI observations and data reductions. X.X thank Keerthi Vasan G. C. for useful comments.

This research is based on observations made with the NASA/ESA Hubble Space Telescope obtained from the Space Telescope Science Institute, which is operated by the Association of Universities for Research in Astronomy, Inc., under NASA contract NAS 5–26555. These observations are associated with programs 11727, 12928, 13017, 13293, 14080, 14168, 14679, 15185, 15626, 15646, 15099, 15840, 16643, and 17526. 

Part of the data presented herein were obtained at Keck Observatory, which is a private 501(c)3 non-profit organization operated as a scientific partnership among the California Institute of Technology, the University of California, and the National Aeronautics and Space Administration. The Observatory was made possible by the generous financial support of the W. M. Keck Foundation. The authors wish to recognize and acknowledge the very significant cultural role and reverence that the summit of Maunakea has always had within the Native Hawaiian community. We are most fortunate to have the opportunity to conduct observations from this mountain.

Part of the data presented were collected from the European Southern Observatory under ESO programs 085.B-0784(A) and 096.B-0192(A).

Additional data analyses were done using Interactive Data Language (IDL), version 8.7.3, L3Harris Geospatial, Boulder, Colorado, USA.

\end{acknowledgements}

\vspace{5mm}


\facilities{HST (COS), Keck (ESI), VLT (X-Shooter)}
\software{
astropy \citep{Astropy22},
CLOUDY \citep{Ferland17},
jupyter \citep{Kluyver16},
MPFIT \citep{Markwardt09}},
ESIRedux \citep{Prochaska07},

\appendix

\section{Spectra Line Fitting for All Galaxies}
\label{app:sec:fit}
\begin{figure*}
\center
	\includegraphics[angle=0,trim={0.2cm 1.75cm 0.0cm 8.5cm},clip=true,width=0.99\linewidth,keepaspectratio]{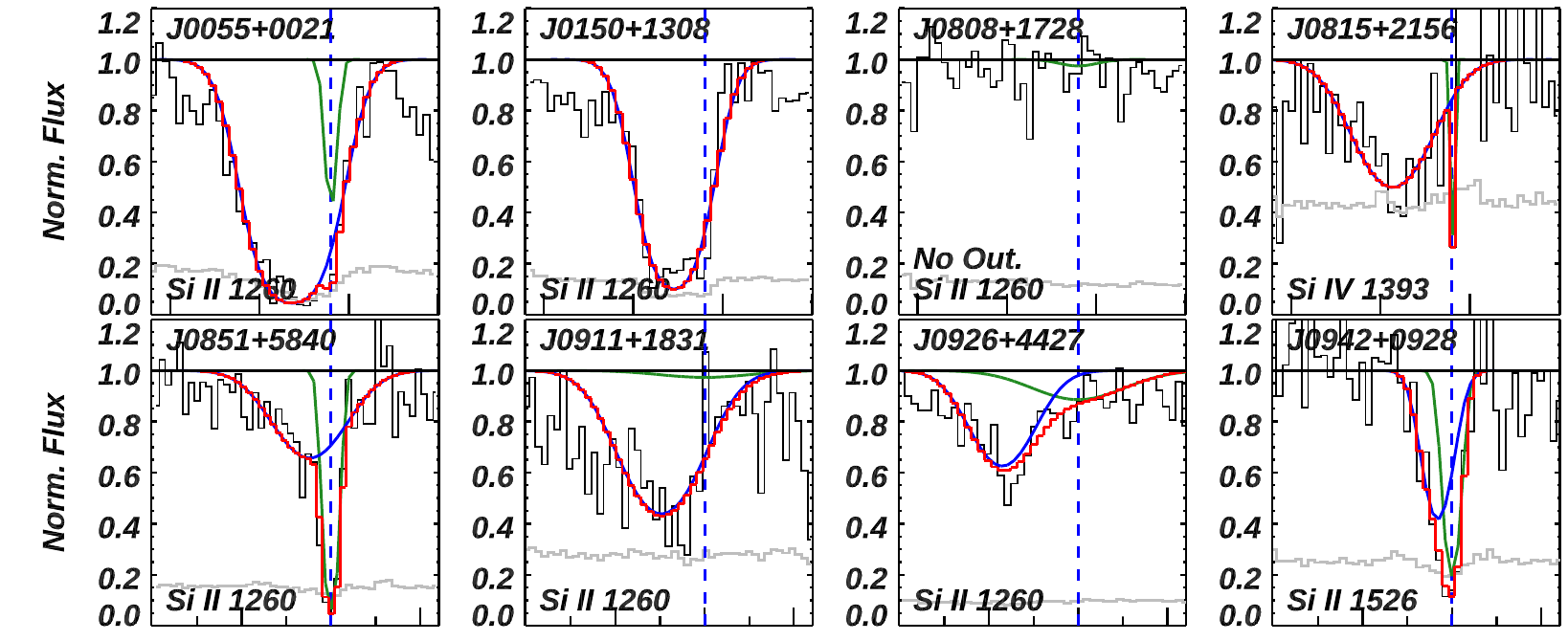}
	
	\includegraphics[angle=0,trim={0.0cm 0.1cm 0.0cm 8.6cm},clip=true,width=1\linewidth,keepaspectratio]{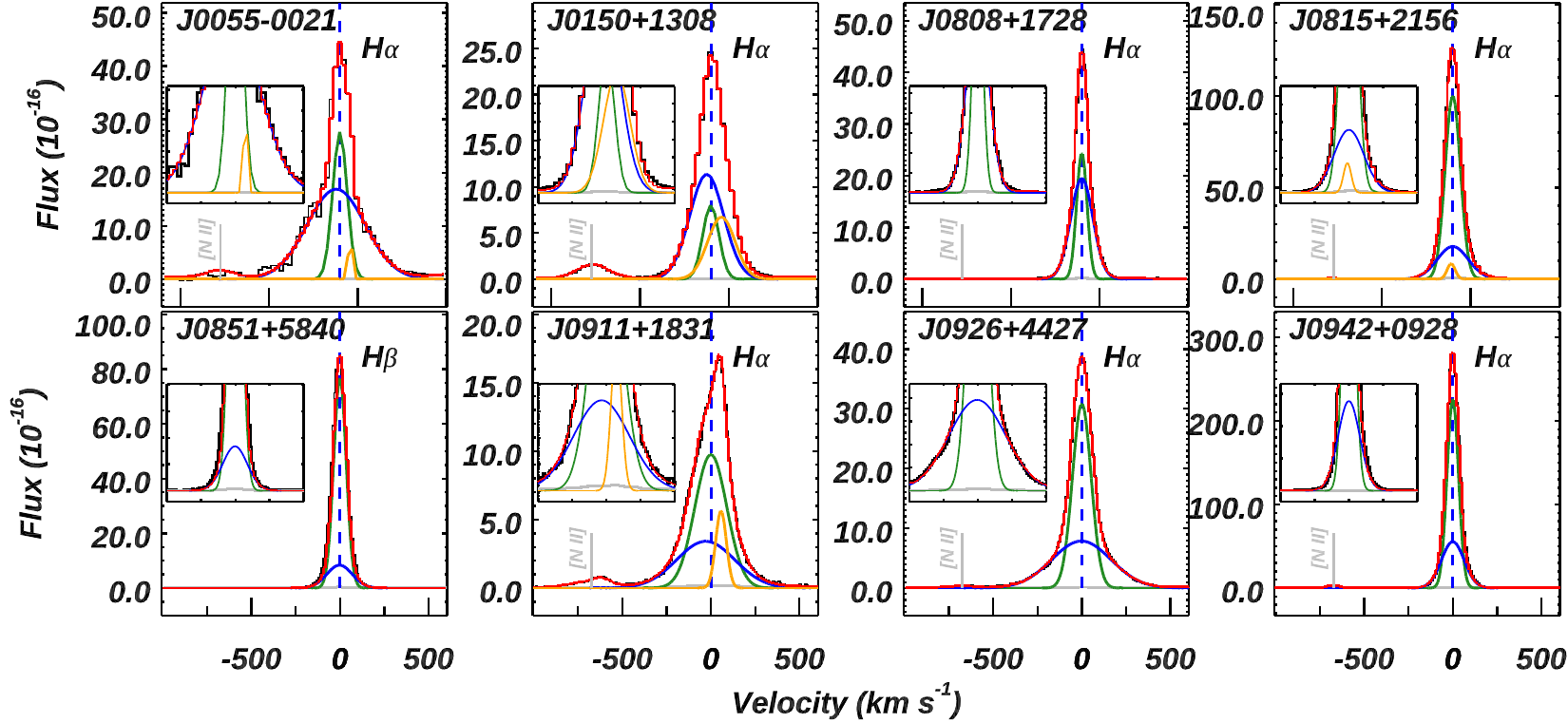}

\caption{\normalfont{Spectral line fitting to for galaxies in our sample. Top two and bottom two rows are for UV-absorption and optical-emission lines, respectively, for the same 8 galaxies. For UV spectra, we show \Siii\ \ly 1260 absorption line unless it is not covered or contaminated. Similarly, for optical spectra, we mainly show the strong lines from \ha\ or \hb. The data and errors are shown in black and gray histograms. We adopt double- to triple-Gaussian profiles to fit the lines. The static ISM and outflow components are shown in green and blue, respectively. For emission lines, the third Gaussian (when necessary) is shown in orange. The red line is the summation of all models. For lines with no outflow signatures, we show only the static ISM component in green and mark it as ``No Out.''. The inset shows a zoom-in view of the broad wings in the emission line. We indicate the locations of Galactic absorption in vertical gray dashed lines and [\nii] \ly 6548 in vertical gray solid lines. See detailed discussion of the fittings in Section \ref{sec:GuassianFit}.} }
\label{fig:spectra-1}
\end{figure*}

\begin{figure*}
\center
	\includegraphics[angle=0,trim={0.2cm 1.75cm 0.0cm 8.5cm},clip=true,width=0.99\linewidth,keepaspectratio]{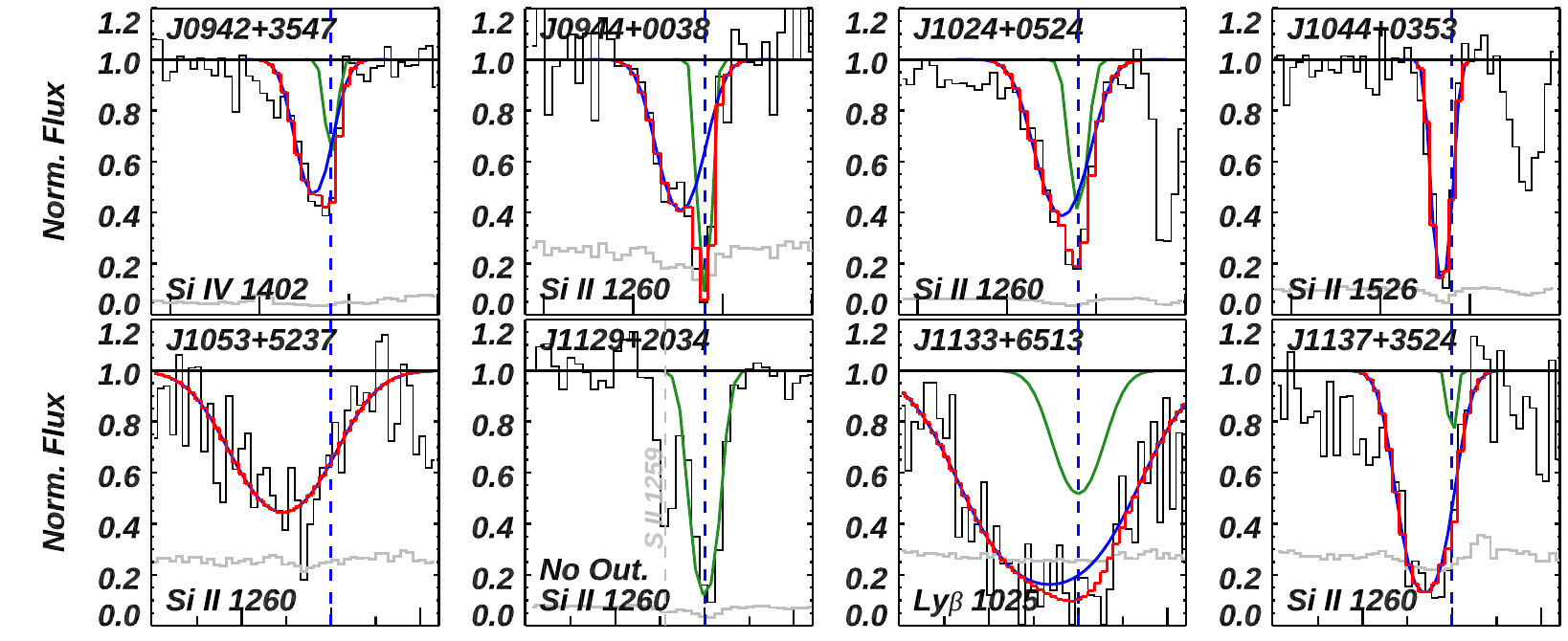}
	
	\includegraphics[angle=0,trim={0.0cm 0.1cm 0.0cm 8.6cm},clip=true,width=1\linewidth,keepaspectratio]{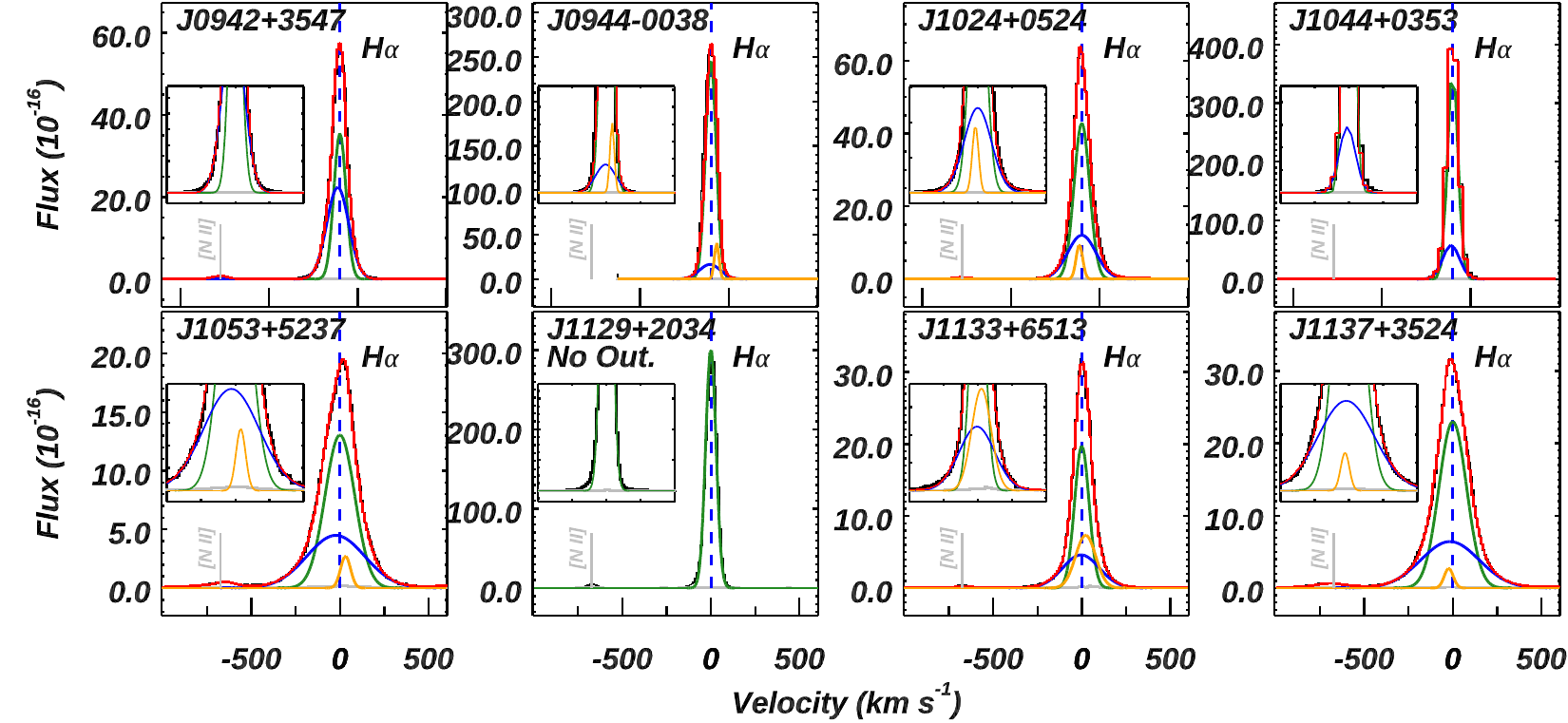}

\caption{\normalfont{Same as Figure \ref{fig:spectra-1} for the other 8 galaxies in our sample.} }
\label{fig:spectra-2}
\end{figure*}

\begin{figure*}
\center
	\includegraphics[angle=0,trim={0.2cm 1.75cm 0.0cm 8.5cm},clip=true,width=0.99\linewidth,keepaspectratio]{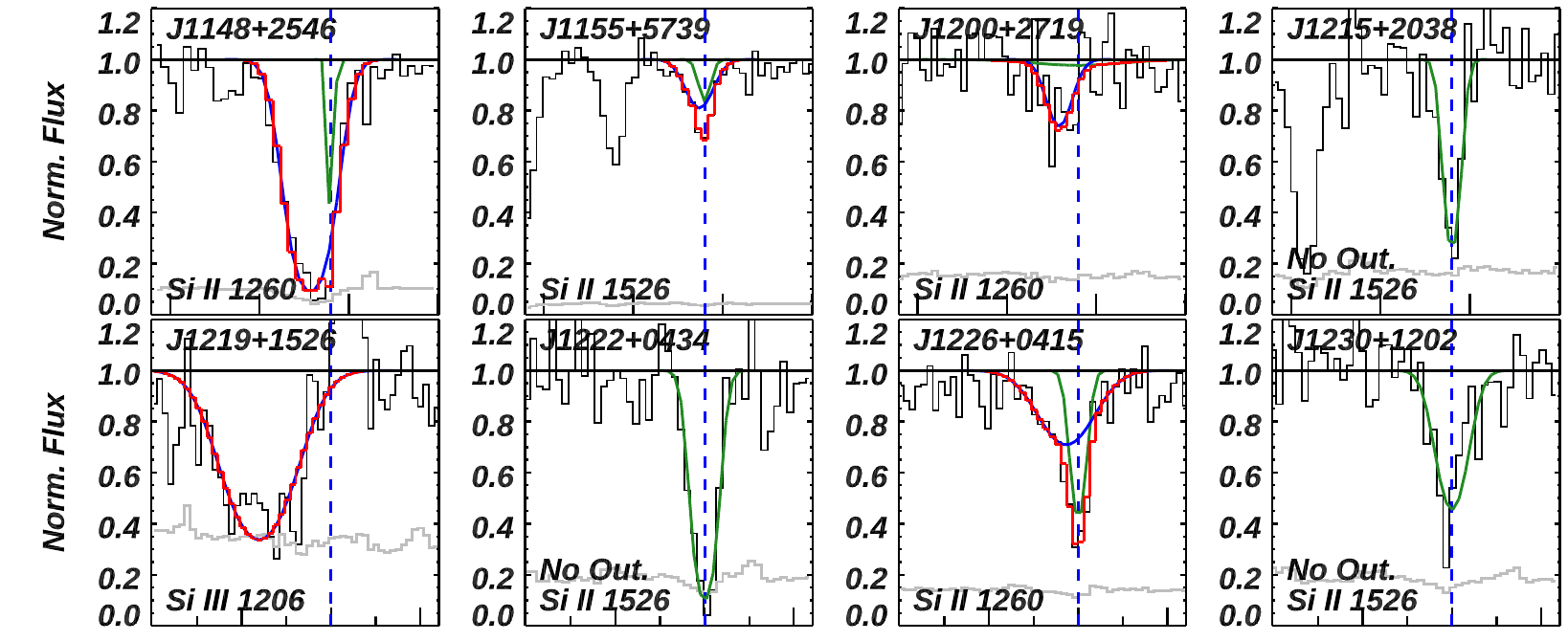}
	
	\includegraphics[angle=0,trim={0.0cm 0.1cm 0.0cm 8.6cm},clip=true,width=1\linewidth,keepaspectratio]{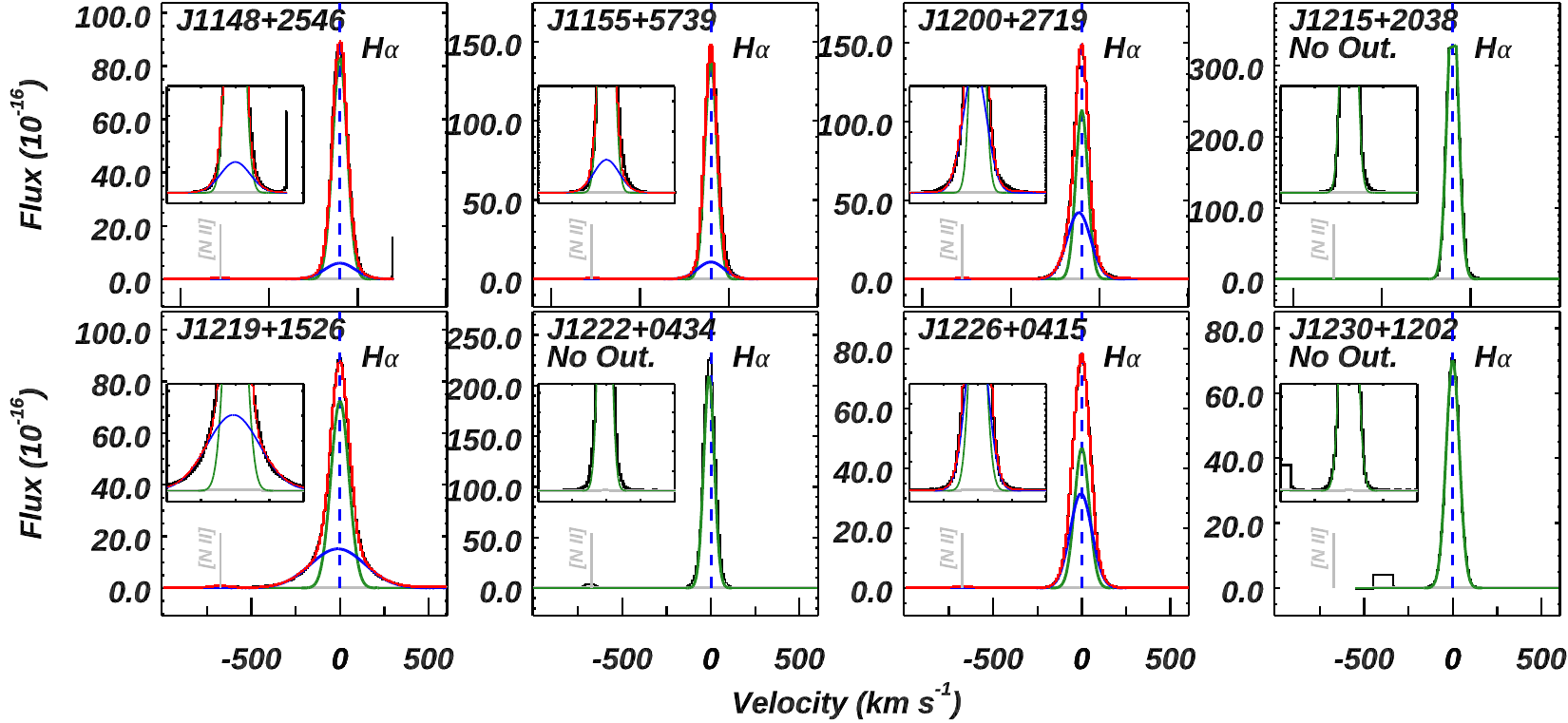}

\caption{\normalfont{Same as Figure \ref{fig:spectra-1} for the other 8 galaxies in our sample. J1148+2546 has a gap at the right side of \ha, but it does not affect our line fit here.} }
\label{fig:spectra-3}
\end{figure*}

\begin{figure*}
\center
	\includegraphics[angle=0,trim={0.2cm 1.75cm 0.0cm 8.5cm},clip=true,width=0.99\linewidth,keepaspectratio]{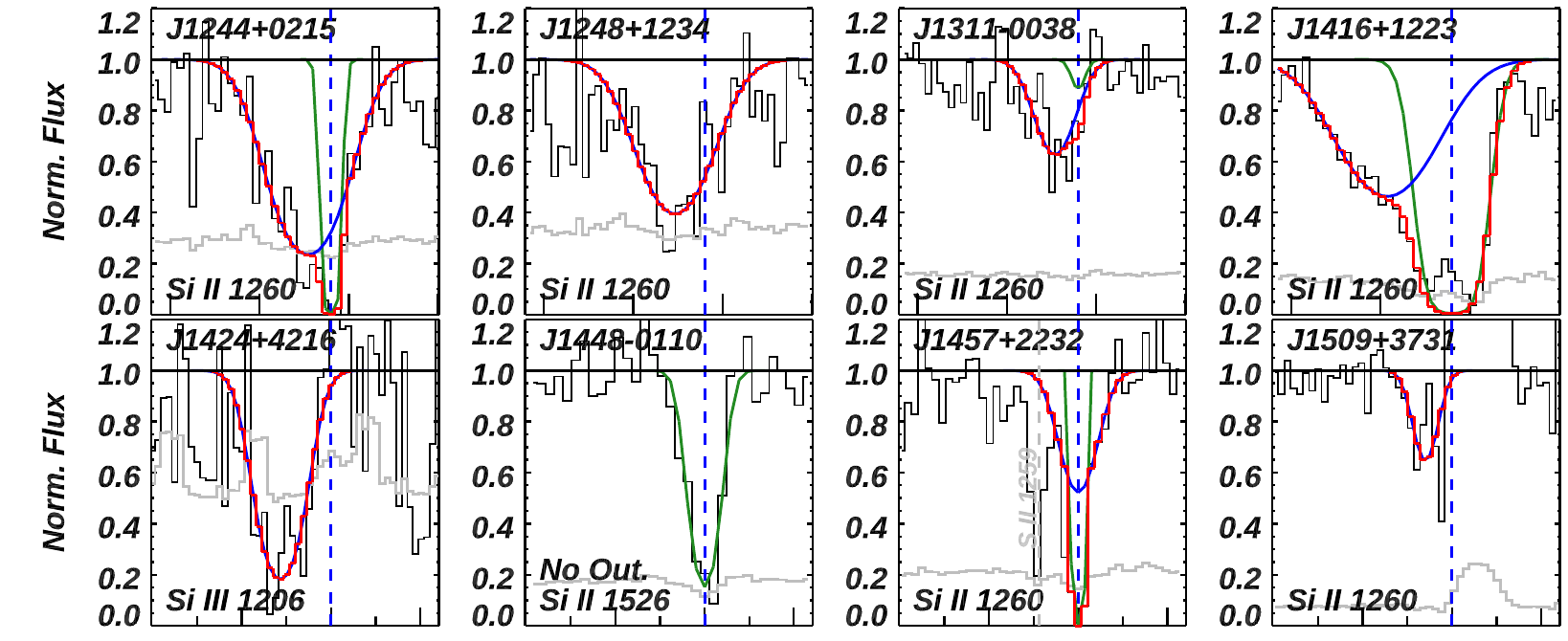}
	
	\includegraphics[angle=0,trim={0.0cm 0.1cm 0.0cm 8.6cm},clip=true,width=1\linewidth,keepaspectratio]{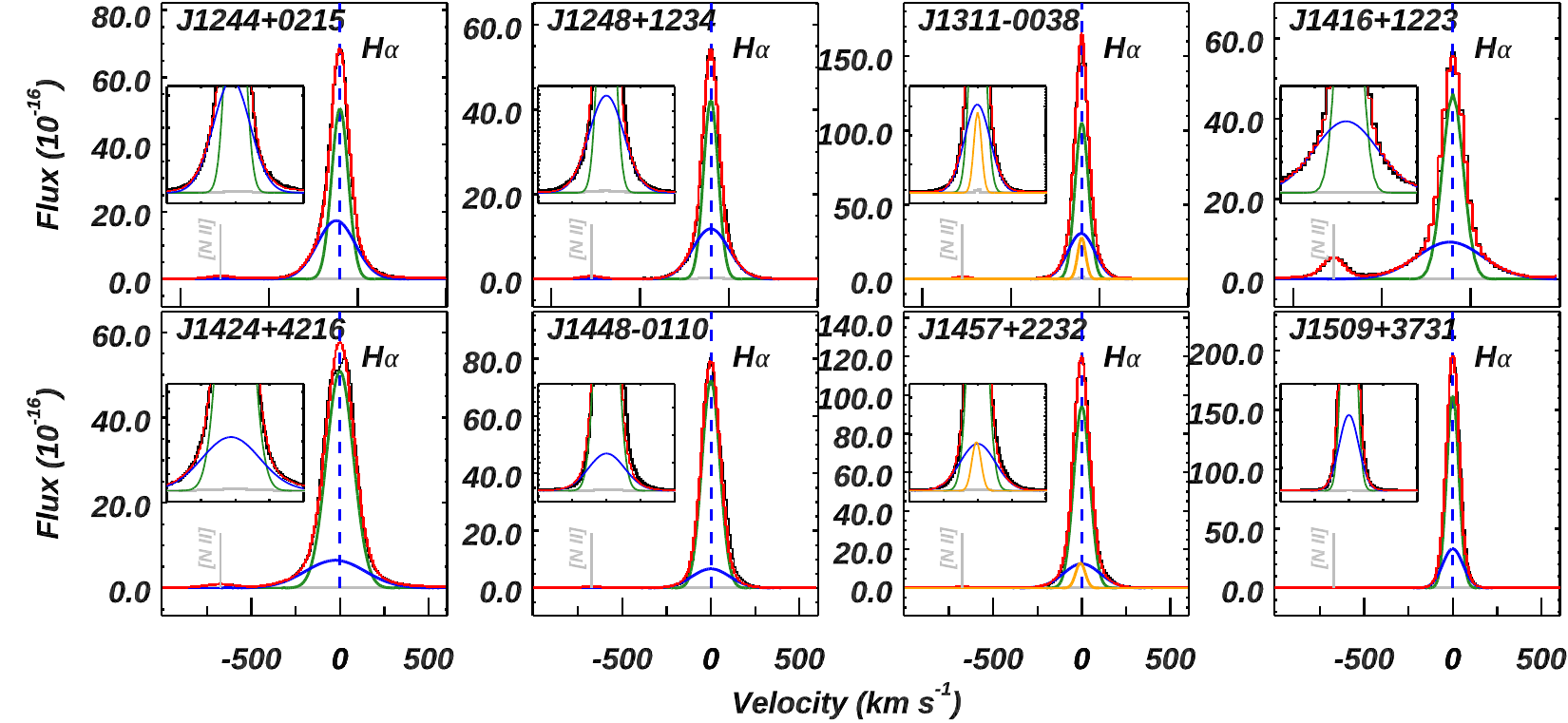}

\caption{\normalfont{Same as Figure \ref{fig:spectra-1} for the other 8 galaxies in our sample.} }
\label{fig:spectra-4}
\end{figure*}

\begin{figure*}
\center
	\includegraphics[angle=0,trim={0.3cm 5.0cm 18.0cm 8.5cm},clip=true,width=0.3\linewidth,keepaspectratio]{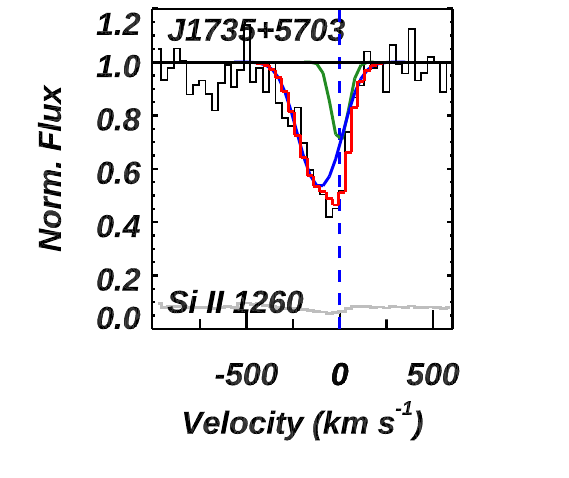}
	\includegraphics[angle=0,trim={0.3cm 5.0cm 18.0cm 8.5cm},clip=true,width=0.3\linewidth,keepaspectratio]{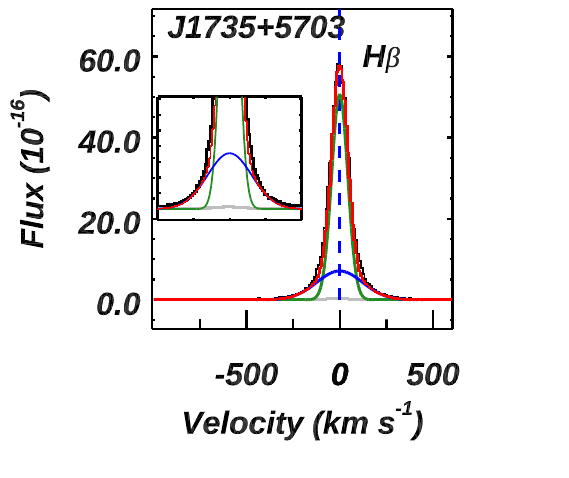}

\caption{\normalfont{Same as Figure \ref{fig:spectra-1} for the last object in our sample.} }
\label{fig:spectra-5}
\end{figure*}

We show fits to UV-absorption and optical-emission lines for all galaxies in our sample in Figures \ref{fig:spectra-1} to \ref{fig:spectra-5}.

\clearpage

\section{SALT Models of Bi-conical Galactic Outflows in Absorption and Emission Lines}
\label{app:sec:SALT}

\begin{figure}
\center

	\includegraphics[page = 1,angle=0,trim={0.5cm 0.2cm 0.1cm 0.0cm},clip=true,width=0.5\linewidth,keepaspectratio]{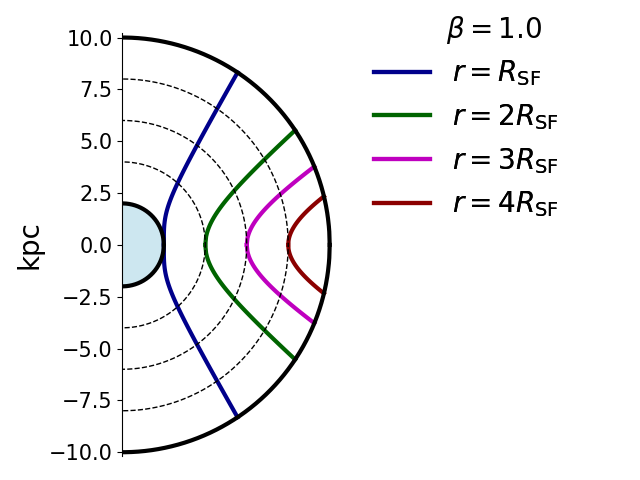}

\caption{\normalfont{ Various intersections (i.e., $\Gamma$) of the surface at constant observed velocity in the outflow with the plane perpendicular to the plane of the sky. We also adopt the $\beta = 1$ velocity field as specified in Equation \ref{eq:v_r}.}  }
\label{fig:Gamma}
\end{figure}

In Semi-analytical Line Transfer (SALT) model, we begin by computing the surfaces of constant observed velocity in the outflow, given the velocity distribution defined by Equation~\ref{eq:v_r}.  Following \cite{Carr23}, it is sufficient to only consider the curve defined by the intersection of the surface with the plane perpendicular to the plane of the sky, i.e., we have:

\begin{eqnarray}
\Gamma(s,\xi) =  
 \resizebox{0.45\textwidth}{!}{$ \left( \frac{x}{y} \frac{1}{1-y^{1/\beta}}, \left[\left(\frac{1}{1-y^{1/\beta}}\right)^2 - \left(\frac{1}{1-y^{1/\beta}}\right)^2 \frac{x^2}{y^2}\right]^{1/2} \right) $},
\end{eqnarray}
where $y = v/v_{\infty}$ and $x = v_{\rm obs}/v_{\infty}$.  We compute several curves for a spherical outflow in Figure~\ref{fig:Gamma} for $\beta = 1.0$.  The velocity field behaves similarly to a power law with index less than one (see the appendix of \citealt{Carr18}).

Next, we compute the Sobolev optical depth \citep{Castor70,Lamers99} from Equation 18 in \cite{Carr23}.  Keeping the same form for the density field as Equation \ref{eq:n_r}, $n=n_0\left(r/R_{\rm SF}\right)^{\gamma}$, we have 

\begin{eqnarray}
    \tau_S = \tau_0 \frac{\left(1-y^{\left(1/\beta\right)}\right)^{\gamma-1}y^{-1}}{1+\left[\beta y^{-1/\beta}-(\beta+1)\right]\left(\frac{x}{y}\right)^2},
\end{eqnarray}
where
\begin{eqnarray}
\tau_0 = \frac{\pi e^2}{m_e c}f\lambda \frac{R_{\rm SF}}{v_{\infty}}n_0.
\end{eqnarray}

We can now calculate the absorption profile resulting from the removal of photons by the surfaces of constant observed velocity.  Adapting Equation 7 in \cite{Carr23}, we have the normalized intensity of absorption lines as:

\begin{eqnarray}
    \frac{I_{\rm abs}}{I_0} = 
     \frac{F_{\lambda}}{F_{c,\lambda}} - \frac{F_{\lambda}}{F_{c,\lambda}} \int_x^{y_1} 2 \frac{x^2\beta-y^{1/\beta}(x^2(\beta+1)+y^2)}{\beta y^3 (y^{1/\beta}-1)^3} 
    \times \left(1-e^{-\tau_S}\right)dy
\end{eqnarray}
where $y_1$ represents the maximum radial velocity a shell can have and still contribute to continuum absorption.  It can be obtained from the relation:
\begin{eqnarray}
x^2 = y_1^2 -y_1^2\left(1-y_1^{1/\beta}\right)^2.
\end{eqnarray}
In a similar fashion, we can compute the emission component by adapting Equations 16 and 17 from \cite{Carr23}.  We have 
\begin{eqnarray}
    \frac{I_{\rm em, blue}}{I_0} = \int_x^1 f_g \frac{dy}{2y}  \nonumber
     \int_{y\cos{\Theta_c}}^y \frac{F(x^{\prime})}{F_c(x)} \frac{2}{\beta} \left[\frac{{x^{\prime}}^2\beta - y^{1/\beta} ({x^{\prime}}^2(\beta+1)+y^2)}{\beta y^3 (y^{1/\beta}-1)^3}\right]
     (1-e^{-\tau}) dx^{\prime}, \nonumber
\end{eqnarray}
and 
\begin{eqnarray}
    \frac{I_{\rm em, red}}{I_0} = \int_{y_1}^1 f_g \frac{dy}{2y}  \nonumber
     \int_{y\cos{\Theta_c}}^y \frac{F(x^{\prime})}{F_c(x)} \frac{2}{\beta} \left[\frac{{x^{\prime}}^2\beta - y^{1/\beta} ({x^{\prime}}^2(\beta+1)+y^2)}{\beta y^3 (y^{1/\beta}-1)^3}\right]
     (1-e^{-\tau}) dx^{\prime}, \nonumber
\end{eqnarray}
where
\begin{eqnarray}
    \Theta_{C} = \arcsin{([1-y^{1/\beta}]^{-1})}
\end{eqnarray}
is the angle subtended by a shell of intrinsic velocity $y$ along the continuum.  Note that we have included the geometric scale factor, $f_g$, from Equation 21 in \cite{Carr23}, which accounts for the geometry of the bi-cone and can be adapted to the new velocity field in the obvious way.  

Next, we develop an emission component emanating from the bi-cone e.g., \citealt{Scuderi92}.  It will be convenient, however, to first solve for the case of a spherical outflow and later scale to the profile of a bi-cone using $f_g$.  We assume an electron density field in the form of:

\begin{eqnarray}
    n_{e} = n_{0,e}\left(\frac{R_{\rm SF}}{r}\right)^{\epsilon},
\end{eqnarray}
where $n_{0,e}$ is the density of electrons at $R_{\rm SF}$ and $\epsilon$ is the power law index. In our calculations, we take $\epsilon = \gamma$ as in Equation \ref{eq:n_r}, but we keep them in different letters in the following equations for interpretation purpose. We adopt the same volume villing factor as in Equation \ref{eq:n_r}.

At a temperature around $10^4\rm\ K$ the contribution to the intrinsic $\rm H\alpha$ emission will be dominated by recombination.  Thus, we compute the intrinsic luminosity of $\rm H\alpha$ emission from a spherical shell of radius, $r$, and thickness, $dr$, as
\begin{eqnarray}
    L_{\rm shell} = h\nu n_e n_{\rm H^0} q_{\rm rec.}^{\rm H\alpha} 4\pi r^2 dr, 
\end{eqnarray}
where $h$ is Plank's constant and $q_{\rm rec.}^{\rm H\alpha}$ is the H$\alpha$ recombination coefficient. Plugging in the definitions of the various density fields and rewriting the expression in terms of the $y$, we have 
\begin{eqnarray}
L_{\rm shell} &=& \frac{4\pi h\nu n_{\rm e,0}n_{\rm H^0,0} q_{\rm rec.}^{\rm H\alpha} R_{SF}^3}{\beta}\\ \nonumber
&& \left(1-y^{1/\beta}\right)^{\gamma+\epsilon-4}y^{\frac{1-\beta}{\beta}}dy.
\end{eqnarray}  

To compute the resulting line profile in terms of the observed velocities, $v_{\rm obs}$, we use the same emission band contour theory developed by \cite{Beals31} and used by \cite{Carr18,Carr23}.  In this context, the flux, $dF_{BC}$, emitted by an arbitrary band contour on the shell (see Figure 4 from \citealt{Carr18}) becomes
 \begin{eqnarray}
 dF_{BC} = L_{\rm{shell}}\frac{dv_{\rm{obs}} }{4\pi r_{\infty}^2 2v},
 \label{emitted_flux}
 \end{eqnarray} 
where $r_{\infty}$ is the distance from the shell to the observer and $v$ is the radial velocity of the shell.  Writing this expression in terms of $y$ and integrating over the shell, we compute the normalized blue emission profile for a bi-conical outflow as

\begin{eqnarray}
\frac{F_{\rm int.em.,blue}}{F_0} =
\int_{x}^{1}f_g \frac{h\nu n_{\rm e,0}n_{\rm H^0,0} q_{\rm rec.}^{\rm H\alpha} R_{SF}^3}{F_c2\beta r_{\infty}^2}\left(1-y^{1/\beta}\right)^{\gamma+\epsilon-4}y^{\frac{1-\beta}{\beta}}dy
\end{eqnarray}
and the normalized red emission profile as
\begin{eqnarray}
    \frac{F_{\rm int.em.,red}}{F_0} = 
    \int_{y_1}^{1}f_g \frac{h\nu n_{\rm e,0}n_{\rm H^0,0} q_{\rm rec.}^{\rm H\alpha} R_{SF}^3}{F_c2\beta r_{\infty}^2}\left(1-y^{1/\beta}\right)^{\gamma+\epsilon-4}y^{\frac{1-\beta}{\beta}}dy
\end{eqnarray}
Note that, once again, we scale by $f_g$ to collapse the solution for a spherical outflow to that of a bi-cone. The results of these profiles given M 82 parameters are shown in Figures \ref{fig:SALT_model_profile}.

\section{Additional Measured Parameters}
\label{app:sec:ancillary}
We summarize other measurements for each galaxy in Table \ref{tab:ancillary}.

\begin{table*}
	\centering
	\caption{Other Measurements for Galaxies in our Sample}
	\label{tab:ancillary}

         \begin{tabular}{lccccc} 
		\hline

		 ID & Log \Mstar\ &	Log SFR  & Emi. Log \Rout\ & Abs. Log \Rout\ & \EBVint  \\
		\hline
	       &  (\Msun)   &   (\Msun\ yr$^{-1}$)         & (kpc) & (kpc) & \\
		\hline
J0055--0021 & 9.7$^{+0.1}_{-0.1}$ & 1.1$^{+0.1}_{-0.1}$ & $<$ --0.2 & 0.6$^{+0.2}_{-0.2}$ & 0.089\\
J0150+1308 & 8.7$^{+0.1}_{-0.1}$ & 0.8$^{+0.1}_{-0.1}$ & 0.2$^{+0.07}_{-0.08}$ & 0.7$^{+0.2}_{-0.2}$ & 0.190\\
J0808+1728 & 8.5$^{+0.1}_{-0.1}$ & --0.5$^{+0.1}_{-0.1}$ & $<$ --0.4 & 0.04$^{+0.01}_{-0.01}$ & 0.000\\
J0815+2156 & 8.6$^{+0.3}_{-0.2}$ & 1.2$^{+0.1}_{-0.1}$ & --0.2$^{+0.09}_{-0.1}$ & 0.5$^{+0.1}_{-0.1}$ & 0.187\\
J0851+5840 & 8.1$^{+0.2}_{-0.1}$ & 0.7$^{+0.1}_{-0.1}$ & --0.2$^{+0.08}_{-0.10}$ & 0.6$^{+0.2}_{-0.2}$ & 0.001\\
J0911+1831 & 9.0$^{+0.1}_{-0.2}$ & 1.2$^{+0.1}_{-0.1}$ & 0.3$^{+0.07}_{-0.09}$ & 0.7$^{+0.2}_{-0.2}$ & 0.094\\
J0926+4427 & 9.8$^{+0.2}_{-0.1}$ & 1.2$^{+0.4}_{-0.1}$ & 0.009$^{+0.06}_{-0.07}$ & 0.7$^{+0.2}_{-0.2}$ & 0.230\\
J0942+3547 & 7.5$^{+0.1}_{-0.1}$ & --1.4$^{+0.1}_{-0.1}$ & --0.2$^{+0.06}_{-0.07}$ & --0.02$^{+0.007}_{-0.007}$ & 0.038\\
J0942+0928 & 7.5$^{+0.1}_{-0.1}$ & --0.8$^{+0.1}_{-0.1}$ & --0.8$^{+0.07}_{-0.08}$ & $>$ 0.01 & 0.253\\
J0944--0038 & 6.7$^{+0.2}_{-0.1}$ & --1.9$^{+0.1}_{-0.1}$ & --0.7$^{+0.06}_{-0.07}$ & $>$ --0.2 & 0.000\\
J1024+0524 & 7.7$^{+0.1}_{-0.2}$ & --0.6$^{+0.2}_{-0.3}$ & --0.1$^{+0.06}_{-0.07}$ & 0.4$^{+0.1}_{-0.1}$ & 0.054\\
J1044+0353 & 7.8$^{+0.1}_{-0.1}$ & --0.8$^{+0.1}_{-0.1}$ & $<$ --1.1 & 0.004$^{+0.001}_{-0.001}$ & 0.000\\
J1053+5237 & 8.9$^{+0.1}_{-0.1}$ & 1.4$^{+0.3}_{-0.1}$ & --0.3$^{+0.2}_{-0.2}$ & 0.7$^{+0.2}_{-0.2}$ & 0.213\\
J1129+2034 & 6.7$^{+0.2}_{-0.1}$ & --1.3$^{+0.1}_{-0.1}$ & \dots & --0.3$^{+0.1}_{-0.1}$ & 0.586\\
J1133+6513 & 9.1$^{+0.2}_{-0.2}$ & 1.1$^{+0.1}_{-0.1}$ & 0.3$^{+0.07}_{-0.09}$ & 0.8$^{+0.2}_{-0.2}$ & 0.065\\
J1137+3524 & 8.9$^{+0.2}_{-0.2}$ & 1.3$^{+0.4}_{-0.1}$ & 0.2$^{+0.07}_{-0.09}$ & 0.7$^{+0.2}_{-0.2}$ & 0.199\\
J1148+2546 & 8.6$^{+0.1}_{-0.1}$ & --0.2$^{+0.1}_{-0.1}$ & --0.2$^{+0.07}_{-0.08}$ & $>$ 0.5 & 0.000\\
J1155+5739 & 7.6$^{+0.1}_{-0.1}$ & --0.9$^{+0.1}_{-0.1}$ & --0.3$^{+0.06}_{-0.07}$ & 0.06$^{+0.02}_{-0.02}$ & 0.001\\
J1200+2719 & 7.9$^{+0.1}_{-0.2}$ & 0.7$^{+0.1}_{-0.1}$ & --0.2$^{+0.08}_{-0.1}$ & 0.5$^{+0.1}_{-0.1}$ & 0.133\\
J1215+2038 & 5.9$^{+0.2}_{-0.1}$ & --2.2$^{+0.1}_{-0.1}$ & \dots & --0.5$^{+0.2}_{-0.2}$ & 0.000\\
J1219+1526 & 9.2$^{+0.2}_{-0.2}$ & 1.6$^{+0.1}_{-0.1}$ & --0.1$^{+0.06}_{-0.07}$ & 0.5$^{+0.1}_{-0.1}$ & 0.214\\
J1222+0434 & 5.3$^{+0.1}_{-0.1}$ & --1.8$^{+0.1}_{-0.1}$ & \dots & $>$ --0.3 & 0.191\\
J1226+0415 & 7.9$^{+0.1}_{-0.1}$ & 0.5$^{+0.2}_{-0.3}$ & --0.1$^{+0.08}_{-0.09}$ & 0.5$^{+0.2}_{-0.2}$ & 0.000\\
J1230+1202 & 6.5$^{+0.1}_{-0.1}$ & --2.5$^{+0.1}_{-0.1}$ & \dots & $>$ --0.3 & 0.000\\
J1244+0215 & 8.8$^{+0.1}_{-0.2}$ & 1.7$^{+0.4}_{-0.1}$ & 0.006$^{+0.10}_{-0.1}$ & 0.9$^{+0.3}_{-0.3}$ & 0.275\\
J1248+1234 & 9.2$^{+0.3}_{-0.2}$ & 1.5$^{+0.3}_{-0.2}$ & 0.05$^{+0.06}_{-0.07}$ & 0.7$^{+0.2}_{-0.2}$ & 0.110\\
J1311--0038 & 7.7$^{+0.1}_{-0.1}$ & 0.8$^{+0.1}_{-0.1}$ & --0.5$^{+0.06}_{-0.07}$ & 0.3$^{+0.09}_{-0.09}$ & 0.191\\
J1416+1223 & 9.2$^{+0.5}_{-0.2}$ & 1.1$^{+0.1}_{-0.1}$ & $<$ --0.2 & 0.4$^{+0.1}_{-0.1}$ & 0.306\\
J1424+4216 & 9.5$^{+0.1}_{-0.1}$ & 1.4$^{+0.1}_{-0.1}$ & $<$ --0.07 & 0.5$^{+0.1}_{-0.1}$ & 0.194\\
J1448--0110 & 8.2$^{+0.1}_{-0.1}$ & --0.5$^{+0.1}_{-0.1}$ & --0.4$^{+0.07}_{-0.08}$ & 0.09$^{+0.03}_{-0.03}$ & 0.129\\
J1457+2232 & 8.6$^{+0.1}_{-0.2}$ & 1.2$^{+0.1}_{-0.1}$ & --0.1$^{+0.09}_{-0.1}$ & 0.6$^{+0.2}_{-0.2}$ & 0.080\\
J1509+3731 & 8.3$^{+0.1}_{-0.1}$ & --0.2$^{+0.1}_{-0.1}$ & --0.2$^{+0.07}_{-0.08}$ & 0.1$^{+0.04}_{-0.04}$ & 0.093\\
J1735+5703 & 8.6$^{+0.1}_{-0.1}$ & 0.07$^{+0.1}_{-0.1}$ & --0.5$^{+0.09}_{-0.1}$ & 0.5$^{+0.1}_{-0.1}$ & 0.001\\

            \hline
            \hline
            
	\multicolumn{6}{l}{%
  	\begin{minipage}{12cm}%
	Note. --\\
    	\textbf{(1)}\ \ From left to right, we list the estimated stellar mass (Section \ref{sec:SED}), star-forming rate (Section \ref{sec:ancillary}), outflow sizes (\Rout) measured from emission lines (Section \ref{sec:size}), \Rout\ measured from absorption lines (Section \ref{sec:size}), and intrinsic dust extinction (Section \ref{sec:ancillary}) for each galaxy.
  	\end{minipage}%
	}\\
	\end{tabular}
	\\ [0mm]
	
\end{table*}

\bibliography{main}{}
\bibliographystyle{aasjournal}



\end{document}